\documentclass[final,3p,times]{elsarticle}

\usepackage{amsmath}
\usepackage{amssymb}
\usepackage{subcaption}
\usepackage{hyperref}
\hypersetup{
     colorlinks   = true,
     citecolor    = blue,
     linkcolor    = blue}
\usepackage{mathtools} 
\usepackage{nomencl} 
\makenomenclature

\usepackage{etoolbox}
\renewcommand\nomgroup[1]{%
  \item[\bfseries
  \ifstrequal{#1}{C}{Coordinates}{%
  \ifstrequal{#1}{V}{Vector spaces}{%
  \ifstrequal{#1}{B}{Basis sets}{%
  \ifstrequal{#1}{N}{Number of basis polynomials}{
  }}}}%
]}

\renewcommand{\v}[1]{\ensuremath{\mathbf{#1}}} 
\newcommand{\gv}[1]{\ensuremath{\mbox{\boldmath$ #1 $}}} 
\newcommand{\uv}[1]{\ensuremath{\mathbf{\hat{#1}}}} 
\newcommand{\abs}[1]{\left| #1 \right|} 
\newcommand{\avg}[1]{\left< #1 \right>} 
\renewcommand{\d}[2]{\frac{d #1}{d #2}} 
\newcommand{\pd}[2]{\frac{\partial #1}{\partial #2}} 
\newcommand{\pdd}[2]{\frac{\partial^2 #1}{\partial #2^2}} 
\newcommand{\grad}[1]{\nabla #1} 
\renewcommand{\div}[1]{\nabla \cdot #1} 
\newcommand{\curl}[1]{\nabla \times #1} 
\let\baraccent=\= 
\renewcommand{\=}[1]{\stackrel{#1}{=}} 
\newcommand{\gradperp}[1]{{{\nabla}_{\perp}} {#1}} 

\newcommand{\vx}{\v{x}}
\newcommand{\vv}{\v{v}}

\newcommand{\vB}{\v{B}}

\newcommand{\cdim}{d_x}

\newcommand{\pdim}{d_p}
\newcommand{\dx}{\mathrm{d}x}

\newcommand{\dv}{\mathrm{d}v}
\newcommand{\dvpar}{\mathrm{d}\vpar}
\newcommand{\dmu}{\mathrm{d}\mu}
\newcommand{\dvv}{\mathrm{d}\v{v}}

\newcommand{\dvR}{\mathrm{d}\v{R}}

\newcommand{\vperp}{v_{\perp}}
\newcommand{\vpar}{v_{\parallel}}
\newcommand{\upar}{u_{\parallel}}
\newcommand{\vt}{v_{t}}
\newcommand{\uparsr}{u_{\parallel sr}}
\newcommand{\vtsr}{v_{t,sr}}
\newcommand{\jacobP}{\mathcal{J}}
\newcommand{\jacobGeo}{J_x}
\newcommand{\jacobVel}{J_v}
\newcommand{\jacobTot}{J_T}
\newcommand{\jacobTots}[1]{J_{T#1}}
\newcommand{\vBs}{\v{B^*}}
\newcommand{\Bspar}{B_\parallel^*}
\newcommand{\bhat}{\uv{b}}
\newcommand{\kperp}{k_\perp}
\newcommand{\kpar}{k_\parallel}
\newcommand{\cmag}{\mathcal{C}}
\newcommand{\dualv}[1]{\v{e^{#1}}}
\newcommand{\tangv}[1]{\v{e_{#1}}}
\newcommand{\xdoti}[1]{\dot{R}^{#1}}
\newcommand{\vpardot}{\dot{\vpar}}
\newcommand{\xdotiSurf}[1]{\dot{R}^{#1}_{\pm}}
\newcommand{\vpardotSurf}{\dot{v}_{\parallel}^{\pm}}
\newcommand{\vpardotSurfj}{\dot{v}_{\parallel j}^{\pm}}

\newcommand{\Tpare}{T_{\parallel e}}
\newcommand{\Tperpe}{T_{\perp e}}
\newcommand{\Tpari}{T_{\parallel i}}
\newcommand{\Tperpi}{T_{\perp i}}

\newcommand{\Dx}{\Delta x}

\newcommand{\Dvpar}{\Delta\vpar}
\newcommand{\Dmu}{\Delta\mu}

\newcommand{\Dt}{\Delta t}
\newcommand{\Nx}{N_x}
\newcommand{\Ny}{N_y}
\newcommand{\Nz}{N_z}
\newcommand{\Nvpar}{N_{\vpar}}
\newcommand{\Nmu}{N_\mu}

\newcommand{\Lz}{L_z}

\newcommand{\Nxi}[1]{N_{#1}}
\newcommand{\Nperp}[1]{N_{\perp #1}}

\newcommand{\pb}[1]{\psi^{(#1)}}
\newcommand{\pbpm}[1]{\psi^{(#1)\pm}}

\newcommand{\dgb}[1]{\varphi^{(#1)}}
\newcommand{\pbSurf}[1]{\xi^{(#1)}}
\newcommand{\febz}[1]{u^{(#1)}}
\newcommand{\febxy}[1]{w^{(#1)}}
\newcommand{\nodb}[1]{\Lambda^{(#1)}}

\newcommand{\dvx}{\mathrm{d}\v{x}}
\newcommand{\cvpar}{\eta}
\newcommand{\cmu}{\kappa}
\newcommand{\xlog}[1]{\chi^{#1}}
\newcommand{\vxlog}{\gv{\xlog{}}}
\newcommand{\vlog}{\zeta}
\newcommand{\mulog}{\lambda}
\newcommand{\Dxi}[1]{\Delta x^{#1}}

\newcommand{\Dcvpar}{\Delta\cvpar}
\newcommand{\Dcmu}{\Delta\cmu}
\newcommand{\dxlog}[1]{\mathrm{d}\xlog{#1}}
\newcommand{\dvxlog}{\mathrm{d}\vxlog}
\newcommand{\dcvpar}{\mathrm{d}\cvpar}
\newcommand{\dcmu}{\mathrm{d}\cmu}
\newcommand{\dvlog}{\mathrm{d}\vlog}
\newcommand{\dmulog}{\mathrm{d}\mulog}

\newcommand{\numbP}{N_b}
\newcommand{\numbPSurf}[1]{N_{b\backslash\{#1\}}}
\newcommand{\numbC}{N_{b,\vx}}

\newcommand{\numbFExy}{N_{b,\overline{x^1x^2}x^3}}
\newcommand{\numbFEz}{N_{b,x^1x^2\overline{x^3}}}
\newcommand{\JTf}{F}

\newcommand{\JTfSurf}{\JTf_{\pm}}
\newcommand{\JTfupwindSurf}{\widehat{\JTf}_{\pm}}
\newcommand{\JTfDvparpUpwindSurf}{\widehat{\frac{\JTfSurf}{\vpar'}}}
\newcommand{\Ghat}{\widehat{G}}
\newcommand{\intnodes}{\v{c}}
\newcommand{\gnodes}{\v{g}}
\newcommand{\hnodes}{\v{h}}
\newcommand{\omegaCFL}{\omega_{\mathrm{CFL}}}
\newcommand{\vparmax}{v_{\parallel\mathrm{max}}}
\newcommand{\mumax}{\mu_\mathrm{max}}

\newenvironment{eqnal}{\equation\aligned}{\endaligned\endequation}
\newcommand{\ignore}[1]{}  

\begin{document}

\begin{frontmatter}



\title{Conservative velocity mappings for  discontinuous Galerkin kinetics}

\author[pppl]{M. Francisquez}
\affiliation[pppl]{organization={Princeton Plasma Physics Laboratory},
            city={Princeton},
            postcode={08540}, 
            state={NJ},
            country={USA}}
            
\author[helm]{P. Cagas}
\affiliation[helm]{organization={Helmholtz-Zentrum Dresden-Rossendorf (HZDR)},
            city={Dresden},
            postcode={D-01328}, 
            country={Germany}}

\author[uta]{A. Shukla}
\affiliation[uta]{organization={Institute for Fusion Studies, University of Texas at Austin},
            city={Austin},
            postcode={78712}, 
            state={TX},
            country={USA}}
            
\author[pppl]{J. Juno}

\author[pppl]{G. W. Hammett}

\begin{abstract}
Continuum computational kinetic plasma models evolve the distribution function of a plasma species $f_s$ on a phase-space grid over time. In many problems of interest the distribution function has limited extent in velocity space; hence, using a uniform, highly refined mesh would be costly and slow. Nonuniform velocity grids can reduce the computational cost by placing more degrees of freedom where $f_s$ is appreciable and fewer where it is not. In this work we introduce a first-of-its kind discontinuous Galerkin approach to nonuniform velocity-space discretization using mapped velocity coordinates. This new method is presented in the context of a gyrokinetic model used to study magnetized plasmas. We create discretizations of collisionless and collisional terms using mappings in a way that exactly conserves particles and energy. Numerical tests of such properties are presented, and we show that this new discretization can reproduce earlier gyrokinetic simulations using grids with up to 6-60 times fewer cells and 22X-60X speed-ups depending on dimensionality, geometry and plasma parameters. 
\end{abstract}







\end{frontmatter}



\section{Introduction} \label{sec:intro}

Modeling high-temperature plasmas can benefit from the use of a kinetic model by which the distribution function $f_s$ of a species $s$ is prescribed throughout position and velocity, i.e. phase space. A rich collection of kinetic models exists, with the Vlasov or Boltzmann equations~\cite{Kardar2007} being some of the most well known. Although constrained by physical principles (e.g. Liouville's theorem), the variation of $f_s$ over $(\v{x},\v{v})$ can be tremendously diverse and complex, exhibiting wildly different profiles from one system to another or from one parameter regime to another. However, something quite ubiquitous is that $f_s$ tends to be of appreciable magnitude only in restricted regions of phase space. Consider one of the most basic distributions in local thermodynamic equilibrium; the Maxwell-Boltzmann distribution decays exponentially with the square of the velocity divided by the thermal speed, so for velocities greater than a few thermal speeds (e.g. 3) $f_s$ is orders of magnitude smaller and possibly negligible.

Despite $f_s$ being localized in phase space, kinetic models are difficult to solve analytically, and thus a large body of numerical methods and codes exists to approximate their solutions~\cite{Filbet2003,Palmroth2018,Nishikawa2021,Califano2023}. Many of these computational approaches are either Lagrangian (e.g. particle-in-cell) or Eulerian, the latter discretizing the kinetic equation on a multidimensional phase space grid in both position and velocity space, and updating $f_s$ at discrete time steps. The discrete kinetic operator can be constructed with a variety of methods, such as finite volume~\cite{vonAlfthan2014,Vogman2018}, finite difference~\cite{Schaeffer1998,Idomura2007}, semi-Lagrangian~\cite{Zaki1988}, or spectral~\cite{Watanabe2004} methods, to name a few. These methods typically use a velocity space mesh that extends sufficiently far such that truncation of the discrete $f_s$ is stable and justified by its smallness there. At the same time, said mesh must be sufficiently fine to reach the desired level of accuracy, resolve sharp gradients, or remain stable. Many continuum kinetic codes meet these constraints with a uniformly spaced velocity grid, effectively employing the same density of degrees of freedom (DOF) throughout. This commonplace strategy is, albeit a simple and acceptable first step, wasteful because one should only need to place a large density of DOF where $f_s$ is appreciable or where strong gradients arise. Nonuniform grids attempt to do precisely that, using more DOF in those parts of the domain that require them and fewer in others.

Nonuniform velocity space grids have been employed by kinetic codes in a variety of contexts before. An early use of geometric velocity-space grids (where cell length increases by a multiplicative constant) facilitated the solution of the bounce-averaged Fokker-Planck equation~\cite{McCoy1981,Killeen1986}. Around the same time, a simple grid with two different cell lengths was used to solve a quasilinear Fokker-Planck equation to study the evolution of the distribution function under radio frequency heating~\cite{Hammett1986}. A similar two-cell length grid was used 30 years later to study kinetic electrostatic electron nonlinear (KEEN) waves~\cite{Afeyan2014}. This technique is still actively developed to, for example, enhance the capabilities of semi-Lagrangian 1D1V kinetic codes~\cite{Bourne2023} and resolve low energies near the wall in the presence of electron-neutral collisions~\cite{Mijin2021}. We should mention that there are other approaches to nonuniform velocity discretization which accomplish the same objective. One is to use a quadrature rule (e.g. Gauss-Laguerre) that places ordinates at unevenly spaced locations in order to maximize the accuracy of the underlying integrals (e.g. to compute moments of the distribution)~\cite{Barnes2010,Barnes2019}. Alternatively, a closely related method is to use a spectral method in velocity space such that, depending on the type and number of basis functions kept, some parts of velocity space are represented with greater accuracy~\cite{Candy2016,Mandell2018,Frei2020}. On the other hand, the number of DOF may be reduced by using sparse grids~\cite{Wang2016,Kormann2016,Schnake2024}. Finally, other techniques to reduce the number of DOF and localize them where phase-structures exist include unstructured meshes and adaptive mesh refinement~\cite{Xu2010,Hittinger2013,Wettervik2017}.

This work offers a new perspective on nonuniform velocity grids for Eulerian kinetic simulations employing the discontinuous Galerkin (DG) method. A DG discretization has a number of benefits~\cite{postECP2020}, including that it is local so it parallelizes well on distributed computers, has high arithmetic intensity for each byte fetched from memory which improves performance on graphical processing units (GPUs)~\cite{Klockner2009}, and supports tailored numerical fluxes at cell boundaries (as done in finite volume methods) to preserve certain properties such as positivity and monotonicity~\cite{Zhang2011}. We thus present a first-of-its-kind DG method to nonuniformly discretize velocity space with the following merits: {\bf a)} it supports user-defined coordinate mappings tailoring the effective grid to each application, {\bf b)} it conserves particles and energy exactly even with curvilinear position-space coordinates, and {\bf c)} it is implemented in a higher-dimensional (5D) GPU-accelerated code. We will describe this novel method in the context of a gyrokinetic model used to study magnetized plasmas (section~\ref{sec:model}), as well as sketch its use for a Vlasov-Poisson solver (\ref{sec:vlasov}). The conservative (gyrokinetic) DG algorithm is described in detail (section~\ref{sec:algo} and~\ref{sec:gk_fac}), including a proof of its conservative properties (\ref{sec:conserv}). We conclude with a series of numerical tests of increasing complexity to demonstrate its use in practical applications (section~\ref{sec:results}), and with a brief summary of this work and future possibilities (section~\ref{sec:conclusion}).


\section{Model} \label{sec:model}

We present a nonuniform velocity-space discretization via conservative discontinuous Galerkin (DG) mappings. As a test-bed for this method, we use a long-wavelength, electrostatic, full-$f$ gyrokinetic model~\cite{Dubin1983,Sugama2000,Brizard2007}. The same technique can be used for an Eulerian DG Vlasov solver~\cite{Juno2018}, as described in~\ref{sec:vlasov}. The gyrokinetic model we focus on is used in several major codes studying magnetized fusion plasmas~\cite{Idomura2009,Dorf2018,Michels2021,gkeyllWeb}; it governs the evolution of the guiding center distribution function $f_s(t,\v{R},\vpar,\mu)$ of species $s$ (with charge $q_s$ and mass $m_s$) in a plasma magnetized by the equilibrium magnetic field $\vB=B\,\bhat$. The time-evolution of $f_s$ as function of guiding center position $\v{R}$, velocity parallel to the magnetic field $\vpar=\bhat\cdot\vv$, and the magnetic moment $\mu=m_s\vperp^2/(2B)$, is given by
\begin{equation} \label{eq:gkeq}
    \pd{\jacobP f_s}{t} + \div{\jacobP \v{\dot{R}} f_s} + \pd{}{\vpar}\jacobP \dot{\vpar} f_s = \jacobP\mathcal{C}_s + \jacobP\mathcal{S}_s + \jacobP\mathcal{D}_s,
\end{equation}
where $\jacobP=\Bspar=\bhat\cdot\vBs$ is the Jacobian of the gyrocenter coordinates and $\vBs=\vB+(m_s\vpar/q_s)\curl{\bhat}$ is the effective magnetic field. The second and third (collisionless) terms on the left-hand side of equation~\ref{eq:gkeq} contain phase-space advection of $f_s$ by the velocities $\v{\dot{R}}$ and $\dot{\vpar}$ (described in~\ref{sec:gk_fac}). The right-hand side contains the actions of collisions ($\mathcal{C}_s$), sources ($\mathcal{S}_s$) and diffusion ($\mathcal{D}_s$). Sources are typically used to account for plasma entering our domain of study, say due to transport from another region not modeled or due to external fueling. Cross-field transport is self-consistently modeled inside our domain by drifts and turbulence (the latter in 3D only). In the absence of anomalous cross-field transport, as is the case in 2D axisymmetric studies, one may use the diffusive term $\mathcal{D}_s$ as a model for such transport. Collisions include the effect of elastic collisions between charged species, and inelastic collisions and reactions (e.g. charge exchange, recombination, ionization, radiation).

We wish to solve equation~\ref{eq:gkeq} in curvilinear coordinates, specifically field-line following coordinates that have been a workhorse of turbulence modeling in the core of tokamaks~\cite{Beer1995} and stellarators~\cite{Maurer2020}, and can be used to study other laboratory devices such as linear machines~\cite{Shi2017} and mirrors~\cite{Francisquez2023}. Such a formulation, together with suitable boundary conditions, enables the solution of these equations in a small volume following a bundle of magnetic field lines (a flux-tube). A flux-tube domain reduces the number of degrees of freedom, supports larger time steps, and thus results in cheaper simulations than, say, global simulations with Cartesian grids. We thus introduce a field-aligned curvilinear coordinate system $\v{x}=\{x^i\}$ (here $i\in[1,\cdim]$ and $\cdim$ is the number of position-space dimensions) with a mapping to the Cartesian gyrocenter position $\v{R}$ given by $\v{R}(\v{x})$. In this new coordinate system we can write equation~\ref{eq:gkeq} as~\cite{Mandell2021thesis}
\begin{equation} \label{eq:gkeq_curv}
    \pd{\jacobP f_s}{t} + \frac{1}{\jacobGeo}\pd{}{x^i}\jacobGeo\jacobP\dualv{i}\cdot\v{\dot{R}} f_s + \pd{}{\vpar}\jacobP \dot{\vpar} f_s = \jacobP\mathcal{C}_s + \jacobP\mathcal{S}_s + \jacobP\mathcal{D}_s,
\end{equation}
where $\tangv{i}=\partial\v{R}/\partial x^i$ and 
$\dualv{i}$
are the tangent (covariant) and dual (contravariant) basis vectors of the $\v{x}$ coordinates. These vectors define the metric tensors $g_{ij}=\tangv{i}\cdot\tangv{j}$ and $g^{ij}=\dualv{i}\cdot\dualv{j}$, and $\jacobGeo=\mathrm{det}(g_{ij})^{1/2} = \mathrm{det}(g^{ij})^{-1/2}$ is the Jacobian~\cite{Dhaeseleer1991} of the $\v{R}(\v{x})$ map. Repeated indices imply summation, unless noted otherwise or explicit summation signs are used. 

\subsection*{Model with mapped velocity coordinates} 

We now map the physical velocity coordinates $(\vpar,\mu)$ from a new pair of computational coordinates $(\cvpar,\cmu)$:
\begin{eqnal} \label{eq:vmap}
    \vpar = \vpar(\cvpar) \qquad \mu = \mu(\cmu)
\end{eqnal}
These transformations can be provided by the user, meeting the specific requirements of their application, and can vary between species. But such mappings must be continuous, monotonically increasing, and univariate. The latter entails that these mappings allow contraction and dilation of each velocity dimension independently, which, while somewhat restrictive, is sufficiently useful in numerous applications. A similar approach was presented for a 1D1V finite difference Vlasov solver~\cite{Hesse2022}. 

When employing the new velocity variables $(\cvpar,\cmu)$, the gyrokinetic equation~\ref{eq:gkeq_curv} becomes
\begin{equation} \label{eq:gkeq_curv_vmap}
    \pd{\jacobP f_s}{t} + \frac{1}{\jacobGeo}\pd{}{x^i}\jacobGeo\jacobP\dualv{i}\cdot\v{\dot{R}} f_s + \frac{1}{\vpar'}\pd{}{\cvpar}\jacobP \dot{\vpar} f_s = \jacobP\mathcal{C}_s + \jacobP\mathcal{S}_s + \jacobP\mathcal{D}_s,
\end{equation}
where $\vpar'=\partial_{\cvpar}\vpar$ and $\mu'=\partial_{\cmu}\mu$. The advection speeds, $\dualv{i}\cdot\v{\dot{R}}$ and $\dot{\vpar}$, are now computed using (see equation~\ref{eq:app_characteristics_curvilinear})
\begin{eqnal} \label{eq:alpha_curv_axisym}
    \dualv{i}\cdot\v{\dot{R}} &= \frac{B^{*i}}{m_sB\vpar'}\pd{H_s}{\cvpar} + \frac{1}{q_sB}\frac{\epsilon^{ijk}}{\jacobGeo}b_j\pd{H_s}{x^k},
    \qquad
    \dot{\vpar} = -\frac{B^{*i}}{m_sB}\pd{H_s}{x^i},
\end{eqnal}
employing the particle Hamiltonian $H_s = m_s\vpar^2/2 + \mu B + q_s\phi$, the Levi-Civitta symbol $\epsilon^{ijk}$, the contravariant components of $\vBs$ ($B^{*i}$, equation~\ref{eq:app_vBs_contra}), and the covariant components of $\bhat$ ($b_j$, equation~\ref{eq:app_bhatcomps}). We also need the potential $\phi$ appearing in the Hamiltonian, which results from solving the quasineutrality equation (see equation~\ref{eq:app_gkPoissonCurvilinearSimp})
\begin{eqnal} \label{eq:gkPoissonCurvSimp}
    -\frac{1}{\jacobGeo}\pd{}{x^i}\jacobGeo\epsilon_\perp g^{ij}\pd{}{x^j}\phi = \sum_sq_sn_s,
    \qquad
    i,j\in\{1,2\},
\end{eqnal}
where $\epsilon_\perp = \sum_s m_sn_{s0}/B_0^2$ for some reference density $n_{s0}$ and magnetic field magnitude $B_0$.

The source and diffusion terms on the right-hand side of equation~\ref{eq:gkeq_curv_vmap} do not require modification due to the transformation in equation~\ref{eq:vmap} because they do not involve differential operators in $\vpar$-$\mu$. One only has to be aware that where $\vpar$-$\mu$ appear therein, these must be interpreted as functions of $\cvpar$-$\cmu$, respectively. The same is true for our ionization, charge-exchange and recombination models in $\mathcal{C}_s$~\cite{Bernard2022}. However, the line radiation model and the Dougherty operator for elastic collisions between charged species (equation~\ref{eq:app_colls}) must be transformed accordingly. The Dougherty operator is expressed in $\cvpar$-$\cmu$ coordinates as
\begin{eqnal} \label{eq:gklbo_vmap}
    \mathcal{C}_s^{\mathrm{el}} = \sum_r\nu_{sr}\,\left\{\frac{1}{\vpar'}\pd{}{\cvpar}\left[\left(\vpar-\uparsr\right)+\frac{\vtsr^2}{\vpar'}\pd{}{\cvpar}\right] + \frac{1}{\mu'}\pd{}{\cmu}2\mu\left(1+\frac{m\vt^2}{B\mu'}\pd{}{\cmu}\right)\right\}f_s,
\end{eqnal}
and the line radiation model is described in a separate publication~\cite{Roeltgen2025}.

\section{Algorithm} \label{sec:algo}

In this section we develop a conservative discontinuous Galerkin (DG) algorithm to discretize the long-wavelength gyrokinetic model presented in section~\ref{sec:model}. 
Of particular interest is a modal implementation of DG~\cite{Atkins1998,Hakim2020b,Mandell2020}, using relatively sparse orthonormalized basis functions that allow analytic evaluation of integrals, thus obviating the need for numerical quadratures and producing an alias-free method. We introduce the tessellation (mesh) $\mathcal{T}$ with cells $K_j\in\mathcal{T}$ labeled by $j=1,\dots,N$, with $N=\Nvpar\Nmu\prod_{i=1}^{\cdim}\Nxi{i}$ meaning the number of cells in a mesh with $(\Nxi{i},\Nvpar,\Nmu)$ cells along $(x^i,\cvpar,\cmu)$, respectively. In each cell we define a set of logical coordinates $(\gv{\xlog{}},\vlog,\mulog)\in[-1,1]^{(\cdim+2)}$ which are related to the computational coordinates $(\v{x},\cvpar,\cmu)$ via
\begin{eqnal} \label{eq:comp2log}
    x^{i} = x^i_j + \frac{\Dx^i}{2}\xlog{i}, \qquad
    \cvpar = \cvpar_j + \frac{\Dcvpar}{2}\vlog, \qquad
    \cmu = \cmu_j + \frac{\Dcmu}{2}\mulog,
\end{eqnal}
where $(\v{x}_j,\cvpar_j,\cmu_j)$ is the cell center of cell $K_j$ and $(\Delta\v{x},\Dcvpar,\Dcmu)$ are its cell lengths along each direction. We also introduce a hybrid space $\mathcal{H}_j^p(\vxlog,\vlog,\mulog)$ of polynomials $\pb{k}_j$ in the set of $(\vxlog,\vlog,\mulog)$ variables, of order up to $p$ in $(\vxlog,\mulog)$ coordinates and up to $\mathrm{max}(p,2)$ in $\vlog$, with cardinality $\abs{\mathcal{H}_j^p}=\numbP$. Such polynomials have compact support $\mathrm{supp}(\pb{k}_j(\v{x},\vpar,\mu)) = K_{j}$ restricted to each cell (i.e. $\pb{k}_j=0$ outside $K_j$), and are orthogonal and normalized such that for $\pb{\ell}_m\in\mathcal{H}^p_{m}$ one has
\begin{equation} \label{eq:orthonorm}
\int_{K_{j}}\dvxlog\,\dvlog\,\dmulog\,\pb{k}_j\pb{\ell}_m = \delta_{jm}\delta_{k\ell}.    
\end{equation}
For example, in one position ($\cdim=1$) and two velocity dimensions (1D2V) the $p=1$ set of polynomials is
\begin{eqnal} \label{eq:basis1x2v}
    \mathcal{H}^{1} \equiv \frac{1}{2^{3/2}} &\left\{1,\sqrt{3}\xlog{1},\sqrt{3}\vlog,\sqrt{3}\mulog,3\xlog{1}\vlog,3\xlog{1}\mulog,3\vlog\mulog,3^{3/2}\xlog{1}\vlog\mulog,3\sqrt{5}\left(\vlog^2-1/3\right), \right.\\
    &\left.~3^{1/2}\sqrt{5}\xlog{1}\left(\vlog^2-1/3\right),3^{3/2}\sqrt{5}\left(\vlog^2-1/3\right)\mulog,9\sqrt{5}\xlog{1}\left(\vlog^2-1/3\right)\mulog\right\}.
\end{eqnal}
We then use these polynomials as a basis set to represent our distribution function $f$ (we drop the species index $s$ for simplicity) in cell $K_j$ as
\begin{eqnal} \label{eq:f_discrete}
    f_j = \sum_{k=1}^{\numbP}f_j^{(k)}\pb{k}_j(\gv{\xlog{}},\vlog,\mulog).
\end{eqnal}
Note the compact support of $\psi^{(k)}_j$ allows us to write the distribution using a global expansion as well:
\begin{eqnal} \label{eq:f_discrete_global}
    f = \sum_{j=1}^{N}\sum_{k=1}^{\numbP}f_j^{(k)}\pb{k}_j(\gv{\xlog{}},\vlog,\mulog).
\end{eqnal}

We also introduce a DG basis set ($\dgb{k}_j$) using elements of space $\mathcal{V}_j^p(q^i)$ of polynomials of order $p$ in $q^i$ variables, whose compact support is cell $K_j$. The dimensionality of $\mathcal{V}_j^p(q^i)$ is the number $q^i$ variables, and its cardinality is denoted with $\abs{\mathcal{V}_j^p(q^i)}=N_{b,q^i}$. Note that if the set $\{q^i\}$ contains a single variable (i.e. a 1D space), then $N_{b,q^1}=p+1$. Let us then use the basis sets $\dgb{k}_j(\vlog)\in\mathcal{V}_j^p(\vlog)$ and $\dgb{k}_j(\mulog)\in\mathcal{V}_j^p(\mulog)$ to represent the velocity transformations in equation~\ref{eq:vmap} as:
\begin{eqnal} \label{eq:velmap_discrete}
    v_{\parallel j} = \sum_{k=1}^{p+1}v_{\parallel j}^{(k)}\dgb{k}_j(\vlog),
    \qquad
    \mu_{j} = \sum_{k=1}^{p+1}\mu_{j}^{(k)}\dgb{k}_j(\mulog).
\end{eqnal}
Although these formulas employ a DG basis, the expansion coefficients $v_{\parallel j}^{(k)}$ and $\mu_{j}^{(k)}$ must be computed such that the representations in equations~\ref{eq:velmap_discrete} are $C^0$ continuous. This continuous discretizations are accomplished by evaluating the $\vpar(\cvpar)$ and $\mu(\cmu)$ maps at Gauss-Lobatto nodes and performing a nodal-to-modal transformation in each cell.

Unlike $\vpar$, we represent the squared of the parallel velocity using a $p\geq2$ basis, $\vartheta^{(k)}_j\in\mathcal{V}_j^{\max(p,2)}(\vlog)$:
\begin{equation}
    v_{\parallel j}^2 = \sum_{k=1}^{\max(p,2)+1}v_{\parallel j}^{2,(k)}\vartheta^{(k)}_j(\vlog).
\end{equation}
Once the discrete representation of $\vpar$ is established, i.e. the $v_{\parallel j}^{(k)}$ coefficients in equation~\ref{eq:velmap_discrete} are known, one can compute the coefficients of $v_{\parallel j}^2$ simply by weak multiplication with the $\vartheta^{(k)}_j$ basis:
\begin{equation}
    v_{\parallel j}^{2,(k)} = \int_{-1}^{1}\dvlog\,\vartheta^{(k)}_j v_{\parallel j}\, v_{\parallel j}.
\end{equation}
Our use of a $p=1$ hybrid basis in phase space, i.e. a basis that is linear in the $\v{x}$ and $\mu$ dimensions but quadratic in $\vpar$ in equations~\ref{eq:basis1x2v}-\ref{eq:f_discrete_global}, and a quadratic basis for $\vpar^2$ helps us avoid some of the complexities needed to conserve energy exactly in the discrete collision operator arising from the use of a purely linear basis~\cite{Francisquez2022,Hakim2020}.

Note that for $p=1$ the use of a piecewise linear representation of $\vpar(\cvpar)$ and $\mu(\cmu)$ (equation~\ref{eq:velmap_discrete}) means that the rate of change in $\vpar$ and $\mu$ with respect to their computational coordinates ($\vpar'$ and $\mu'$), appearing in equations~\ref{eq:gkeq_curv_vmap}-\ref{eq:gklbo_vmap}, are cellwise constants:
\begin{eqnal}
    v_{\parallel j}' = \pd{v_{\parallel j}}{\cvpar} = \frac{2}{\Dcvpar}\pd{v_{\parallel j}}{\vlog} = \frac{\sqrt{6}}{\Dcvpar}v_{\parallel j}^{(1)}, \qquad
    \mu'_j = \pd{\mu_j}{\cmu} = \frac{2}{\Dcmu}\pd{\mu_j}{\mulog} = \frac{\sqrt{6}}{\Dcmu}\mu_j^{(1)}.
\end{eqnal}
In principle one can use a $p\geq2$ basis to represent $\vpar(\cvpar)$ and $\mu(\cmu)$, but in that case $\vpar'$ and $\mu'$ are no longer cellwise constants and the complexity and cost of the algorithm grows. Furthermore, while it may seem that the $p=1$ choice for the velocity maps limits our accuracy, we believe that accuracy is dominated by the basis order used to represent $f_s$. However we did not quantify such statement and left this task for future work.

Given the definitions thus far, we can proceed to discretize the collisionless and collisional terms in our full-$f$ gyrokinetic equation~\ref{eq:gkeq_curv_vmap}. 

\subsection{Dicrete collisionless terms} \label{sec:algo_collisionless}

In this section we discretize the collisionless terms in equation~\ref{eq:gkeq_curv_vmap} and thus temporarily set its right-hand side to zero. The first step is to name the Jacobian of the transformation in equation~\ref{eq:vmap}:
\begin{equation}
    \jacobVel = \vpar'\mu',
\end{equation}
define a ``total'' Jacobian
\begin{equation}
    \jacobTot = \jacobGeo\jacobVel\jacobP,
\end{equation}
and introduce a scaled distribution
\begin{equation}
    \JTf = \jacobTot f,
\end{equation}
whose discrete representation in cell $K_j$ is obtained via weak multiplication:
\begin{equation} \label{eq:JTf_discrete}
    \JTf_j = \sum_{k=1}^{\numbP}\JTf_j^{(k)}\pb{k}_j, \qquad \JTf_j^{(k)} = \int_{-1}^{1}\dvxlog\,\dvlog\,\dmulog\,\pb{k}_jJ_{T,j} f_j.
\end{equation}
Next, we multiply the (collisionless, sourceless and diffusionless) gyrokinetic equation by $\jacobGeo\jacobVel$ to obtain:
\begin{equation} \label{eq:gkeq_curvilinear_vmap_conserv}
    %
    \pd{\JTf}{t} + \pd{}{x^i}\xdoti{i}\JTf + \pd{}{\cvpar} \vpardot\frac{\JTf}{\vpar'} = 0,
\end{equation}
where we used the notation $\xdoti{i}=\dualv{i}\cdot\v{\dot{R}}$. This way the gyrokinetic equation in field-aligned and mapped velocity coordinates appears in conservative form. The discrete form arises from projecting it onto the phase-space basis $\pb{k}_j$ in cell $K_j$ and integrating by parts:
\begin{eqnal} \label{eq:gkeq_weak}
    &\int_{K_j}\dvx\,\dcvpar\,\dcmu\,\pb{\ell}_j\pd{\JTf}{t} + \oint_{\partial K_j}\mathrm{d}\v{S_i}\,\dcvpar\,\dcmu\,\pbpm{\ell}_{j}\Ghat_{x^i,j}^{\pm} + \oint_{\partial K_j}\dvx\,\dcmu\,\pbpm{\ell}_{j}\Ghat_{\vpar,j}^{\pm} - \int_{K_j}\dvx\,\dcvpar\,\dcmu\,\left(\pd{\pb{\ell}_j}{x^i}\xdoti{i} + \pd{\pb{\ell}_j}{\cvpar}\frac{\vpardot}{\vpar'}\right)\JTf = 0.
\end{eqnal}
where $\mathrm{d}\v{S_i}$
is a differential surface element perpendicular to the $i$-th direction (pointing outward from cell $K_j$). The superscripts $\pm$ denote a quantity at the upper ($+$) or lower ($-$) surface perpendicular to $x^i$ (in the second term) or to $\vpar$ (in the third term). The quantities $\Ghat_{x^i,j}^{\pm}$ and $\Ghat_{\vpar,j}^{\pm}$ represent numerical fluxes that depend on the values of their respective arguments on either side of a cell boundary; for example:
\begin{eqnal}
    \Ghat_{x^i,j}^+ = \Ghat_{x^i,j}^+\left(\xdoti{i+}_{j},\JTf_{j}^+,\JTf_{j+1}^-\right),
    \qquad
    \Ghat_{\vpar,j}^+ = \Ghat_{\vpar,j}^+\left(\vpardot^+_{j},\vpar'^+,\vpar'^-,\JTf_{j}^+,\JTf_{j+1}^-\right).
\end{eqnal}
These numerical fluxes are constructed such that they are consistent~\cite{LeVeque1992}, conserve total number of particles and total energy, and avoid the evaluation of geometric quantities at cell corners (to avoid geometric singularities such as the magnetic axis or the X-point of a diverted tokamak). Use the relationship between computational and logical coordinates, equation~\ref{eq:comp2log}, to rewrite equation~\ref{eq:gkeq_weak} in terms of $(\vxlog,\vlog,\mulog)$:
\begin{eqnal} \label{eq:gkeq_weak_log}
    &\int_{-1}^{1}\dvxlog\,\dvlog\,\dmulog\,\pb{\ell}_j\pd{\JTf}{t} + \frac{2}{\Dxi{i}}\int_{-1}^{1}\mathrm{d}\v{S_i}\,\dvlog\,\dmulog\,\pbpm{\ell}_{j}\Ghat^{\pm}_{x^i,j} + \frac{2}{\Dcvpar}\int_{-1}^{1}\dvxlog\,\dmulog\,\pbpm{\ell}_{j}\Ghat^{\pm}_{\vpar,j} \\
    &\quad- \int_{-1}^{1}\dvxlog\,\dvlog\,\dmulog\,\left(\frac{2}{\Dxi{i}}\pd{\pb{\ell}_j}{\xlog{i}}\xdoti{i} + \frac{2}{\Dcvpar}\pd{\pb{\ell}_j}{\vlog}\frac{\vpardot}{\vpar'}\right)\JTf = 0.
\end{eqnal}
We can employ the orthonormality relation, equation~\ref{eq:orthonorm}, and substitute the expansion of $\JTf$, equation~\ref{eq:JTf_discrete}, in the first term to make it explicit that this equation provides a temporal evolution for each expansion coefficient of $\JTf$:
\begin{eqnal} \label{eq:gkeq_weak_log_coeff}
    &\pd{F_j^{(\ell)}}{t} + \frac{2}{\Dxi{i}}\int_{-1}^{1}\mathrm{d}\v{S_i}\,\dvlog\,\dmulog\,\pbpm{\ell}_{j}\Ghat^{\pm}_{x^i,j} + \frac{2}{\Dcvpar}\int_{-1}^{1}\dvxlog\,\dmulog\,\pbpm{\ell}_{j}\Ghat^{\pm}_{\vpar,j} - \int_{-1}^{1}\dvxlog\,\dvlog\,\dmulog\,\left(\frac{2}{\Dxi{i}}\pd{\pb{\ell}_j}{\xlog{i}}\xdoti{i} + \frac{2}{\Dcvpar}\pd{\pb{\ell}_j}{\vlog}\frac{\vpardot}{\vpar'}\right)\JTf = 0.
\end{eqnal}
Two ingredients remain in order to fully prescribe the discretization of collisionless terms: we must construct the discrete advection speeds $\xdoti{i}$ and $\vpardot$ used in the volume term (i.e. last term in equation~\ref{eq:gkeq_weak_log_coeff}), and we must define the numerical fluxes $\Ghat^{\pm}_{x^i,j}$ and $\Ghat^{\pm}_{\vpar,j}$ employed in the surface terms (i.e. the second and third terms in equation~\ref{eq:gkeq_weak_log_coeff}).

The discrete advection speeds $\xdoti{i}$ and $\vpardot$ for the volume term of equation~\ref{eq:gkeq_weak_log_coeff} are simply calculated by projecting their definition (equation~\ref{eq:alpha_curv_axisym}) onto the phase-space basis $\pb{k}_j$, using the discrete Hamiltonian in cell $K_j$
\begin{equation} \label{eq:HamiltonianDiscrete}
    H_j = \frac{1}{2}mv_{\parallel j}^2 + \mu_j B_j + q\mathcal{P}(\phi)_j,
\end{equation}
where $\mathcal{P}(A)$ is an operator which makes $A$ $C^0$-continuous along $x^3$; the continuity of $\phi$ is necessary as particle potential energy should be single-valued at cell boundaries, and the continuity at $x^1$-$x^2$ cell boundaries will be enforced by the finite element (FE) solution of the quasineutrality equation~\ref{eq:gkPoissonCurvSimp}. This continuity requirement is analogous to how DG Vlasov-Poisson solvers need a $\phi$ continuous in all directions~\cite{Hakim2019}, a property guaranteed by a FE solution of a $\cdim$-dimensional Poisson equation. In our case, the ($\cdim$-$1$)-dimensional quasineutrality equation returns a $\phi$ that is continuous only in $x^1$-$x^2$, thus needing an additional operation to enforce continuity in $x^3$. 

For the purposes of the volume terms, the discrete advection speeds in cell $K_j$ are 
\begin{eqnal} \label{eq:xdot_vpardot}
    \xdoti{i}_j = \sum_{k=1}^{\numbP} \dot{R}_j^{i,(k)} \pb{k}_j,
    \qquad
    \vpardot_j = \sum_{k=1}^{\numbP} \dot{\vpar}_j^{(k)} \pb{k}_j,
\end{eqnal}
\ignore{
whose expansion coefficients are obtained as follows:
\begin{eqnal} \label{eq:xdotik_vpardotk}
    \dot{R}^{i,(k)}_j &= \int_{K_j}\dvxlog\,\dvlog\,\dmulog\,\pb{k}_j\left(\frac{B^{*i}}{m_sB\vpar'}\frac{2}{\Dcvpar}\pd{H_j}{\vlog} + \frac{\epsilon^{i\ell p}}{q_s}\frac{b_\ell}{\jacobGeo B}\frac{2}{\Dxi{p}}\pd{H_s}{\xlog{p}}\right), \\
    \dot{\vpar}^{(k)}_j &= \int_{K_j}\dvxlog\,\dvlog\,\dmulog\,\pb{k}_j\left(-\frac{B^{*i}}{m_sB}\frac{2}{\Dxi{i}}\pd{H_j}{\xlog{i}}\right).
\end{eqnal}
Incorporating the definition of $B^{*i}$ and the fact that $\vpar=[2/(\vpar'\Dcvpar)]\partial_{\vlog}H_j$, these coefficients become:
}
whose expansion coefficients, using equations~\ref{eq:alpha_curv_axisym},~\ref{eq:app_bmag} and~\ref{eq:app_vBs_contra}, and the fact that $\vpar=(\partial_{\cvpar}H_j)/(m\vpar')$, are:
\begin{eqnal} \label{eq:xdotik_vpardotk_v2}
    B^{*i} &= \dualv{i}\cdot\vBs = B^3\delta^i_3+\frac{m_s\vpar}{q_s}\dualv{i}\cdot\curl{\bhat}, \\
    %
    %
    %
    %
    \dot{R}^{i,(k)}_j &= \int_{K_j}\dvxlog\,\dvlog\,\dmulog\,\pb{k}_j\left[\left(g_{33}^{-1/2}\delta^i_3+\frac{m_s\vpar}{q_s}\frac{\dualv{i}\cdot\curl{\bhat}}{B}\right)\vpar + \frac{\epsilon^{i\ell p}}{q_s}\frac{b_\ell}{\jacobGeo B}\frac{2}{\Dxi{p}}\pd{H_s}{\xlog{p}}\right], \\
    %
    %
    %
    %
    \dot{\vpar}^{(k)}_j &= \int_{K_j}\dvxlog\,\dvlog\,\dmulog\,\pb{k}_j\left[-\frac{1}{m_s}\left(g_{33}^{-1/2}\delta^i_3+\frac{m_s\vpar}{q_s}\frac{\dualv{i}\cdot\curl{\bhat}}{B}\right)\frac{2}{\Dxi{i}}\pd{H_j}{\xlog{i}}\right],
\end{eqnal}
where $\delta^{i}_j$ is the Kronecker delta. We can compute these integrals analytically with computer algebra if we use a discrete (polynomial) representation of the geometric quantities $g_{33}^{-1/2}$ and $\dualv{i}\cdot(\curl{\bhat})/B$. These geometric quantities are discretized by evaluating them at Gauss-Legendre nodes interior to a cell $\intnodes_k(\vxlog{},\vlog,\mulog)$, and effecting a nodal-to-modal transformation. The choice of Gauss-Legendre nodes is primarily motivated by our desire to avoid evaluation at cell corners, where a magnetic X- or O-point may be located. For this purpose we introduce a nodal basis $\nodb{k}_j(\vxlog{},\vlog,\mulog)$ local to cell $K_{j}$ ($\mathrm{supp}(\nodb{k}_j(\vxlog,\vlog,\mulog)) = K_{j}$) that takes a value of 1 at the $k$-th node $\intnodes_k$ and 0 at the other nodes. Such nodal basis can be used to represent these geometric coefficients in terms of their nodal values; for example
\begin{eqnal} \label{eq:sqrtg33inv_nod}
    \left(g_{33}^{-1/2}\right)_{j} = \sum_{k=1}^{N_{b,{\scriptsize\vxlog}}} \left(g_{33}^{-1/2}\right)_{j}(\intnodes_k) ~\nodb{k}_j(\vxlog,\vlog,\mulog), \qquad
    \text{nodal form},
\end{eqnal}
where the expansion coefficients are simply $g_{33}^{-1/2}$ evaluated at the $k$-th Gauss-Legendre node $\intnodes_k$ in the $j$-th cell. Equivalently, one can write its modal representation:
\begin{eqnal} \label{eq:sqrtg33inv_mod}
    \left(g_{33}^{-1/2}\right)_{j} = \sum_{k=1}^{N_{b,{\scriptsize\vxlog}}} \left(g_{33}^{-1/2}\right)^{(k)}_{j}~\dgb{k}_j(\vxlog,\vlog,\mulog), \qquad
    \text{modal form},
\end{eqnal}
whose DG expansion coefficients result from projecting the nodal form in equation~\ref{eq:sqrtg33inv_nod} onto the modal basis:
\begin{eqnal} \label{eq:n2m_conf}
    \left(g_{33}^{-1/2}\right)^{(k)}_{j} = \int_{-1}^{1}\dvxlog\,\dvlog\,\dmulog\,\dgb{k}_j(\vxlog,\vlog,\mulog)~ \sum_{\ell=1}^{N_{b,{\scriptsize\vxlog}}} \left(g_{33}^{-1/2}\right)_{j}(\intnodes_\ell) ~\nodb{\ell}_j(\vxlog,\vlog,\mulog).
\end{eqnal}
A similar treatment is used to discretize $\dualv{i}\cdot(\curl{\bhat})/B$, allowing us to perform the integrals in equation~\ref{eq:xdotik_vpardotk_v2} analytically, once, with computer algebra, avoiding numerical quadratures and aliasing errors~\cite{Hakim2020b}.

We turn to the calculation of the surface terms of equation~\ref{eq:gkeq_weak_log_coeff}. Recall that we wish the numerical fluxes $\Ghat_{x^i,j}^{\pm}$ and $\Ghat_{\vpar,j}^{\pm}$ to: {\bf a)} be consistent, {\bf b)} conserve particles and energy, and {\bf c)} avoid evaluation of geometric quantities at cell corners. Let us then introduce a space of polynomials $\mathcal{H}^p_j(\backslash\{z\})$ with cardinality $\abs{\mathcal{H}_j^p(\backslash\{z\})}=\numbPSurf{z}$ which is a subset of $\mathcal{H}^p_j(\vxlog,\vlog,\mulog)$ but is independent of $z$, and use it to form the basis set $\pbSurf{k}_j(\backslash\{z\})\in\mathcal{H}^p_j(\backslash\{z\})$ which can  represent quantities on the surface (of a cell) perpendicular to $z$. The numerical fluxes on a cell surface are then:
\begin{eqnal} \label{eq:gk_flux_surf}
    \Ghat_{x^i,j}^{\pm} = \sum_{k=1}^{\numbPSurf{\xlog{i}}} \Ghat_{x^i,j}^{\pm,(k)} \pbSurf{k}_j(\backslash\{\xlog{i}\}),
    \qquad
    \Ghat_{\vpar,j}^{\pm} = \sum_{k=1}^{\numbPSurf{\vlog}} \Ghat_{\vpar,j}^{\pm,(k)} \pbSurf{k}_j(\backslash\{\vlog\}),
    \qquad
    \text{modal form}.
\end{eqnal}
In order to compute the DG coefficients in these expansions, we evaluate the fluxes at the set of $\numbPSurf{\xlog{i}}$ or $\numbPSurf{\vlog}$ Gauss-Legendre nodes, $\gnodes_k(\backslash\{\xlog{i}\})$ or $\hnodes_k(\backslash\{\vlog\})$, on the cell surface perpendicular to $x^i$ or $\vpar$, respectively. These evaluations, denoted $\Ghat_{x^i,j}^{\pm}(\gnodes_k)$ and $\Ghat_{\vpar,j}^{\pm}(\hnodes_k)$, can be used to represent the numerical fluxes as an expansion in 
a nodal basis $\nodb{k}_j(\backslash\{z\})$ local to cell $K_{j}$ and independent of the variable $z$ (e.g. at a $\vpar$ surface $\mathrm{supp}(\nodb{m}_k(\backslash\{\vlog\}))=\left[\v{x}_{k-1/2},\v{x}_{k+1/2}\right]\times\left[\cmu_{k-1/2},\cmu_{k+1/2}\right]$):
\begin{eqnal} \label{eq:gk_flux_surf_nod}
    \Ghat_{x^i,j}^{\pm} = \sum_{k=1}^{\numbPSurf{\xlog{i}}} \Ghat_{x^i,j}^{\pm}(\gnodes_k) ~\nodb{k}_j(\backslash\{\xlog{i}\}),
    \qquad
    \Ghat_{\vpar,j}^{\pm} = \sum_{k=1}^{\numbPSurf{\vlog}} \Ghat_{\vpar,j}^{\pm}(\hnodes_k) ~\nodb{k}_j(\backslash\{\vlog\}),
    \qquad
    \text{nodal form}.
\end{eqnal}
If we have the nodal value of the surface fluxes (coefficients in equation~\ref{eq:gk_flux_surf_nod}), we can obtain the modal expansion coefficients in equation~\ref{eq:gk_flux_surf} simply by projecting the nodal representation onto the modal basis:
\begin{eqnal} \label{eq:gk_flux_surf_n2m}
    \Ghat_{x^i,j}^{\pm,(k)} &= \oint_{\partial K_j}\mathrm{d}\v{S_i}\,\dvlog\,\dmulog\, \pbSurf{k}_j(\backslash\{\xlog{i}\}) \sum_{k=1}^{\numbPSurf{\xlog{i}}} \Ghat_{x^i,j}^{\pm}(\gnodes_k) ~\nodb{k}_j(\backslash\{\xlog{i}\}),
    \\
    \Ghat_{\vpar,j}^{\pm,(k)} &= \oint_{\partial K_j}\dvxlog\,\dmulog\, \pbSurf{k}_j(\backslash\{\vlog\}) \sum_{k=1}^{\numbPSurf{\vlog}} \Ghat_{\vpar,j}^{\pm}(\hnodes_k) ~\nodb{k}_j(\backslash\{\vlog\}).
\end{eqnal}
Now we simply need to evaluate the numerical fluxes at Gauss-Legendre nodes on cell surfaces, and we can do so in a Lax-Friedrichs fashion~\cite{LeVeque1992}:
\begin{eqnal} \label{eq:lf_fluxes}
    %
    %
    \Ghat_{x^i,j}^{\pm}(\gnodes_k) &= \frac{1}{2}\xdoti{i\pm}_j(\gnodes_k)\left(\JTf_{j\pm1}^{\mp}(\gnodes_k) +\JTf_j^{\pm}(\gnodes_k)\right) \mp \frac{1}{2}\abs{\xdoti{i\pm}_j(\gnodes_k)}\left(\JTf_{j\pm1}^{\mp}(\gnodes_k) -\JTf_j^{\pm}(\gnodes_k)\right), \\
    \Ghat_{\vpar,j}^{\pm}(\hnodes_k) &= \frac{1}{2}\vpardotSurfj(\hnodes_k)\left(\frac{\JTf_{j\pm1}^{\mp}(\hnodes_k)}{v_{\parallel j\pm1}'} +\frac{\JTf_j^{\pm}(\hnodes_k)}{v_{\parallel j}'}\right) \mp \frac{1}{2}\abs{\vpardotSurfj(\hnodes_k)}\left(\frac{\JTf_{j\pm1}^{\mp}(\hnodes_k)}{v_{\parallel j\pm1}'} -\frac{\JTf_j^{\pm}(\hnodes_k)}{v_{\parallel j}'}\right).
\end{eqnal}
Lax-Friedrichs fluxes inherently upwind the discontinuous distribution function $\JTf$, introducing some numerical diffusion proportional to the discontinuity at cell boundaries, and helping avoid spurious oscillations. The last ingredient for the calculation of the surface terms in equation~\ref{eq:gkeq_weak_log_coeff} is the evaluation of the advection speeds at cell-surface Gauss-Legendre nodes appearing in equation~\ref{eq:lf_fluxes}. This simply involves evaluation of the advection speeds in equation~\ref{eq:alpha_curv_axisym} at their respective nodes. After using equations~\ref{eq:app_bmag} and~\ref{eq:app_vBs_contra}, and the fact that $\vpar=(\partial_{\cvpar}H_j)/(m\vpar)$, these values are:
\begin{eqnal} \label{eq:xdotik_vpardotk_surf}
    \dot{R}^{i\pm}_j(\gnodes_k) &= \left[g_{33}^{-1/2}(\gnodes_k)\delta^i_3+\frac{m_s\vpar(\gnodes_k)}{q_s}\left(\frac{\dualv{i}\cdot\curl{\bhat}}{B}\right)(\gnodes_k)\right]\vpar(\gnodes_k) + \frac{\epsilon^{i\ell p}}{q_s}\left(\frac{b_\ell}{\jacobGeo B}\right)(\gnodes_k)~\frac{2}{\Dxi{p}}\left(\pd{H_s}{\xlog{p}}\right)(\gnodes_k), \\
    \dot{\vpar}^{\pm}_j(\hnodes_k) &= -\frac{1}{m_s}\left[g_{33}^{-1/2}(\hnodes_k)\delta^i_3+\frac{m_sv_{\parallel j}^{\pm}}{q_s}\left(\frac{\dualv{i}\cdot\curl{\bhat}}{B}\right)(\hnodes_k)\right]\frac{2}{\Dxi{i}}\left(\pd{H_j}{\xlog{i}}\right)(\hnodes_k).
\end{eqnal}

Equations~\ref{eq:HamiltonianDiscrete}-\ref{eq:xdotik_vpardotk_surf} provide the means to compute the surface and volume collisionless terms of equation~\ref{eq:gkeq_weak_log_coeff}. This recipe, and its consistency and conservative properties, relies on the Halmitonian in equation~\ref{eq:HamiltonianDiscrete} being $C^0$ continuous. Continuity of the Hamiltonian, as well as single-valued nodal geometric quantities appearing in the surface advection speeds (equation~\ref{eq:xdotik_vpardotk_surf}) ensure that the advection speeds, and the Lax-Friedrichs fluxes in equation~\ref{eq:lf_fluxes}, are $C^0$ continuous, thus ensuring that the same number particles that leave a cell enter its adjacent cell. The continuity of the Hamiltonian is ensured by using a continuous representation of $\vpar(\cvpar)$ and $\vpar(\cmu)$ (see equation~\ref{eq:velmap_discrete}). It is also necessary to use a continuous representation of the magnetic field amplitude $B_j$, which we construct by evaluating $B$ at Gauss-Lobatto nodes and performing a nodal-to-modal transformation in each cell. Lastly, we must guarantee that the electrostatic potential $\phi$ is also continuous, as described next.

\subsubsection{Discrete quasineutrality equation} \label{sec:quasineut_discrete}

Following past work on DG methods for Hamiltonian systems with continuous stream functions or potentials~\cite{Liu2000,Shi2017thesis,Mandell2020}, we solve equation~\ref{eq:gkPoissonCurvSimp} such that $\phi$ is continuous. We start by multiplying the quasineutrality equation by $\jacobGeo$ 
\begin{eqnal} \label{eq:gkPoissonCurvilinearSimpJmul}
    -\pd{}{x^i}\varepsilon^{ij}\pd{}{x^j}\phi = \mathcal{P}\left(\varrho\right) = \overline{\varrho}, \qquad i,j\in\{1,2\},
\end{eqnal}
and employing the notation $\varepsilon^{ij}=\jacobGeo\epsilon_\perp g^{ij}$ for the polarization density weight including geometric factors, and $\varrho=\jacobGeo\rho=\jacobGeo\sum_sq_sn_s$ for the charge density including the position-space Jacobian. We have also introduced the continuous-along-$x^3$ charge density, $\overline{\varrho}$, produced by the projection operator $\mathcal{P}$ acting on $\varrho$. We use $\overline{\varrho}$ instead of $\varrho$ because $\mathcal{P}(\phi)$ appeared in the discrete Hamiltonian (equation~\ref{eq:HamiltonianDiscrete}) in order to have continuous and conservative numerical fluxes through cell surfaces perpendicular to $x^3$, and self-adjointness will allow the same operation on $\rho$.

The projection operator $\mathcal{P}\left(g\right) = \bar{g}$ takes a DG scalar field $g$ and produces a field $\bar{g}$ which is continuous in $x^3$. There are various ways to define the $\mathcal{P}$ operator, and we here choose to use a projection onto a FE basis that is continuous only in the $x^3$ direction. That is, introduce the space $\mathcal{F}^{p}_{i,j,\Omega_{x^3}}\equiv\left\{\febz{\ell}_{i,j}(\xlog{1},\xlog{2},x^3)~|~\mathrm{supp}_{x^1,x^2}(\febz{\ell}_{i,j})=E_{i,j},~C^0(x^3)\right\}$ of polynomials $\febz{\ell}_{i,j}$ with compact support along $x^1$ and $x^2$ limited to the cell $E_{i,j}\equiv\left[x^1_{i-1/2},x^1_{i+1/2}\right]\times\left[x^2_{j-1/2},x^1_{j+1/2}\right]$ (here $i\in[1,\Nx]$ and $j\in[1,\Ny]$), yet continuous and nonlocal (i.e. they span more than 1 cell) along $x^3$. We denote the cardinality of this vector space as $\abs{\mathcal{F}^{p}_{i,j,\Omega_{x^3}}}=\numbFEz$. We can use these polynomials as a basis set in which to expand a continuous-in-$x^3$ scalar field $g$ in cell $E_{i,j}$ and along $x^3$:
\begin{equation}
    \bar{g}_{i,j} = \sum_{\ell=1}^{\numbFEz}\bar{g}_{i,j}^{(\ell)}\febz{\ell}_{i,j}(\xlog{1},\xlog{2},x^3).
\end{equation}
Simply put, this quantity is discontinuous in $x^1$-$x^2$ and continuous in $x^3$. Our projection operator $\mathcal{P}$ that computes this field is thus a projection of the discontinuous field $g$ onto the $\febz{\ell}_{i,j}$ basis
\begin{eqnal} \label{eq:fem_parproj}
    \bar{g}_{i,j} = \mathcal{P}(g_{i,j}) \quad\Rightarrow\quad
    \int_{E_{i,j}}\mathrm{d}x^1\,\mathrm{d}x^2\int_{\Omega_{x^3}}\mathrm{d}x^3~\febz{\ell}_{i,j}\bar{g}_{i,j} = \int_{E_{i,j}}\mathrm{d}x^1\,\mathrm{d}x^2\int_{\Omega_{x^3}}\mathrm{d}x^3~\febz{\ell}_{i,j} g_{i,j}.
\end{eqnal}
As it is typically done in a FE method, we can split the global integral over $\Omega_{x^3}$ into local contributions from each of the $N_z$ cells of $\Omega_{x^3}$, employing the restriction of $\febz{\ell}_{i,j}$ to cell $\left[x^3_{k-1/2},x^3_{k+1/2}\right]$ for $k\in[1,N_z]$, which we denote $\febz{\ell}_{i,j,k}(\vxlog)$. The discrete form of the projection $\mathcal{P}$ operator is thus
\begin{eqnal} \label{eq:fem_parproj_local}
    \int_{E_{i,j}}\mathrm{d}x^1\,\mathrm{d}x^2\sum_{k=1}^{\Nz}\int_{x^3_{k-1/2}}^{x^3_{k+1/2}}\mathrm{d}x^3~\febz{\ell}_{i,j,k} \sum_{m=1}^{\numbC}\bar{g}_{i,j,k}^{(m)}\febz{m}_{i,j,k} = \int_{E_{i,j}}\mathrm{d}x^1\,\mathrm{d}x^2\sum_{k=1}^{\Nz}\int_{x^3_{k-1/2}}^{x^3_{k+1/2}}\mathrm{d}x^3~\febz{\ell}_{i,j,k} \sum_{m=1}^{\numbC}g_{i,j,k}^{(m)}\dgb{m}_{i,j,k}(\vxlog).
\end{eqnal}
where we inserted the DG representation of $g$ employing the set of $\numbC$ piecewise discontinuous polynomials $\dgb{m}_{i,j,k}(\vxlog)\in\mathcal{V}_j^p(\vxlog)$. Equation~\ref{eq:fem_parproj_local} can be written in terms of logical coordinates (see equation~\ref{eq:comp2log}) as:
\begin{eqnal} \label{eq:fem_parproj_local_log}
    \int_{-1}^{1}\mathrm{d}\xlog{1}\,\mathrm{d}\xlog{2}\sum_{k=1}^{N_z}\int_{-1}^{1}\mathrm{d}\xlog{3}~\febz{\ell}_{i,j,k} \sum_{m=1}^{\numbC}\bar{g}_{i,j,k}^{(m)}\febz{m}_{i,j,k} = \int_{-1}^{1}\mathrm{d}\xlog{1}\,\mathrm{d}\xlog{2}\sum_{k=1}^{N_z}\int_{-1}^{1}\mathrm{d}\xlog{3}~\febz{\ell}_{i,j,k} \sum_{m=1}^{\numbC}g_{i,j,k}^{(m)}\dgb{m}_{i,j,k}(\vxlog).
\end{eqnal}
This FE projection along $x^3$ results in a linear system of equations that can be written in matrix form $A x = b$ and solved with standard direct or iterative methods. Since these systems are small (e.g. $A$ is sparse and $\left(\numbC N_z\right)^2$ elements large), we typically employ a direct sparse solver. Lastly we note that the adjoint of this operator $\mathcal{P}^\dagger$, defined by
\begin{eqnal} \label{eq:fem_parproj_adjoint}
    \int_{E_{i,j}}\mathrm{d}x^1\,\mathrm{d}x^2\int_{\Omega_{x^3}}\mathrm{d}x^3~f\mathcal{P}(g) = \int_{E_{i,j}}\mathrm{d}x^1\,\mathrm{d}x^2\int_{\Omega_{x^3}}\mathrm{d}x^3~\mathcal{P}^\dagger(f)g,
\end{eqnal}
can be shown to be equal to the operator $\mathcal{P}$ itself, i.e. $\mathcal{P}=\mathcal{P}^\dagger$ is self-adjoint. Note that because $\mathcal{P}$ is local in $x^1$-$x^2$, we can promote this self-adjointness relationship to a global one by summing over all cells in $x^1$ and $x^2$:
\begin{eqnal} \label{eq:fem_parproj_adjoint_global}
    \sum_{i=1}^{\Nx}\sum_{j=1}^{\Ny}\int_{E_{i,j}}\mathrm{d}x^1\,\mathrm{d}x^2\int_{\Omega_{x^3}}\mathrm{d}x^3~f\mathcal{P}(g) &= \sum_{i=1}^{\Nx}\sum_{j=1}^{\Ny}\int_{E_{i,j}}\mathrm{d}x^1\,\mathrm{d}x^2\int_{\Omega_{x^3}}\mathrm{d}x^3~\mathcal{P}^\dagger(f)g, \\
    \int_{\Omega_{\vx}}\dvx~f\mathcal{P}(g) &= \int_{\Omega_{\vx}}\dvx~\mathcal{P}^\dagger(f)g.
\end{eqnal}

We can now use the projection operator $\mathcal{P}$ to compute a continuous-in-$x^3$ charge density ($\overline{\varrho}=\mathcal{P}\left(\jacobGeo\rho\right)$) to solve the quasineutrality equation~\ref{eq:gkPoissonCurvilinearSimpJmul}. We use a discontinuous representation of the geometric quantity $\jacobGeo g^{ij}$, constructed by evaluating it at Gauss-Legendre nodes interior to a cell and effecting a nodal-to-modal transformation. This approach avoids evaluation of geometric quantities that may blow up at a magnetic O- or X-point (which our meshing ensures only exist at cell corners). Lastly, we will solve the quasineutrality equation locally in $x^3$ and globally in $x^1$-$x^2$, for which we introduce the function space $\mathcal{G}_{\Omega_{x^1,x^2},k}^p\equiv\left\{\febxy{m}_k(x^1,x^2,\xlog{3})~|~\mathrm{supp}_{x^3}(\febxy{m}_k)=E_k,~C^0(x^1,x^2)\right\}$ with cardinality $\abs{\mathcal{G}_{\Omega_{x^1,x^2},k}^p}=\numbFExy$ of polynomials $\febxy{m}_{k}$ with compact support along $x^3$ limited to the $k$-th cell $E_k\equiv\left[x^3_{k-1/2},x^3_{k+1/2}\right]$ (where $k\in\left[1,\Nz\right]$). These polynomials are continuous ($C^0(x^1,x^2)$) and nonlocal along $x^1$ and $x^2$. We can project the quasineutrality equation onto $\febxy{m}_{k}$ in each cell along $x^3$:
\begin{eqnal}
    -\int_{x^3_{k-1/2}}^{x^3_{k+1/2}}\dx^3\int_{\Omega_{x^1,x^2}}\dx^1\,\dx^2\,\febxy{m}_k\pd{}{x^p}\varepsilon_k^{pq}\pd{}{x^q}\phi_k = \int_{x^3_{k-1/2}}^{x^3_{k+1/2}}\dx^3\int_{\Omega_{x^1,x^2}}\dx^1\,\dx^2\,\febxy{m}_k\overline{\varrho}_k, \qquad p,q\in\{1,2\}
\end{eqnal}
which after integration by parts becomes
\begin{eqnal} \label{eq:gkPoissonCurvilinearSimpJmulWeak}
    &-\int_{x^3_{k-1/2}}^{x^3_{k+1/2}}\dx^3\int_{\Omega_{x^2}}\dx^2\,\febxy{m}_k\varepsilon_k^{1q}\pd{\phi_k}{x^q}\Bigg|_{x^1=x^1_{\min}}^{x^1=x^1_{\max}} -\int_{x^3_{k-1/2}}^{x^3_{k+1/2}}\dx^3\int_{\Omega_{x^1}}\dx^1\,\febxy{m}_k\varepsilon_k^{2q}\pd{\phi_k}{x^q}\Bigg|_{x^2=x^2_{\min}}^{x^2=x^2_{\max}} \\
    &\quad+\int_{x^3_{k-1/2}}^{x^3_{k+1/2}}\dx^3\int_{\Omega_{x^1,x^2}}\dx^1\,\dx^2\,\pd{\febxy{m}_k}{x^p}\varepsilon_k^{pq}\pd{\phi_k}{x^q} = \int_{x^3_{k-1/2}}^{x^3_{k+1/2}}\dx^3\int_{\Omega_{x^1,x^2}}\dx^1\,\dx^2\,\febxy{m}_k\overline{\varrho}_k, \qquad i,j\in\{p,q\}
\end{eqnal}

As done for the projection operator $\mathcal{P}$, we can write this global (in $x^1$-$x^2$) weak form in terms of local contributions from each cell, using the restriction of the basis function $\febxy{m}(x^1,x^2,\xlog{3})$ to cell $E_{i,j,k}$ denoted as $\febxy{m}_{i,j,k}(\vxlog)$. Then the weak quasineutrality equation~\ref{eq:gkPoissonCurvilinearSimpJmulWeak} becomes
\begin{eqnal} \label{eq:gkPoissonCurvilinearSimpJmulWeakLocal}
    &- \sum_{j=1}^{\Ny}\int_{x^3_{k-1/2}}^{x^3_{k+1/2}}\dx^3\int_{x^2_{j-1/2}}^{x^2_{j+1/2}}\dx^2~\febxy{m}_{i,j,k}\varepsilon_{i,j,k}^{1q}\pd{\phi_{i,j,k}}{x^q}\Bigg|_{x^1=x^1_{\min}}^{x^1=x^1_{\max}} - \sum_{i=1}^{\Nx}\int_{x^3_{k-1/2}}^{x^3_{k+1/2}}\dx^3\int_{x^1_{i-1/2}}^{x^1_{i+1/2}}\dx^1~\febxy{m}_{i,j,k}\varepsilon_{i,j,k}^{2q}\pd{\phi_{i,j,k}}{x^q}\Bigg|_{x^2=x^2_{\min}}^{x^2=x^2_{\max}} \\
    &\quad+ \sum_{i,j=1}^{\Nx,\Ny}\int_{x^3_{k-1/2}}^{x^3_{k+1/2}}\dx^3\int_{x^1_{i-1/2}}^{x^1_{i+1/2}}\int_{x^2_{j-1/2}}^{x^2_{j+1/2}}\dx^1\,\dx^2~\pd{\febxy{m}_{i,j,k}}{x^p}\varepsilon_{i,j,k}^{pq}\pd{\phi_{i,j,k}}{x^q} = \sum_{i,j=1}^{\Nx,\Ny}\int_{x^3_{k-1/2}}^{x^3_{k+1/2}}\dx^3\int_{x^1_{i-1/2}}^{x^1_{i+1/2}}\int_{x^2_{j-1/2}}^{x^2_{j+1/2}}\dx^1\,\dx^2~\febxy{m}_{i,j,k}\overline{\varrho}_{i,j,k}.
\end{eqnal}
Furthermore, we effect the computational-to-logical coordinate transformation (equation~\ref{eq:comp2log}) to obtain
\begin{eqnal} \label{eq:gkPoissonCurvilinearSimpJmulWeakLocalLog}
    &- \frac{2}{\Dxi{1}}\sum_{j=1}^{\Ny}\int_{-1}^{1}\dxlog{3}\dxlog{2}~\febxy{m}_{i,j,k}\varepsilon_{i,j,k}^{1q}\frac{2}{\Dxi{j}}\pd{\phi_{i,j,k}}{\xlog{q}}\Bigg|_{i=1,~\xlog{1}=-1}^{i=\Nx,~\xlog{1}=1} - \frac{2}{\Dxi{2}}\sum_{i=1}^{\Nx}\int_{-1}^{1}\dxlog{3}\dxlog{1}~\febxy{m}_{i,j,k}\varepsilon_{i,j,k}^{2q}\frac{2}{\Dxi{q}}\pd{\phi_{i,j,k}}{\xlog{j}}\Bigg|_{j=1,~\xlog{2}=-1}^{j=\Ny,~\xlog{2}=1} \\
    &\quad+ \sum_{i,j=1}^{\Nx,\Ny}\int_{-1}^{1}\dvxlog~\pd{\febxy{m}_{i,j,k}}{x^i}\varepsilon_{i,j,k}^{ij}\pd{\phi_{i,j,k}}{x^j} = \sum_{i,j=1}^{\Nx,\Ny}\int_{-1}^{1}\dvxlog~\febxy{m}_{i,j,k}\overline{\varrho}_{i,j,k}.
\end{eqnal}

In order to use a typical Galerkin FE method we can represent $\varepsilon^{ij}$ and $\overline{\varrho}$ in terms of the local restrictions of the FE basis, $\febxy{\ell}_{i,j,k}$, such as
\begin{eqnal} \label{eq:vareps_rho_fe_exp}
    \varepsilon^{pq}_{i,j,k} &= \sum_{\ell=1}^{\numbC} \hat{\varepsilon}^{pq,(\ell)}_{i,j,k}\febxy{\ell}_{i,j,k}(\vxlog), 
    \qquad
    \overline{\varrho} = \sum_{\ell=1}^{\numbC} \hat{\overline{\varrho}}^{(\ell)}_{i,j,k}\febxy{\ell}_{i,j,k}(\vxlog), \\
\end{eqnal}
whose expansion coefficients are obtained by evaluating their DG representation at the FE nodes. After inserting these and the FE representation of $\phi$
\begin{equation} \label{eq:phiFE}
    \phi_{i,j,k} = \sum_{n=1}^{\numbC}
    \hat{\phi}_{i,j,k}^{(n)}\febxy{n}_{i,j,k}(\vxlog),
\end{equation}
into equation~\ref{eq:gkPoissonCurvilinearSimpJmulWeakLocalLog}, a discrete system of equation ensues. We note a subtlety in the last two equations that arises from mixing DG and FE representations: while $\phi$ evaluated at the same node using equation~\ref{eq:phiFE} from two adjacent cells give equal values, doing so for $\varrho$ using equation~\ref{eq:vareps_rho_fe_exp} from two adjacent cells give different contributions to the right-hand side vector of the ensuing matrix form $Ax=b$ because $\varrho$ is discontinuous along $x^1$-$x^2$. Finally, the construction of the discrete FE problem is completed by incorporating boundary conditions. Typically we use periodic, Dirichlet ($\phi(x^i=x^i_{\min,\max})=\phi_b$) or Neumann
$(\varepsilon^{iq}\partial_{x^q}\phi)(x^i=x^i_{\min,\max}) = \phi'_b$
boundary conditions (where $\phi_b$ and $\phi'_b$ are externally provided values). The final linear system of equations can be solved with standard methods. These linear problems are global in $x^1$-$x^2$ and local in $x^3$, with a sparse matrix $A$ of approximately $\numbFExy^2$ elements, and can be solved cheaply (compared to the 5D gyrokinetic update) with a direct solver. 

The solution to equation~\ref{eq:gkPoissonCurvilinearSimpJmulWeakLocalLog} yields an FE representation of $\phi$ in each cell. We can reconstruct the modal representation within a cell by performing a nodal-to-modal transformation exactly using our computer algebra implementation, minimizing aliasing errors. An equivalent way to compute the modal DG coefficients in the $(i,j,k)$-th cell is to perform projections of the nodal representation onto the corresponding modal DG basis. For example, the modal DG coefficients in the $(i,j,k)$ cell would be
\begin{eqnal}
    \phi_{i,j,k}^{(m)} = \int_{-1}^{1}\dvxlog\,\dgb{m}_{i,j,k}(\vxlog) \sum_{n=1}^{\numbC}
    \hat{\phi}_{i,j,k}^{(n)}\febxy{n}_{i,j,k}(\vxlog).
\end{eqnal}
The procedure thus far produces a potential $\phi$ that is continuous in $x^1$-$x^2$, but since it was carried out independently in each $x^3$ cell it is discontinuous in that direction. Hence we finalize the solution of the quasineutrality equation by applying the parallel FE projection operator to the potential: $\mathcal{P}(\phi)$. The final outcome is a potential that is continuous in all three directions, allowing us to construct a continuous Hamiltonian, advection speeds and fluxes.

\subsubsection{Time integration stability constraint} \label{sec:algo_collisionless_cfl}

The phase-space discretization of the collisionless terms in section~\ref{sec:algo_collisionless} is paired with a suitable time integration scheme. We employ an explicit strong stability preserving (SSP) third-order Runge-Kutta scheme~\cite{Gottlieb2001}. The size of the time step $\Dt$ is calculated every time step according to the Courant-Friedrichs-Lewy (CFL) stability constraint
\begin{eqnal} \label{eq:cflConstraintDef}
    \omega_{\mathrm{CFL}}\Dt = f_{\mathrm{CFL}}\sim 1,
\end{eqnal}
where $f_{\mathrm{CFL}}$ is some factor typically set to 1 and $\omega_{\mathrm{CFL}}$ is the largest frequency produced by our gyrokinetic system~\cite{Francisquez2020}. We use a conservative estimate of this frequency by adding advection speeds divided by cell lengths in each direction:
\begin{eqnal}
    \omega_{\mathrm{CFL}} = \max_{\Omega}\left(\sum_{i=1}^{\cdim}\frac{(2p+1)\,\xdoti{i}}{\Dx^i} + \frac{(2\max(p,2)+1)\,\vpardot}{\Dvpar}\right).
\end{eqnal}
The $p$-dependent factors account for the higher-order discretization within each cell, and the $\max_{\Omega}$ operation is done over the entire phase-space domain. However, a technical detail is that $\Dvpar$ does not actually appear in our equations or in their implementation given the velocity map in equation~\ref{eq:vmap}, and that two versions of the advection speeds are computed, one for the surface and one for the volume terms in equation~\ref{eq:gkeq_weak_log_coeff}. We choose to compute a CFL frequency using a sum of the maxima of the advection speeds at Gauss-Legendre nodes on each cell-surface:
\begin{eqnal} \label{eq:omega_CFL_collisionless}
    \omega_{\mathrm{CFL},j} &= \sum_{i=1}^{\cdim}\frac{(2p+1)}{\Dx^i}\left[\max\left(\abs{\dot{R}^{i-}_j(\gnodes_k)}\right)+\max\left(\abs{\dot{R}^{i+}_j(\gnodes_k)}\right)\right] + \frac{(2\max(p,2)+1)}{\Dcvpar}\left[\frac{\max\left(\abs{\vpardot^{-}_j(\hnodes_k)}\right)}{\min\left(v_{\parallel j-1}',v_{\parallel j}'\right)}+\frac{\max\left(\abs{\vpardot^{+}_j(\hnodes_k)}\right)}{\min\left(v_{\parallel j}',v_{\parallel j+1}'\right)}\right].
\end{eqnal}
The final CFL frequency is just the maximum of equation~\ref{eq:omega_CFL_collisionless} over all cells:
\begin{eqnal}
    \omegaCFL = \max_{j=1}^{N}\left(\omega_{\mathrm{CFL},j}\right).
\end{eqnal}

\subsection{Discrete collisional terms} \label{sec:algo_collisions}

The collision terms on the right-hand side of the gyrokinetic equation~\ref{eq:gkeq_curv_vmap} also need to be adapted to incorporate the velocity maps. For ionization, charge exchange and recombination the changes are trivial as they only require evaluating the $\vpar$ and $\mu$ maps~\cite{Bernard2022}. The radiation operator does involve more substantial changes, presented in a separate publication~\cite{Roeltgen2025}. The essence of such changes is captured by the algorithm for the Dougherty operator modeling elastic collisions between charged particles, equation~\ref{eq:gklbo_vmap}. Multiplying across by $\jacobTot$, this operator becomes
\begin{eqnal} \label{eq:lbo_vmap_Jtot}
    \jacobTot\mathcal{C}_s^{\mathrm{el}} = \sum_r\nu_{sr}\,\left[\pd{}{\cvpar}\left(\frac{\vpar-\uparsr}{\vpar'}\JTf_s+\frac{\vtsr^2\jacobVel}{\vpar'^2}\pd{}{\cvpar}\frac{\JTf_s}{\jacobVel}\right) + \pd{}{\cmu}\frac{2\mu}{\mu'}\left(\JTf_s+\frac{m\vtsr^2\jacobVel}{B\mu'}\pd{}{\cmu}\frac{\JTf_s}{\jacobVel}\right)\right].
\end{eqnal}

In order to simplify notation we consider only like-species collisions ($r=s$) and drop the species index; the multi-species case uses previously reported models~\cite{Francisquez2022}, and its algorithm is equal to that presented here. The algorithm for this operator, extending previous works~\cite{Francisquez2020,Francisquez2022}, stems from projecting the operator onto our DG basis $\pb{\ell}_j$ in cell $K_j$:
\begin{eqnal}
    \int_{K_j}\dvx\,\dcvpar\,\dcmu\,\pb{\ell}_j\jacobTot\mathcal{C}^{\mathrm{el}} &= \int_{K_j}\dvx\,\dcvpar\,\dcmu\,\pb{\ell}_j\nu\left[\pd{}{\cvpar}\left(\frac{\vpar-\upar}{\vpar'}\JTf+\frac{\vt^2\jacobVel}{\vpar'^2}\pd{}{\cvpar}\frac{\JTf}{\jacobVel}\right) + \pd{}{\cmu}\frac{2\mu}{\mu'}\left(\JTf+\frac{m\vt^2\jacobVel}{B\mu'}\pd{}{\cmu}\frac{\JTf}{\jacobVel}\right)\right].
\end{eqnal}
We can use integration by parts to convert this equation into the form
\begin{eqnal} \label{eq:gklbo_nonuni_weak_1ibp}
    &\int_{K_j}\dvx\,\dcvpar\,\dcmu\,\pb{\ell}_j\jacobTot\mathcal{C}^{\mathrm{el}} = \\
    &\quad\oint_{\partial K_j}\dvx\,\dcmu\,\pb{\ell}_{j\pm}\nu\left[\left(\vpar-\upar\right)_{\pm}\widehat{\frac{\JTf}{\vpar'}} + v_t^2\frac{\jacobVel}{\vpar'^2}\pd{}{\cvpar}\widetilde{\frac{\JTf}{\jacobVel}}\Bigg|_{\pm}\right] - \int_{K_j}\dvx\,\dcvpar\,\dcmu\,\pd{\pb{\ell}_j}{\cvpar}\nu\left[\left(\vpar-\upar\right)\frac{\JTf}{\vpar'} + \frac{v_t^2\jacobVel}{\vpar'^2}\pd{}{\cvpar}\frac{\JTf}{\jacobVel}\right] \\
    &\quad+ \oint_{\partial K_j}\dvx\,\dcvpar\,\pb{\ell}_{j\pm}\nu 2\mu_{\pm}\left(\widehat{\frac{\JTf}{\mu'}} + \frac{m\vt^2}{B}\frac{\jacobVel}{\mu'^2}\pd{}{\cmu}\widetilde{\frac{\JTf}{\jacobVel}}\Bigg|_{\pm}\right) - \int_{K_j}\dvx\,\dcvpar\,\dcmu\,\pd{\pb{\ell}_j}{\cmu}\nu 2\mu\left(\frac{\JTf}{\mu'} + \frac{m\vt^2}{B}\frac{\jacobVel}{\mu'^2}\pd{}{\cmu}\frac{\JTf}{\jacobVel}\right),
\end{eqnal}
and we can integrate the diffusion volume terms by parts a second time to arrive at
\begin{eqnal} \label{eq:gklbo_nonuni_weak_2ibp}
    &\int_{K_j}\dvx\,\dcvpar\,\dcmu\,\pb{\ell}_j\jacobTot\mathcal{C}^{\mathrm{el}} = \oint_{\partial K_j}\dvx\,\dcmu\,\nu\left\{\pb{\ell}_{j\pm}\left[\left(\vpar-\upar\right)_{\pm}\widehat{\frac{\JTf}{\vpar'}} + v_t^2\frac{\jacobVel}{\vpar'^2}\pd{}{\cvpar}\widetilde{\frac{\JTf}{\jacobVel}}\Bigg|_{\pm}\right] - \pd{\pb{\ell}_j}{\cvpar}\frac{v_t^2\jacobVel}{\vpar'^2}\widetilde{\frac{\JTf}{\jacobVel}}\Bigg|_{\pm}\right\} \\
    &\quad- \int_{K_j}\dvx\,\dcvpar\,\dcmu\,\nu\left[\pd{\pb{\ell}_j}{\cvpar}\left(\vpar-\upar\right)\frac{\JTf}{\vpar'} - \pdd{\pb{\ell}_j}{\cvpar}\frac{v_t^2\jacobVel}{\vpar'^2}\frac{\JTf}{\jacobVel}\right] \\
    &\quad+ \oint_{\partial K_j}\dvx\,\dcvpar\,\nu\left[\pb{\ell}_{j\pm} 2\mu_{\pm}\left(\widehat{\frac{\JTf}{\mu'}} + \frac{m\vt^2}{B}\frac{\jacobVel}{\mu'^2}\pd{}{\cmu}\widetilde{\frac{\JTf}{\jacobVel}}\Bigg|_{\pm}\right) - \pd{\pb{\ell}_j}{\cmu}2\mu\frac{m\vt^2}{B}\frac{\jacobVel}{\mu'^2}\widetilde{\frac{\JTf}{\jacobVel}}\Bigg|_{\pm}\right] \\
    &\quad- \int_{K_j}\dvx\,\dcvpar\,\dcmu\,\nu \left[\pd{\pb{\ell}_j}{\cmu}2\mu\frac{\JTf}{\mu'} - \pd{}{\cmu}\left(\pd{\pb{\ell}_j}{\cmu}2\mu\right)\frac{m\vt^2}{B}\frac{\jacobVel}{\mu'^2}\frac{\JTf}{\jacobVel}\right].
\end{eqnal}
Following the notation used in equation~\ref{eq:gkeq_weak} for the collisionless terms, the $\pm$ subscripts refer to a quantity at the upper ($+$) and lower ($-$) surface of a cell perpendicular to $\vpar$ (first term) or to $\mu$ (third term). The quantities $\widehat{\JTf/\vpar'}$ and $\widehat{\JTf/\mu'}$ indicate that the quantities $\JTf/\vpar'$ and $\JTf/\mu'$ are to be upwinded according to the sign of their accompanying advection speed. Lastly, the expression $\widetilde{\JTf/\jacobVel}$ refers to a two-cell recovered representation of $\JTf/\jacobVel$~\cite{Francisquez2020,Hakim2020,VanLeer2005}, needed to produce continuous surface terms that help us guarantee exact conservation.

We can write equation~\ref{eq:gklbo_nonuni_weak_2ibp} in terms of logical coordinates using equation~\ref{eq:comp2log}:
\begin{eqnal} \label{eq:gklbo_nonuni_weak_2ibp_log}
    &\int_{K_j}\dvxlog\,\dvlog\,\dmulog\,\pb{\ell}_j\jacobTot\mathcal{C}^{\mathrm{el}} = \frac{2}{\Dcvpar}\oint_{\partial K_j}\dvxlog\,\dmulog\,\nu\left\{\pb{\ell}_{j\pm}\left[\left(\vpar-\upar\right)_{\pm}\widehat{\frac{\JTf}{\vpar'}} + v_t^2\mu'\pd{}{\vpar}\widetilde{\frac{\JTf}{\jacobVel}}\Bigg|_{\pm}\right] - \frac{2}{\Dcvpar}\pd{\pb{\ell}_j}{\vlog}\frac{v_t^2\jacobVel}{\vpar'^2}\widetilde{\frac{\JTf}{\jacobVel}}\Bigg|_{\pm}\right\} \\
    &\quad- \int_{K_j}\dvxlog\,\dvlog\,\dmulog\,\nu\left[\frac{2}{\Dcvpar}\pd{\pb{\ell}_j}{\vlog}\left(\vpar-\upar\right)\frac{\JTf}{\vpar'} - \left(\frac{2}{\vpar'\Dcvpar}\right)^2\pdd{\pb{\ell}_j}{\vlog}v_t^2\JTf\right] \\
    &\quad+ \frac{2}{\Dcmu}\oint_{\partial K_j}\dvxlog\,\dvlog\,\nu\left\{\pb{\ell}_{j\pm} 2\mu_{\pm}\left(\widehat{\frac{\JTf}{\mu'}} + \frac{m\vt^2}{B}\vpar'\pd{}{\mu}\widetilde{\frac{\JTf}{\jacobVel}}\Bigg|_{\pm}\right) - \frac{2}{\Dcmu}\pd{\pb{\ell}_j}{\mulog}2\mu\frac{m\vt^2}{B}\frac{\jacobVel}{\mu'^2}\widetilde{\frac{\JTf}{\jacobVel}}\Bigg|_{\pm}\right\} \\
    &\quad- \int_{K_j}\dvxlog\,\dvlog\,\dmulog\,\nu \left[\frac{2}{\Dcmu}\pd{\pb{\ell}_j}{\mulog}2\mu\frac{\JTf}{\mu'} - \left(\frac{2}{\mu'\Dcmu}\right)^2\pd{}{\mulog}\left(\pd{\pb{\ell}_j}{\mulog}2\mu\right)\frac{m\vt^2}{B}\JTf\right].
\end{eqnal}
Note that in the second term of each surface term (in curly braces) we have reverted to a derivative of the recovered $\JTf/\jacobVel'$ with respect to $\vpar$ and $\mu$, rather than the computational $\cvpar$ and $\cmu$, respectively. This approach was favored because we opted to perform the recovery of $\JTf/\jacobVel'$ in $\vpar$- and $\mu$-space at $\vpar$ and $\mu$ boundaries, respectively, using a variant of the two-cell recovery previously reported~\cite{Francisquez2020,Hakim2020,VanLeer2005} and extended to the case in which cells have different lengths.

The discretization of the Dougherty operator given in equation~\ref{eq:gklbo_nonuni_weak_2ibp_log} is accompanied by mixed boundary conditions at the edge of velocity space. We zero out the flux in advection-like discrete terms:
\begin{eqnal} \label{eq:lbo_zeroflux_BCs}
    \nu\left[\left(\vpar-\upar\right)_{\pm}\widehat{\frac{\JTf}{\vpar'}} + v_t^2\mu'\pd{}{\vpar}\widetilde{\frac{\JTf}{\jacobVel}}\Bigg|_{\pm}\right] &= 0 \qquad \text{at}~\cvpar=\cvpar_{\min},\,\cvpar_{\max} \\
    \nu2\mu_{\pm}\left(\widehat{\frac{\JTf}{\mu'}} + \frac{m\vt^2}{B}\vpar'\pd{}{\mu}\widetilde{\frac{\JTf}{\jacobVel}}\Bigg|_{\pm}\right) &= 0 \qquad \text{at}~\cmu=\cmu_{\min},\,\cmu_{\max}.
\end{eqnal}
The last surface terms originating from the second integration by parts of the diffusion terms do not use a 2-cell recovery and are evaluated using the $\JTf/\jacobVel$ in the cell abutting the boundary instead:
\begin{eqnal}
    \nu\frac{2}{\Dcvpar}\pd{\pb{\ell}_j}{\vlog}\frac{v_t^2\jacobVel}{\vpar'^2}\widetilde{\frac{\JTf}{\jacobVel}}\Bigg|_{\pm} &= \nu\frac{2}{\Dcvpar}\pd{\pb{\ell}_j}{\vlog}\frac{v_t^2\jacobVel}{\vpar'^2}\frac{\JTf}{\jacobVel}\Bigg|_{\pm} \qquad \text{at}~\cvpar=\cvpar_{\min},\,\cvpar_{\max} \\
    \nu\frac{2}{\Dcmu}\pd{\pb{\ell}_j}{\mulog}2\mu\frac{m\vt^2}{B}\frac{\jacobVel}{\mu'^2}\widetilde{\frac{\JTf}{\jacobVel}}\Bigg|_{\pm} &= \nu\frac{2}{\Dcmu}\pd{\pb{\ell}_j}{\mulog}2\mu\frac{m\vt^2}{B}\frac{\jacobVel}{\mu'^2}\frac{\JTf}{\jacobVel}\Bigg|_{\pm} \qquad \text{at}~\cmu=\cmu_{\min},\,\cmu_{\max}.
\end{eqnal}
The above discretization, and choice of surface terms and boundary conditions, yields an algorithm that conserves particle, momentum and energy density exactly, provided that $\upar$ and $\vt^2$ satisfy discrete constraints~\cite{Francisquez2020,Francisquez2022}. 

\subsubsection{Time integration stability constraint} \label{sec:algo_collisions_cfl}

The collision operator imposes a restriction on the time step size $\Dt$, in addition to that from the collisionless terms described in section~\ref{sec:algo_collisionless_cfl}. This time the CFL frequency ($\omegaCFL$) contains an advection or drag contribution ($\omega_{\mathrm{CFL},\mathrm{drag}}$) and a diffusion contribution ($\omega_{\mathrm{CFL},\mathrm{diff}}$), and the drag portion is evaluated differently compared to the collisionless terms
\begin{eqnal} \label{eq:lbo_cfl_drag}
    \omega_{\mathrm{CFL},\mathrm{drag},j} = \left|-\frac{2\max(p,2)+1}{\Dcvpar}\nu_j\frac{v_{\parallel j}-u_{\parallel j}}{v_{\parallel j}'} - \frac{2p+1}{\Dcmu}\nu_j\frac{2\mu_j}{\mu'_j}\right|_{\vxlog=\vlog=\mulog=0},
\end{eqnal}
that is, we evaluate the drag at cell centers rather than at cell surface Gauss-Legendre nodes. The diffusion $\omegaCFL$ is
\begin{eqnal} \label{eq:lbo_cfl_diff}
    \omega_{\mathrm{CFL},\mathrm{diff},j} = \left|\left[2\frac{\max(p,2)+1}{\Dcvpar}\right]^2\nu_j\frac{v_{t,j}}{v_{\parallel j}'^2} + \left(2\frac{p+1}{\Dcmu}\right)^2\nu_j\frac{m}{B_j}\frac{2\mu_jv_{t,j}}{\mu_{j}'^2}\right|_{\vxlog=\vlog=\mulog=0}.
\end{eqnal}
The final time step constraint due to the Dougherty operator is then
\begin{equation} \label{eq:lbo_cfl}
    \omegaCFL = \max_{j=1}^{N}\left(\omega_{\mathrm{CFL},\mathrm{drag},j}+\omega_{\mathrm{CFL},\mathrm{diff},j}\right).
\end{equation}

\section{Test results} \label{sec:results}

This section presents tests verifying the properties and implementation (in Gkeyll~\cite{gkeyllWeb}) of the algorithms described in section~\ref{sec:algo}. We show simulations using collisionless and collisional terms independently, followed by simulations using both in 1D, 2D and 3D. Input files and the version of Gkeyll used to generate these results can be found in \url{https://github.com/ammarhakim/gkyl-paper-inp/tree/master/2025_JCP_nonuniformV}. Unless stated otherwise, these simulations used a real electron-ion mass ratio. 

\subsection{Collisionless terms} \label{sec:res_collisionless}

\begin{figure}[h]
  \centering
  \begin{subfigure}[b]{0.45\textwidth}
    \includegraphics[width=\textwidth]{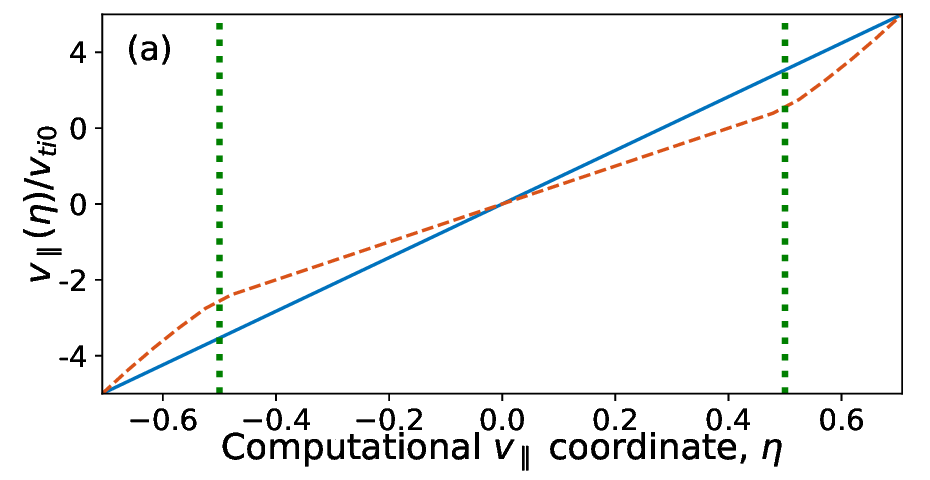}
  \end{subfigure}
  \begin{subfigure}[b]{0.45\textwidth}
    \includegraphics[width=\textwidth]{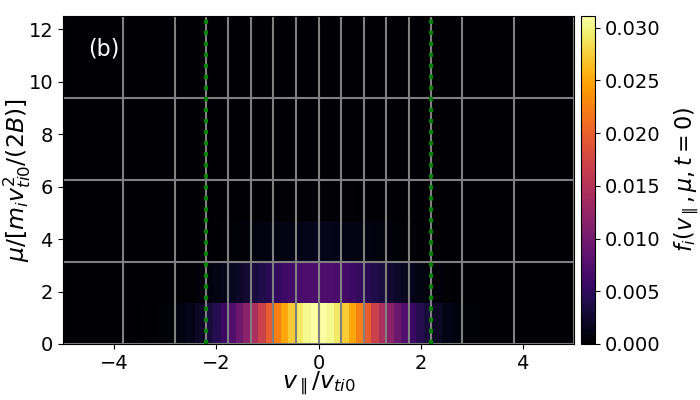}
  \end{subfigure}
  \caption{(a) A uniform $\vpar(\cvpar)$ map (solid blue), and the map in equation~\ref{eq:linquad_vpar_map} (dashed orange). (b) Initial ion distribution at $z=0$, as well as the nonuniform velocity grid (grey mesh). Both figures show demarcate the region with a linear $\vpar(\cvpar)$ map using dotted green lines.}
  \label{fig:iadamping_map}
\end{figure}

As a demonstration that the collisionless terms are discretized correctly with velocity space mappings we perform simulations of collisionless Landau damping of an ion acoustic wave with adiabatic electrons. Previously we had shown that the gyrokinetic solver in Gkeyll reproduces analytically-estimated damping rates for this problem when using a uniform velocity space grid~\cite{Francisquez2020}. Here we reproduce this problem keeping a uniform discretization of $\mu\in[0,m_i(5v_{ti0})^2/(2B)]$, while introducing a slightly nonuniform discretization of $\vpar\in[-5v_{ti0},5v_{ti0}]$. Specifically, $\vpar$ is uniform for $\abs{\vpar}\leq 2.5v_{ti0}$ and quadratic for $2.5v_{ti0}<\abs{\vpar}$ (here $v_{ti0}=\sqrt{T_{i0}/m_i}$ is the ion thermal speed):
\begin{eqnal} \label{eq:linquad_vpar_map}
\vpar = \begin{cases}
    \vparmax \cvpar &\quad \abs{\cvpar}\leq1/2, \\
    2\vparmax\,\mathrm{sign}\left(\cvpar\right)\cvpar^2 &\quad \abs{\cvpar}>1/2,
\end{cases}
\qquad
\mu = \mumax\,\cmu.
\end{eqnal}
The $\vpar(\cvpar)$ mapping is visualized in figure~\ref{fig:iadamping_map}(a).
The mapping in equation~\ref{eq:linquad_vpar_map}, for  $(\cvpar,\cmu)\in[-2^{-1/2},2^{-1/2}]\times\left[0,1\right]$, produces the $(\vpar,\mu)$ grid in figure~\ref{fig:iadamping_map}(b). This discretization employed $32\times16\times4$ cells in $(z,\vpar,\mu)$, where $z\in[-\pi/(\kpar\rho_i),\pi/(\kpar\rho_i)]$. We use a proton-electron mass ratio ($m_i/m_e\approx1836$) and the temperature ratio $T_{e0}/T_{i0}=1$.

\begin{figure}[h]
  \centering
  \begin{subfigure}[h]{0.45\textwidth}
    \includegraphics[width=\textwidth]{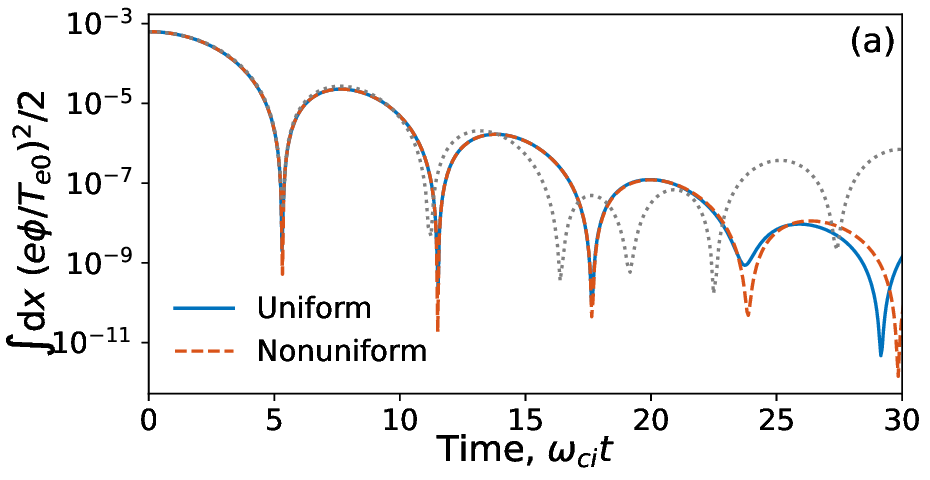}
  \end{subfigure}
  \begin{subfigure}[h]{0.45\textwidth}
    \includegraphics[width=\textwidth]{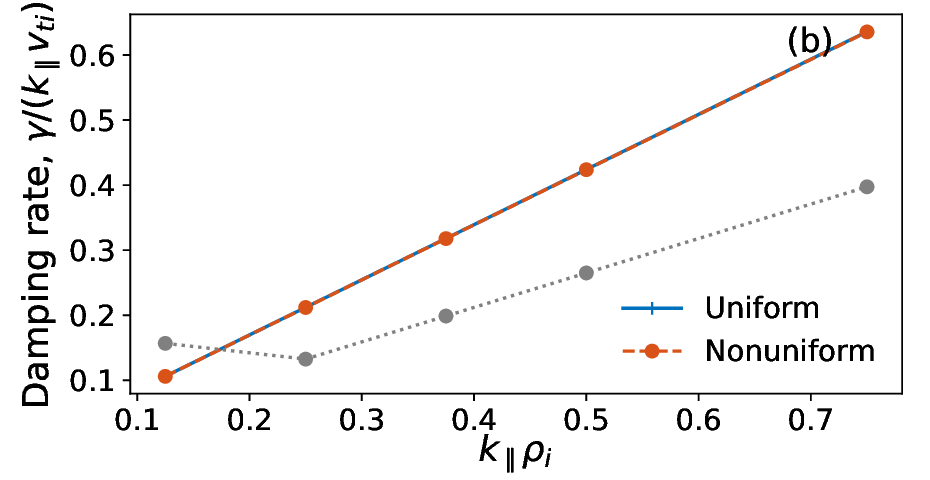}
  \end{subfigure}
\caption{\label{fig:iadamping}(a) Field energy as a function of time (normalized to the ion cyclotron frequency $\omega_{ci}$) of ion acoustic wave simulation exhibiting collisionless Landau damping with uniform and nonuniform velocity mappings for $\kpar\rho_i=0.25$. (b) Damping rate as a function of wave number for uniform and nonuniform velocity mappings. Grey dotted lines show a case in which only the $\abs{\eta}<1/2$ part velocity-space is used.}
\end{figure}

A small sinusoidal perturbation (of relative amplitude $10^{-3}$ and wavenumber $\kpar\rho_i$) is seeded in the density at $t=0$, which launches a wave that collisionlessly damps as particles with velocities close to the phase speed of the wave resonate with and draw energy from the wave. This damping can be observed in time traces of the field energy, for example, as shown for $\kpar\rho_i=0.25$ in figure~\ref{fig:iadamping}(a). In that plot we show that the time evolution of the field energy for both uniform and non-uniform velocity space grids is nearly identical, with the exception of a slightly earlier onset of numerical recurrence~\cite{Canosa1974,Pezzi2016} for the case of a uniform discretization (solid blue). Both of these simulations ran in 0.57 s on a single core of a 2023 Apple MacBookPro with an M3 Pro processor. We also made sure that the value of the damping rate continues to be well-reproduced with nonuniform velocity mappings for a range of $\kpar\rho_i$ wavenumbers, as demonstrated in figure~\ref{fig:iadamping}(b). Note that although both simulations have a uniform velocity discretization in $\abs{\vpar}\leq 2.5v_{ti0}$, the $2.5v_{ti0}<\abs{\vpar}$ region is essential to produce correct results; in figure~\ref{fig:iadamping} a dotted grey line shows that including only the $\abs{\eta}\leq1/2$ region leads to earlier recurrence and erroneous damping rates. 

\begin{figure}[h]
  \centering
  \includegraphics[width=\textwidth]{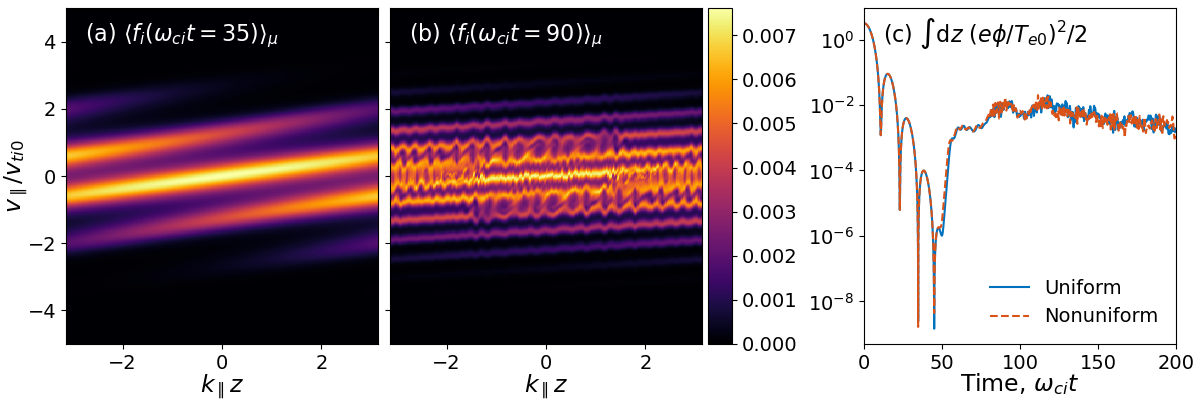}
\caption{\label{fig:ialargepert}
Ion distribution function averaged along $\mu$ at $\omega_{ci}t=35$ (a) and $\omega_{ci}t=90$ (b); these look nearly the same in both uniform and nonuniform velocity simulations. (c) Normalized field energy for simulations with uniform (solid blue) and nonuniform velocity maps (dashed orange).}
\end{figure}

By increasing the amplitude of the perturbation to 0.5 we can drive the system nonlinear, generating larger amplitude structures in velocity-space that further stress test the correctness of the nonuniform velocity map method. We carried out two simulations with this large amplitude and $\kpar\rho_i=0.125$, one using a uniform velocity grid and $128\times214\times4$ cells, the other using the velocity maps in equation~\ref{eq:linquad_vpar_map} with $128\times108\times4$ cells. Both simulations covered the velocity space $[-5v_{ti0},5v_{ti0}]\times[0,m_i(5v_{ti0})^2/(2B_0)]$.Figures~\ref{fig:ialargepert}(a-b) display two snapshots of the ion distribution function averaged along $\mu$; these snapshots look very similar in the uniform and nonuniform velocity simulations. Figure~\ref{fig:ialargepert}(c) portrays the normalized field energy for these simulations. Together these figures show the qualitative reproduction of velocity-space structures and the quantitative match in evolution of the field energy when using nonuniform velocity maps (relative to uniform $\vpar(\cvpar)$). These simulations ran on 64 cores of a 2.9 GHz Intel Cascade Lake chip in 174.56 s (uniform) and 82.8 s (nonuniform).

\subsection{Collisional terms} \label{sec:res_collision}

\begin{figure}[h]
  \centering
  \begin{subfigure}[b]{0.45\textwidth}
    \includegraphics[width=\textwidth]{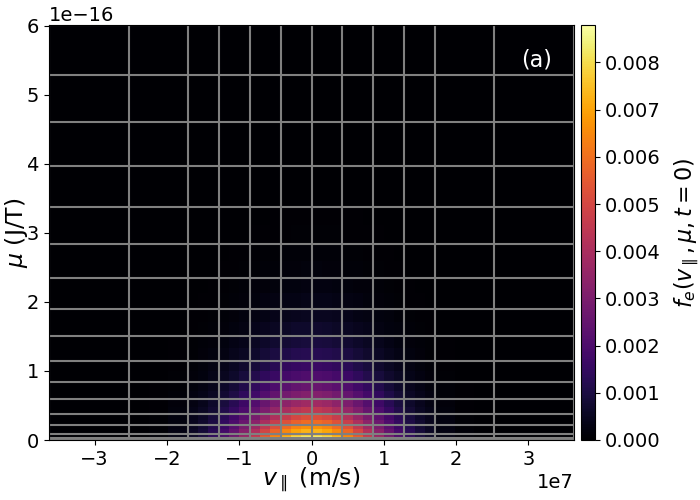}
  \end{subfigure}
  \begin{subfigure}[b]{0.45\textwidth}
    \includegraphics[width=\textwidth]{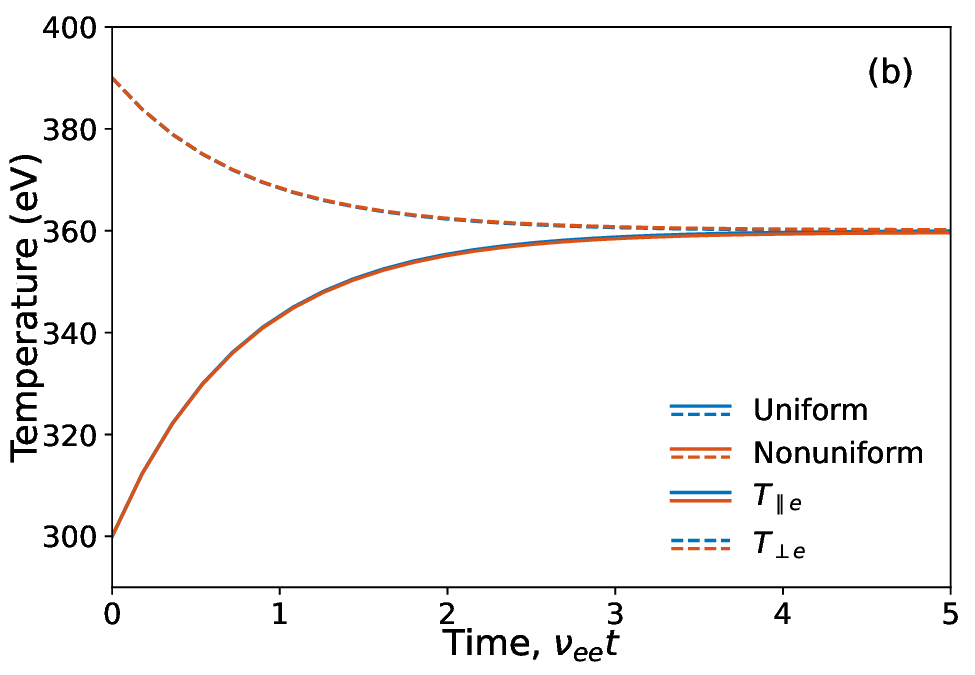}
  \end{subfigure}
  \caption{(a) Initial ion distribution function in velocity space with the nonuniform grid overlaid (grey mesh). (b) Parallel ($\Tpare$, solid) and perpendicular ($\Tperpe$, dashed) temperatures as a function of time (normalized to the electron collision frequency $\nu_{ee}$) for collisional relaxation simulations with uniform (blue) and nonuniform (orange) velocity space grids.}
  \label{fig:lboTempRelax}
\end{figure}

The discrete Dougherty operator in Gkeyll with uniform velocity grid has been shown to conserve particles, momentum and energy to machine precision independently of resolution, for both like-species~\cite{Francisquez2020} and unlike-species~\cite{Francisquez2022} collisions. The operator was also shown to produce expected isotropization rates (e.g. compared to the Fokker-Planck operator)~\cite{Francisquez2022}. Given the algorithm's description in section~\ref{sec:algo_collisions}, we would expect these properties to be preserved when using nonuniform velocity grids. In order to demonstrate this, we initialize an anisotopic electron population as was done in~\cite{Francisquez2022}, using the bi-Maxwellian
\begin{equation} \label{eq:lbo_ICs}
    f_e(t=0) = \frac{n_0}{\left(2\pi\right)^{3/2}v_{t\parallel e0}v_{t\perp e0}^2}\exp\left[-\frac{\left(\vpar-u_{\parallel e0}\right)^2}{2v_{t\parallel e0}^2}-\frac{\mu B/m}{v_{t\perp e0}^2}\right]
\end{equation}
on a $(\cvpar,\cmu)\in[-2^{-1/2},2^{-1/2}]\times\left[0,1\right]$ computational grid, with a mapping that is quadratic in $\cmu$ and linear in $\cvpar$ for $\abs{\vpar}\leq\vparmax/2$ and quadratic above that:
\begin{eqnal} \label{eq:linquad_map}
\vpar = \begin{cases}
    \vparmax \cvpar &\quad \abs{\cvpar}\leq1/2, \\
    2\vparmax\,\mathrm{sign}\left(\cvpar\right)\cvpar^2 &\quad \abs{\cvpar}>1/2,
\end{cases}
\qquad
\mu = \mumax\,\cmu^2.
\end{eqnal}
The velocity-space grid is visualized in figure~\ref{fig:lboTempRelax}(a) with $12\times16$ cells. For this test we use the reference temperatures $T_{\parallel e0} = 300$ eV and $T_{\perp e0}/T_{\parallel e0} = 1.3$, which define the squared thermal speed $v_{te0}^2=T_{e0}/m_e=\left(T_{\parallel e0}+2T_{\perp e0}\right)/(3m_e)$, and the velocity extents $\vparmax=5v_{te0}$ and $\mumax=m_e\vparmax^2/(2 B)$, where we use a constant $B=1$ T magnetic field. The initial condition in equation~\ref{eq:lbo_ICs} uses the parallel drift speed $u_{\parallel e0}=0.5\sqrt{m_e/m_i}v_{te0}$, where $m_i$ is the deuterium ion mass.

\begin{figure}
  \centering
  \includegraphics[width=0.95\textwidth]{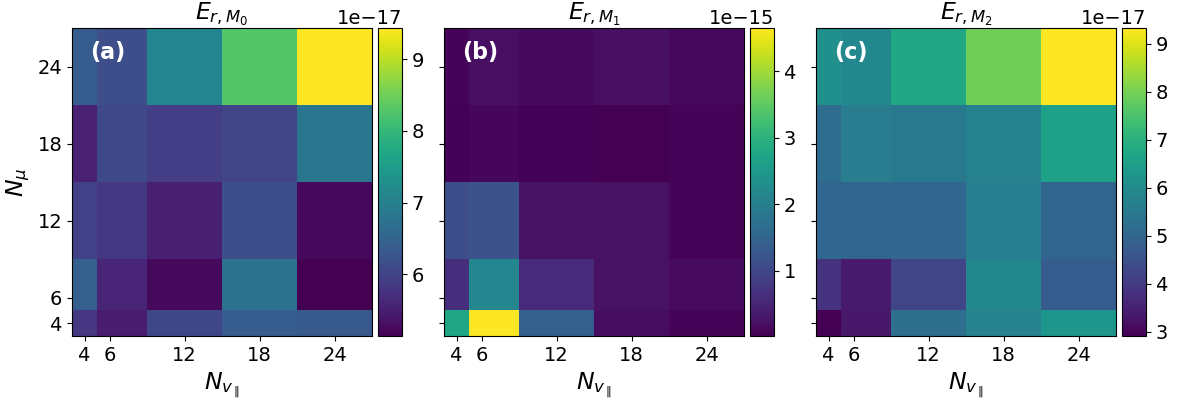}
  \caption{Relative error in integrated particle number (a), momentum (b) and kinetic energy (c) density as a function of the number of cells along $\vpar$ ($\Nvpar$) and along $\mu$ ($\Nmu$). The error is of order of the machine precision in all cases.}
  \label{fig:lboConservMError}
\end{figure}

The initial condition evolves solely under the action of elastic electron-electron collisions, so the collisionless, source and diffuson terms in equation~\ref{eq:gkeq} are turned off. As seen in figure~\ref{fig:lboTempRelax}(b) for $12\times16$ cells in velocity space, the electron distribution function isotropizes at the same rate with nonuniform velocity maps as it did with the previous, well-tested uniform velocity grid implementation. We further scanned velocity space resolution, computing the per-time-step error in the first three velocity-moments of the distribution ($M_k$) as
\begin{equation}
    E_{r,M_k} = \frac{1}{N_t}\frac{\avg{M_{k}}(t=N_t\Delta t)-\avg{M_{k}}(t=0)}{\avg{M_{k}}(t=0)},
    \qquad
    k\in\{0,1,2\},
\end{equation}
where $\avg{\cdot}$ indicates a volume average and $N_t$ is the number of time steps taken to reach 5 collision periods. Figure~\ref{fig:lboConservMError} demonstrates that the error was of order of machine precision independently of resolution, supporting the claim that the collisional algorithm is conservative and that it is implemented in such a manner in Gkeyll.

\subsection{Test with collisionless \& collision terms} \label{sec:res_full}

This section presents tests with both collisionless and collisional terms in one, two and three dimensions.

\subsubsection{1D HTS mirror} \label{sec:res_full_1x}

The use of uniform velocity grids can make gyrokinetic modeling of high-temperature superconducting (HTS) mirrors particularly challenging. As described in previous work, the temperature and density can decrease by orders of magnitude between the core of the confined plasma and the expander region on the other side of the mirror coils~\cite{Francisquez2023}. This variation means that the velocity-space width of the appreciable distribution function diminishes rapidly between these two regions, and the uniform velocity-space grid that seems sufficient in the core is too coarse in the expander or viceversa, a grid that seems suitable in the expander is too narrow or over resolved for the core. Previous 1D simulations of an HTS mirror, including the core plasma and those of the expander, and approximating the Wisconsin HTS Axisymmetric Mirror (WHAM)~\cite{Endrizzi2023}, employed a grid that compromised between the requirements of the core and the expander, as well as the numerical requirements of the previous version of Gkeyll~\cite{Francisquez2023} (since then, Gkeyll was rewritten, adding support for GPUs and improving the algorithms). We restrict this discussion to Boltzmann electron simulations, where $f_e$ is not evolved~\cite{Francisquez2023}, since they are sufficient to describe three challenges, and how nonuniform velocity maps can help.
The first of these challenges is that $f_i$ was only marginally resolved in parts of the domain, such as the mirror throat and the expander wall. Second, despite the Boltzmann electron approximation these simulations still took a long time (days on hundreds of cores to reach $t=8\nu_{ee}^{-1}/3=48~\mu$s, where $\nu_{ee}$ is the electron collision frequency). And third, the largest velocity extent we were able to afford was $\vparmax=3.75v_{ti0}$, but this small $\vparmax$ caused an artificial accumulation of $f_i$ near the boundary and eventually made simulations unstable.

\begin{figure}[ht]
  \centering
  \includegraphics[width=0.9\textwidth]{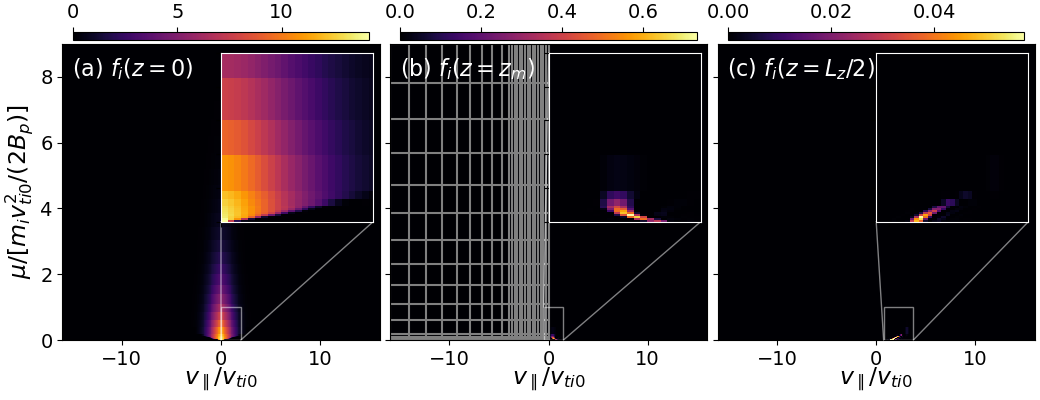}
  \caption{\label{fig:adiabaticfi}Ion distribution function in Boltzmann electron simulation at $t=48\,\mu\mathrm{s}=8\nu_{ee}^{-1}/3$ at three locations along the field line: (a) the center, (b) the mirror throat (with the mesh plotted for the $\vpar<0$ part of the grid in grey), (c) the sheath entrance in the expander. Each figure's inset zooms into the appreciable part of the distribution function.}
\end{figure}

The combination of GPUs and nonuniform velocity maps lets us tackle all three of these challenges. We thus revisit such HTS mirror simulations, with a position space domain $z\in\left[-\Lz/2,\Lz/2\right]$ covering a field line of length $\Lz$ across a 5 m mirror. We use a velocity space mapping resulting in a linear $\vpar$ grid up to $4v_{ti0}$ and cubic above that, and similarly the $\mu(\cmu)$ mapping yields a linear $\mu$ grid up to $0.012\mumax$ and a quadratic grid above that. Specifically, the computational velocity space $(\cvpar,\cmu)=\left[-1/4^{2/3},1/4^{2/3}\right]\times\left[0,1\right]$ employed the mapping
\begin{eqnal}
\vpar = \begin{cases}
    \vparmax \cvpar &\quad \abs{\cvpar}\leq1/4, \\
    16\vparmax\,\mathrm{sign}\left(\cvpar\right)\abs{\cvpar}^3 &\quad \abs{\cvpar}>1/4,
\end{cases}
\qquad
\mu = \begin{cases}
    \frac{g}{w}\,\mumax\,\cmu, &\quad \cmu\leq w\\
    \mumax\,\left(\cmu^2\frac{g-1}{w^2-1}+\frac{w^2-g}{w^2-1}\right), &\quad \cmu>w,
\end{cases}
\end{eqnal}
with $w=0.3$, $g=0.012$, $\vparmax=16v_{ti0}$, $\mumax=m_i\left(3v_{ti0}\right)^2/(2B_p)$, $v_{ti0}^2=T_{i0}/m_i$, $T_{i0}=8.361$ keV and the central field $B_p=0.53$. This mapping, for a grid with $192\times48\times16$ cells, is visualized in figure~\ref{fig:adiabaticfi}(b). The linearly mapped regions in the $\vpar$ and $\mu$ directions have 30 and 4 cells, respectively. This resolution was found to be, through a resolution convergence study, the smallest one that was stable and produced physical results.

Using this nonuniform velocity discretization, and running on 2 NVIDIA A100 GPUs, Gkeyll can reach $t=5.56\nu_{ee}^{-1}=100~\mu$s in 2 hours. The ion distribution function at $t=8\nu_{ee}^{-1}/3=48~\mu$s, for example, can be seen in figure~\ref{fig:adiabaticfi} at the center (a), throat (b) and expander wall (b) of the mirror; in all these cases $f_i$ is very similar to the uniform velocity grid simulations carried out previously (see figure 7 of~\cite{Francisquez2023}). In fact, the distribution function at small $(\vpar,\mu)$ near the throat and in the expander is better resolved than it was with uniform velocity discretizations. Additionally, because nonuniform velocities allow us to use a much higher $\vparmax$ at little cost, we are able to push the $\vpar$ boundary outwards to avoid $f_i$ accumulation there (see figure 7c in~\cite{Francisquez2023}).

\begin{figure}[h]
  \centering
  \begin{subfigure}[h]{0.35\textwidth}
    \includegraphics[width=\textwidth]{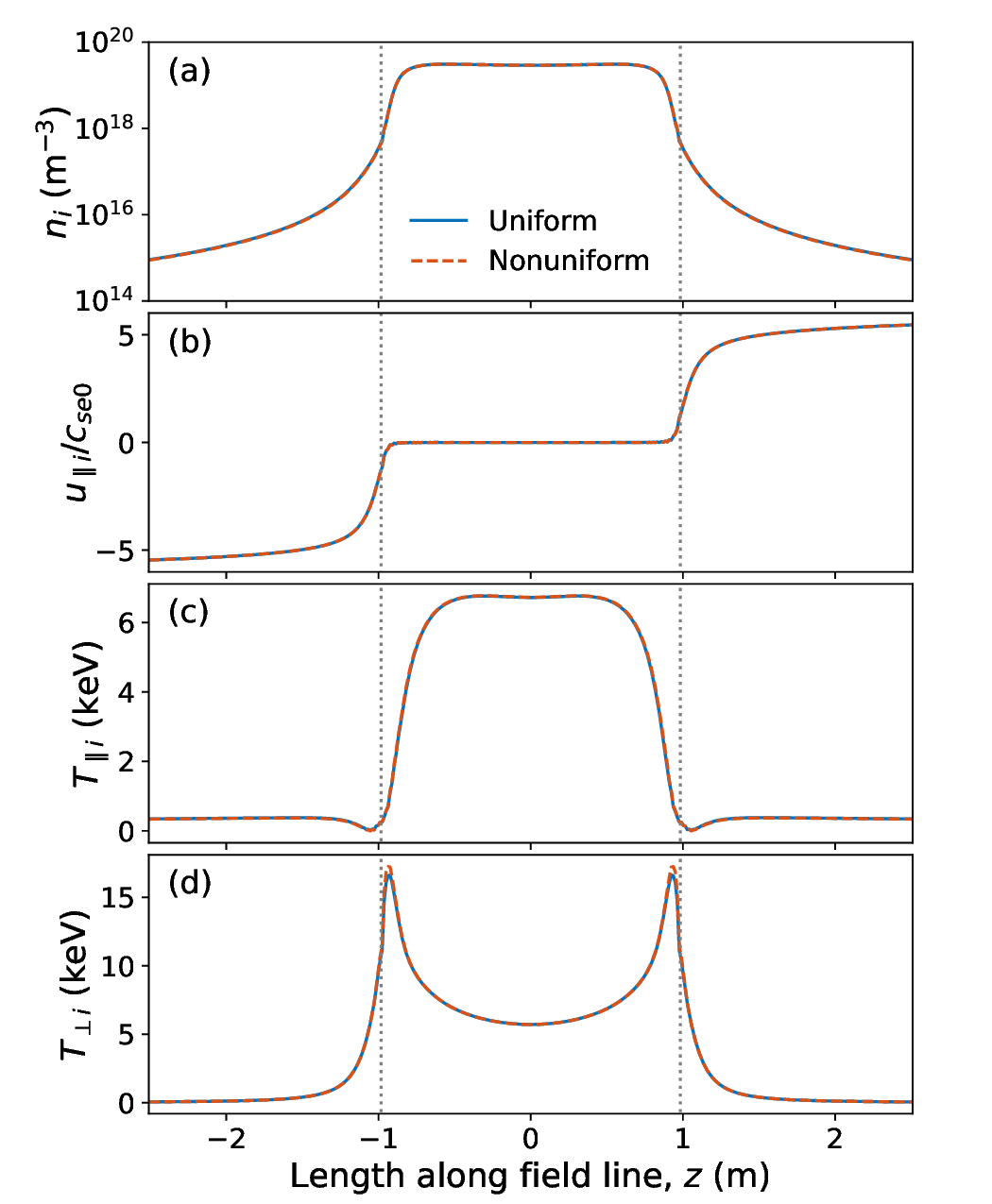}
  \end{subfigure}
  \begin{subfigure}[h]{0.4\textwidth}
    \includegraphics[width=\textwidth]{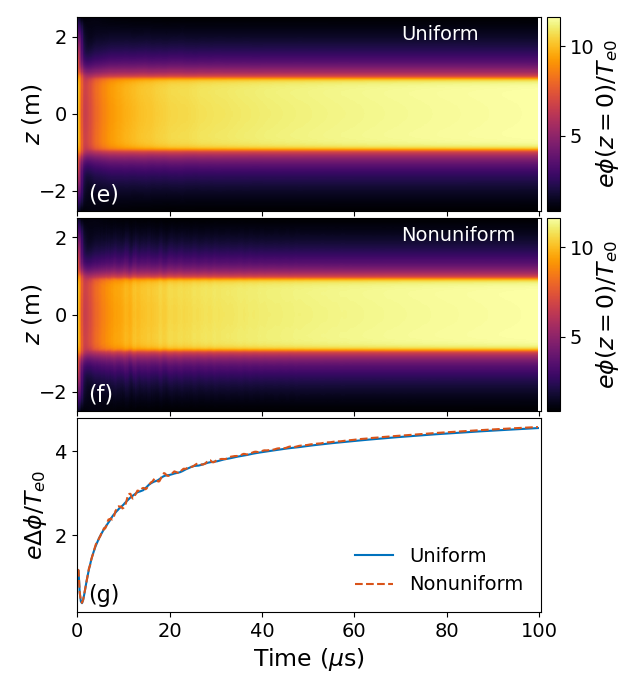}
  \end{subfigure}
\caption{\label{fig:adiabaticMomsPhiComp}Left: Ion guiding center density ($n_i$), (b) mean drift speed ($u_{\parallel i}$) normalized to $c_{se0}=\sqrt{T_{e0}/m_i}$, and (c) parallel and (d) perpendicular temperatures ($\Tpari$, $\Tperpi$) at $t=48\,\mu\mathrm{s}\approx(8/3)\nu_{ee}^{-1}$ with uniform (solid blue) and nonuniform (dashed orange) velocity-space discretization. Vertical dotted lines indicate the mirror throat locations. Right: Spatio-temporal evolution of the electrostatic potential in uniform (e) and nonuniform (f) velocity-space discretizations. The potential drop between the mirror center and throats ($\Delta\phi$) normalized by the central electron temperature is shown as a function of time (g).}
\end{figure}

In order to more cleanly evaluate the accuracy and performance of the nonuniform velocity maps, without conflating the latter with speed-ups provided by GPUs, we also ran two simulations with uniform velocity grids. First, we attempted to include only the region that was linearly mapped in the nonuniform map simulation, i.e. $\abs{\vpar}\leq4v_{ti0}$, but this suffered from the accumulation of $f_i$ at the boundary described at length in previous work~\cite{Francisquez2023}, becoming unstable and unable to proceed. We thus performed a second uniform velocity grid simulation, this time with the same $\vparmax$ and $\mumax$ used in the nonuniform velocity map simulation, but increasing the number of cells so that $\Dvpar$ and $\Dmu$ have the same size as the smallest of these in the nonuniform velocity map simulation, $192\times120\times400$ cells. Using 2 NVIDIA A100 GPUs, it took 100 hours
to reach $t=5.56\nu_{ee}^{-1}=100~\mu$s when using this uniform velocity grid. Figures~\ref{fig:adiabaticMomsPhiComp}(a-d) compares moments of the uniform and nonuniform velocity-space distributions. The density, parallel flow (normalized to $c_{se0}=\sqrt{T_{e0}/m_i}$), and the parallel $T_{\parallel i}$ and perpendicular $T_{\perp i}$ temperatures are all essentially the same in uniform and nonuniform velocity grid simulations. Furthermore, the original physics intent of this exercise was to check whether this model produces a Pastukhov-level ambipolar potential, which helps confine electrons and reduce parallel electron heat losses. As demonstrated in figures~\ref{fig:adiabaticMomsPhiComp}(e-g), the evolution of the potential drop across the core and the mirror throat ($e\Delta \phi/T_{e0}$) is the same in both simulations. The potential profiles in uniformly and nonuniformly discretized simulations are also essentially the same (not shown, but it can be interpreted from figures~\ref{fig:adiabaticMomsPhiComp}(e-f)).

The nonuniform velocity map simulation used 62X fewer velocity-space cells and took only 2 hours on 2 GPUs when the uniform velocity grid simulation took 100 hours on the same hardware, signifying a 50X speed-up in time-to-solution. We remark that it may be possible to use even fewer cells along $z$; for example, with 128 cells along $z$ (a 94X reduction in overall number of cells) the potential drop $e\Delta\phi/T_{e0}$, $n_i$, $u_{\parallel i}$ and $T_{\perp i}$ all look nearly identical to those from the nonuniform simulation presented here, but $T_{\parallel i}$ drops below zero in the expander, which is why we did not include it here. Other methods, such as enforcing positivity of $f_s$, are being explored in order to make these coarser simulations more robust.

\subsubsection{2D axisymmetric ASDEX-Upgrade} \label{sec:res_full_2x}

\begin{figure}[h]
  \centering
  \begin{subfigure}[h]{0.32\textwidth}
    \includegraphics[width=\textwidth]{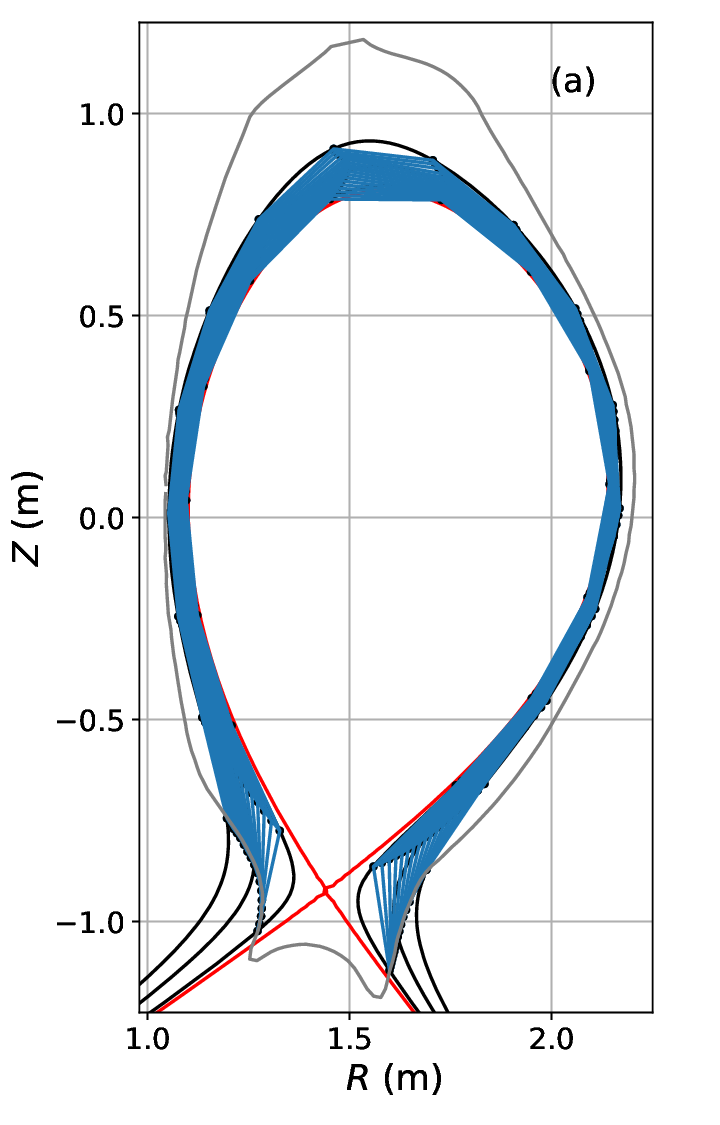}
  \end{subfigure}
  \begin{subfigure}[h]{0.35\textwidth}
    \includegraphics[width=\textwidth]{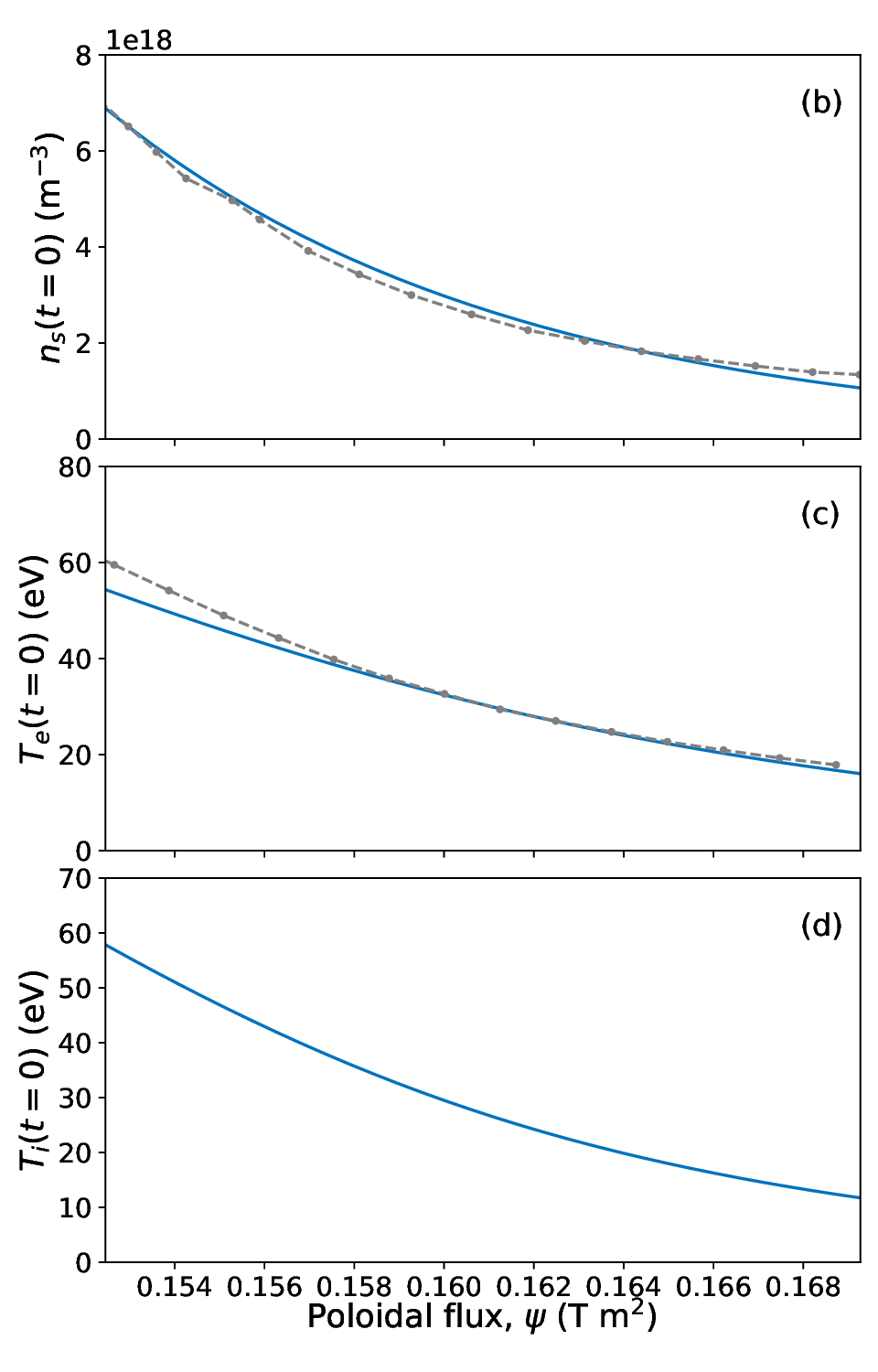}
  \end{subfigure}
\caption{\label{fig:augSetup}(a) Simulation domain of AUG 36190 SOL (blue), lower, middle and upper flux surfaces of the domain (black), the separatrix (red) and the vacuum vessel (gray). (b-d) Initial velocity-moments of the distribution functions for the AUG 36190 SOL simulation, near the outboard midplane (OMP), with the experimental profiles shown in dashed gray (experimental $T_i$ was not available for this shot)~\cite{Michels2022}.}
\end{figure}

Field-aligned coordinates $\vx=\left(\psi,-\alpha,\theta\right)$ where $\psi$ is the poloidal flux, $\theta$ is the normalized poloidal arc length, and $\alpha$ is the coordinate binormal to $\psi$ and $\theta$~\cite{Beer1995}, can be employed for simulating tokamaks in the perfectly axisymmetric limit, where we assume that all quantities are independent of $\alpha$. In this case the gyrokinetic model reduces to a 2D2V problem, and due to the absence of self-consistent anomalous (turbulent) cross-field transport, we employ the diffusion term to mimic cross-field transport (see equations~\ref{eq:gkeq} and~\ref{eq:app_diffModel_final}). We here demonstrate that axisymmetric tokamak simulations are well-behaved with the curvilinear, nonuniform velocity algorithm presented in this work. For this purpose we approximate the ASDEX-Upgrade discharge 36190 that was previously explored with 3D2V turbulent simulations~\cite{Michels2022}. We do not cover the entire poloidal annulus and instead focus on the scrape-off layer (SOL) only (newer capabilities allow including the X-point and a full poloidal annulus~\cite{Francisquez2025}). A diagram showing the grid employed is given in figure~\ref{fig:augSetup}(a). This grid covers the flux surfaces $\psi=\left[0.1522,0.1695\right]$. Velocity space is discretized with the same linear-quadratic hybrid mapping in equation~\ref{eq:linquad_map}, using the reference electron and deuterium temperatures $T_{e0}=37.5$ eV and $T_{i0}=38$ eV, respectively, with $\vparmax=6v_{ts0}$, $\mumax=m_s\left(4v_{ts0}\right)^2/(2B_0)$ and $B_0=2.934$ T. Position space is meshed with $16\times16$ cells, while velocity space is meshed with $16\times8$ cells (we performed a resolution convergence study going up to $64\times64$ cells in position space, and $32\times16$ cells in velocity space). Neutrals~\cite{Bernard2022} are not included, although they have improved agreement with experiment in turbulent fluid simulations~\cite{Zholobenko2021b}.

Both deuterium and electrons are initialized as being Maxwellian everywhere, with velocity moments that approximate experimentally measured profiles~\cite{Michels2022}. These profiles, at the outboard midplane, are shown in figures~\ref{fig:augSetup}(b-d). We mimic radial transport from the core, and compensate for particle losses to the wall (which would physically be compensated by external fueling or recycling), including an explicit source
\begin{equation}
    \mathcal{S}_s = \frac{\dot{n}_{\mathrm{src}}}{\left(2\pi v_{t,\mathrm{src}}^2\right)^{3/2}} f_{Ms}\left(\vpar,\mu;v_{t,\mathrm{src}}^2\right)
\end{equation}
with Maxwellian shape ($f_M$) in velocity space, 
temperature $T_{s,\mathrm{src}}=m_sv_{t,\mathrm{src}}^2=4.485\,T_{s0}$ and particle source rate
\begin{equation}
    \dot{n}_{\mathrm{src}} = \frac{4.2\,n_0c_{se0}}{L_c/4}\exp\left[-\frac{\left(\psi-\psi_{\mathrm{src}}\right)^2}{2\sigma_{\psi,\mathrm{src}}}\right] ~
    \exp\left[-\frac{1}{2}\left(\frac{\theta-\theta_{\mathrm{src}}}{\sigma_{\theta,\mathrm{src}}}\right)^6\right],
\end{equation}
where $c_{se0}^2=T_{e0}/m_i$, $n_0=0.4\times10^{19}$ m$^{-3}$, $L_c=67$ m, $\psi_{\mathrm{src}}=0.1522$, $\sigma_\psi=0.14\left(\psi_{\max}-\psi_{\min}\right)$, $\theta_{\mathrm{src}}=-1.42$ rad and $\sigma_\theta=1$ rad. A Maxwellian equal to the initial condition is also placed in the inner-most radial ghost cell, which provides an additional (weaker) source wherever the $\grad{B}$ drift points into our domain and compensates the losses in parts of the domain where $\grad{B}$ points out of our domain. Conducting sheath BCs are used at the divertor plates~\cite{Shi2017thesis}, and $\phi=0$ boundary conditions are employed at $\psi=\psi_{\min},\,\psi_{\max}$. Lastly, given the absence of self-consistent anomalous (i.e. turbulent) transport in these 2D simulations, we employ the diffusive transport model in equation~\ref{eq:app_diffModel_final}. Note that this model does not allow us to directly set the heat diffusivity, which is simply $\chi_s = (3/2)D_s$.

\begin{figure}[h]
  \centering
  \includegraphics[width=0.85\textwidth]{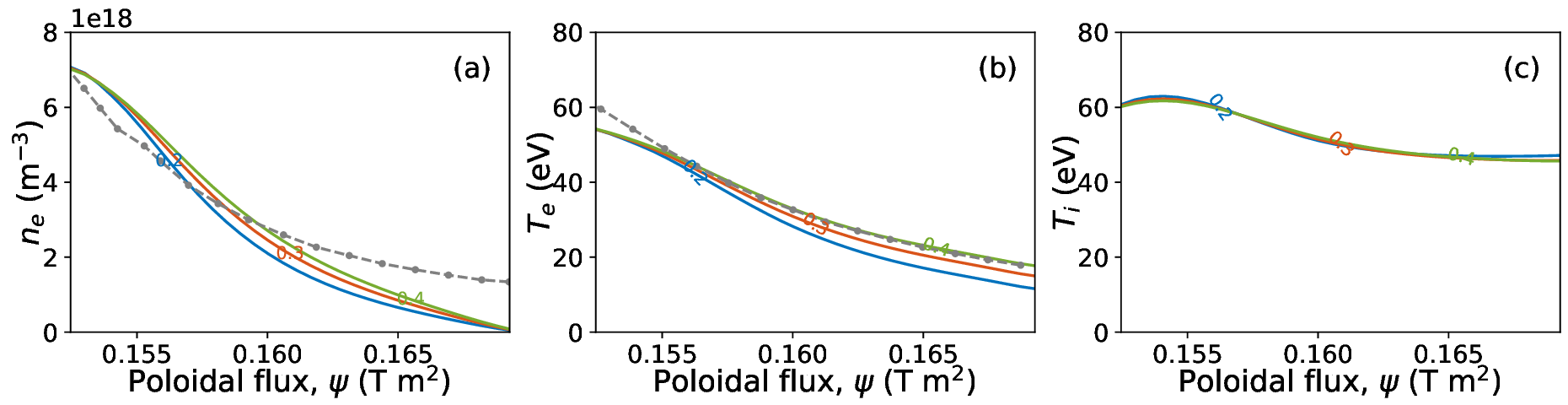}
\caption{\label{fig:augFinalProf} Radial profiles of electron density (a), and electron (b) and ion (c) temperature at the outboard midplane in ASDEX simulation for three different levels of anomalous diffusive coefficient: $D=0.2$ m$^2$/s (solid blue), $D=0.3$ m$^2$/s (solid orange), $D=0.4$ m$^2$/s (solid green). The experimental profiles are shown in dashed gray~\cite{Michels2022}.}
\end{figure}

\begin{figure}[h]
  \centering
  \begin{subfigure}[h]{0.35\textwidth}
    \includegraphics[width=\textwidth]{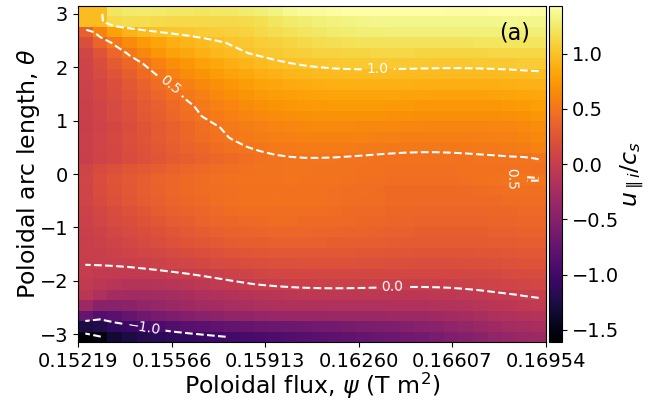}
  \end{subfigure}
  \begin{subfigure}[h]{0.35\textwidth}
    \includegraphics[width=\textwidth]{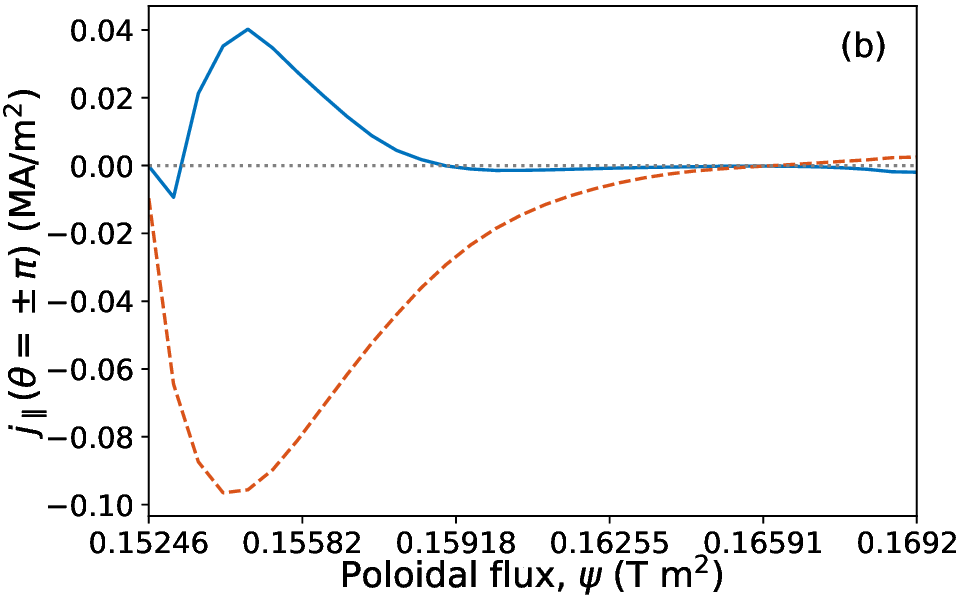}
  \end{subfigure}
\caption{\label{fig:augFlows}(a) Ion parallel drift speed ($u_{\parallel i}$) normalized to the local sound speed ($c_s^2=(T_e+T_i)/m_i$). (b) Parallel current at the inboard (dashed orange) and outboard (solid blue) divertor plates.}
\end{figure}

As the simulation progresses the sources and sinks evolve towards a steady state. Once this steady state is reached ($t\approx 4$ ms) we extract radial, outboard midplane profiles such as those in figure~\ref{fig:augFinalProf}. These results were produced for three choices of the anomalous transport coefficient $D_s$. Note that compared to the experimental profile (dashed grey) the electron density reached by Gkeyll is quite close, except for the far SOL ($\psi>0.165$). 
The electron temperature (figure~\ref{fig:augFinalProf}(b)) also exhibits a good agreement with the experiment, at least for $D_s=0.4$ m$^2$/s, with a relatively small level of disagreement near the core boundary ($\psi_{\min}$), but this location is very heavily impacted by our choice of BCs; in particular, more realistic near-SOL values likely require simulating the core and there is ongoing work to do so~\cite{Francisquez2025}. This level of agreement is fortuitous given our limited model for the heat diffusivity $\chi_s=(3/2)D_s$ and other approximations (e.g. omitting neutrals).
The ion temperature in figure~\ref{fig:augFinalProf}(c) is not compared to an experimental measurement since such data was unavailable for this discharge, but its increased magnitude relative to $T_e$ is likely a consequence of the slower parallel ion heat loss.

A unique feature in Gkeyll is its conducting sheath BCs. These BCs do not explicitly enforce a Bohm criterion~\cite{Stangeby2000} or a vanishing current~\cite{Shi2015,Boesl2019,Geyko2023} at the sheath entrance near the divertor plates. Yet, as shown in figure~\ref{fig:augFlows}(a), the simulation can self-consistently produce regions satisfying the $u_{\parallel i}\geq c_s$ Bohm criterion, although such result is not guaranteed. We can extract the parallel current at the divertor plates (figure~\ref{fig:augFlows}(b)) and note that they do trend towards ambipolarity in the far SOL, but in the near SOL finite currents into or out of the wall can persist.

This ASDEX SOL simulation employing nonuniform velocity maps ran to $t=4$ ms in 13.5 hours on 2 NVIDIA A100 GPUs. We attempted to run two uniform velocity grid simulations for a direct comparison as we did for the mirror simulation in section~\ref{sec:res_full_1x}, one with the same $\mumax$ but $\vparmax=3v_{ts0}$ (i.e. the linearly mapped part of the nonuniform velocity simulation) and $32\times64$ cells, and another with the same velocity limits and $66\times64$ cells in velocity space. Both cases were not feasible because the time step became 46 time smaller. The reason for this drop is found in the CFL constraint for the diffusive part of the Dougherty collision operator, equation~\ref{eq:lbo_cfl_diff}. Given that the diffusion term along $\mu$ has an effective diffusivity proportional to $\mu$, the time step required by this term scales like $\Delta t\sim(\Dmu)^2/\mumax$. So although we ensured that both uniform and nonuniform map simulations had the same minimum $\Dmu$, this cell length at $\mumax$ was much larger in the nonuniform map simulation. Despite this limitation we evolved the uniform velocity grid case until $t=0.37$ ms, at which point it exhibited nearly the same radial profiles as the nonuniform case (not shown because it is not the steady state solution). This uniform grid simulation took 50 hours on 2 A100 NVIDIA GPUs, meaning that nonuniform velocity maps provided a 60X speed-up in this case.

\subsubsection{3D LAPD turbulence} \label{sec:res_full_3x}

\begin{figure}[t]
  \centering    \includegraphics[width=\textwidth]{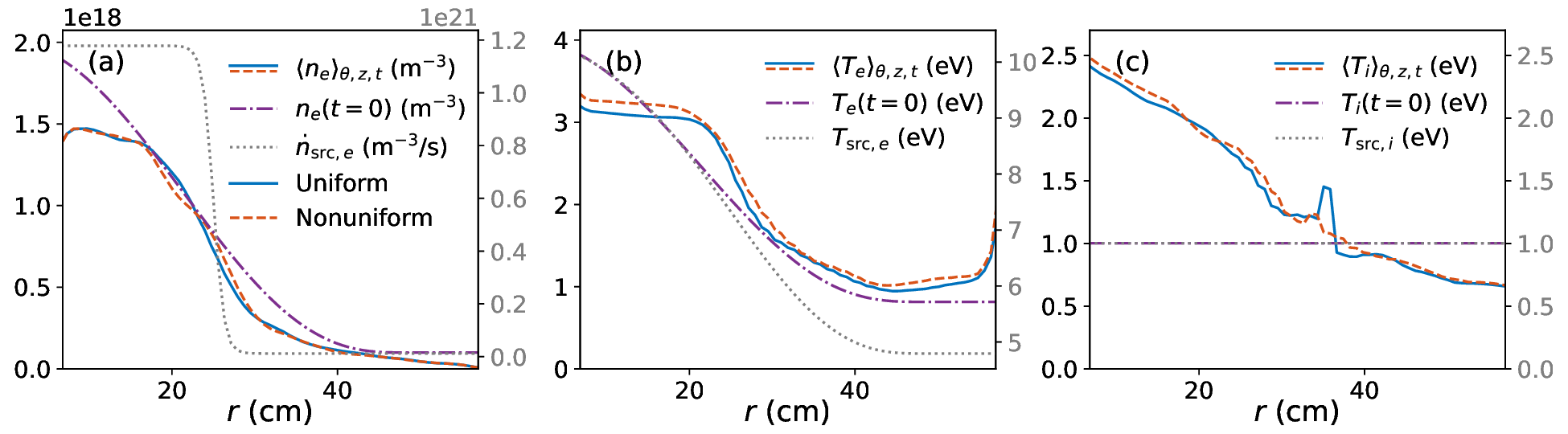}
\caption{\label{fig:lapdProfiles} Electron density (a) and temperature (b), and ion temperature (c) averaged over $\theta\in[-\pi,\pi]$, $z\in[-4.5~\mathrm{m},4.5~\mathrm{m}]$ and $t\in[4~\mathrm{ms},6~\mathrm{ms}]$ in the uniformly (solid blue) and nonuniformly (dashed orange) discretized velocity-space simulations. Purple dash-dot line shows the initial profiles, and grey dotted lines provide the density rate (a) and temperatures (b-c) of the sources.}
\end{figure}

The Large Plasma Device (LAPD) at the University of California Los Angeles offers a useful testbed for turbulence codes as it has some of the basic ingredients of a SOL: open field lines, sheaths, background gradients, competition of parallel and perpendicular dynamics, and sheared flows. This platform has been used for many years to test fluid~\cite{Fisher2015} and gyrokinetic~\cite{Shi2017,Frei2024}  turbulence codes. Here we also use this machine as a testbed for the new nonuniform velocity-space discretization. We use parameters similar to those used previously~\cite{Shi2015}, i.e. $n_0=10^{18}$ m$^{-3}$, $T_{e0}=2$ eV, $T_{i0}=1$ eV, $B=0.0398$ T, $m_{i}=m_{He}$, $m_e=m_i/400$, and do not include neutrals~\cite{Bernard2022}. The initial density and temperatures are given by blue dashed lines in figure~\ref{fig:lapdProfiles}. Unlike previous simulations in a Cartesian box, we use a cylindrical domain to further stress test our algorithms with both curvilinear position coordinates and nonuniform velocity maps. We do not presently include the cylindrical axis (to avoid numerical complexities associated with computing fluxes and $\Dt$ becoming very small at the center point), so we place a boundary at 5\% of the maximum radius, and at this boundary particles are absorbed and the quasineutrality equation is solved with a $\phi=0$ BC (the outer radial boundary also uses a $\phi=0$ BC, as well as zero-flux BCs for the collisionless terms, while the $z$ boundaries use conducting sheath BCs~\cite{Shi2017thesis}). Furthermore, we employ a radially centered and peaked particle and heat source (see grey dotted lines in figure~\ref{fig:lapdProfiles}), with an amplitude of $\dot{n}_{\mathrm{src}}=1.08\,2^{3/2}\,n_0\,c_{se0}/L_z$ (here $L_z=18$ m is the length of LAPD) and a temperature of $T_{\mathrm{src},e}=12$ eV and $T_{\mathrm{src},i}=1$ eV. These simulations were carried out using $36\times36\times8\times8\times4$ cells to discretize a computational domain $(r,\theta,z,\cvpar,\cmu)\in\left[r_{\min},r_{\max}\right]\times\left[-\pi,\pi\right]\times\left[-\Lz/2,\Lz/2\right]\times\left[-2^{-1/2},2^{-1/2}\right]\times\left[0,1\right]$, $r_{\min}=0.03$ m, $r_{\max}=0.6$ m and the velocity-space mapping is again the linear-quadratic hybrid in equation~\ref{eq:linquad_map}, using $\vparmax=6v_{t0}$ and $\mumax=m_s\left(6v_{t0}\right)^2/(2B)$. A resolution convergence study ensured that this resolution produced essentially the same results compared to using $128\times128$ cells in $r$-$\theta$ or increasing the number of cells along $z$ to 24.

As the plasma is sourced and the profiles steepen, a layer with steep pressure and velocity gradients form near the edge of the source at $r\approx25$ cm. Eventually instabilities drawing energy from these gradients grow to large amplitudes and nonlinearly interact to form a turbulent state. We run these simulations well into steady state, until $t=6$ ms. In the last 2 ms of the simulation, density and temperature profiles are averaged over time, $\theta$ and $z\in\left[-4.5\,\mathrm{m}\,,4.5\,\mathrm{m}\right]$; these steady state profiles are shown with solid lines in figure~\ref{fig:lapdProfiles}. The density and electron temperature profile are in qualitative agreement with those obtained from similar experiments~\cite{Schaeffer1998} and ealier simulations using Cartesian domains~\cite{Shi2017}. Artifacts of the inner radial boundary do appear; for example, $n_e$ dips near the inner radial boundary. At the outer radial boundary Gkeyll's density drops by $10^3$ compared to the core and there is an increase in $T_e$, both of which may be a symptom of lacking neutral interactions as wall recycling and electron-impact ionization would both increase the plasma density and cool the electrons near the edge. A detailed validation with experiments is beyond the scope of this work and likely requires including neutral interactions, as well as reexamining the appropriateness of our heating and fueling method, reduced mass ratio, and magnitude of $B$ (the experimental $B$ was 4X higher).

\begin{figure}[h]
  \centering    \includegraphics[width=\textwidth]{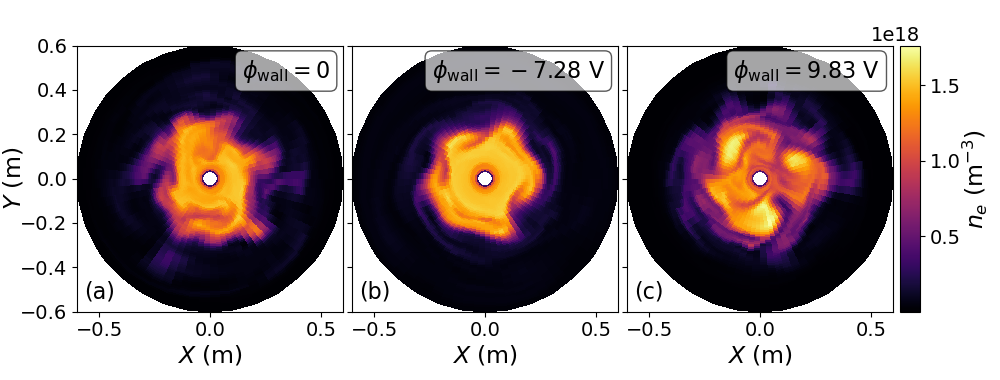}
\caption{\label{fig:lapdM0comp}Snapshot of the electron density at the center of the device ($z=0$) in the unbiased case at $t=6$ ms (a), and -7.275 V (b) and 9.825 V (c) biasing cases 3 ms after the biasing was turned on.}
\end{figure}

\begin{figure}[h]
  \centering
  \includegraphics[width=0.45\textwidth]{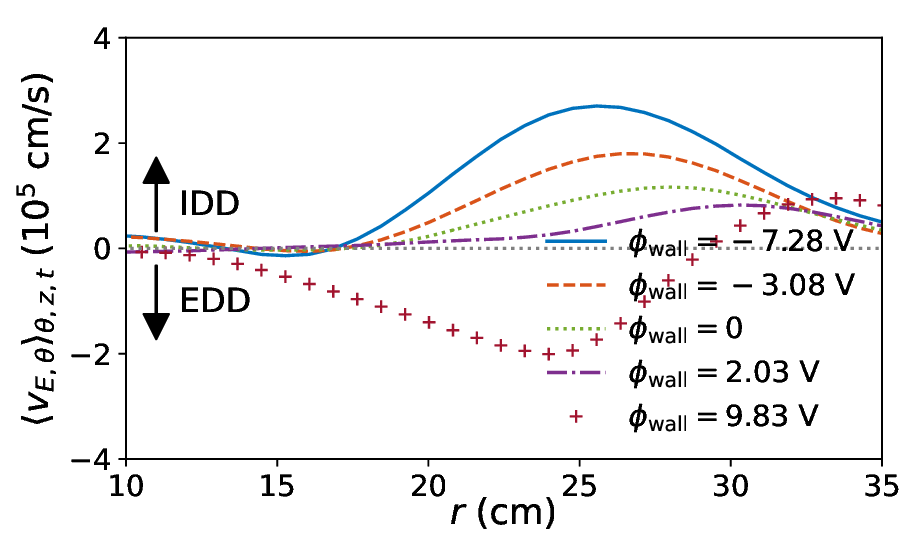}
\caption{\label{fig:lapdvEcomp}Azimuthal component of $E\times B$ drift, averaged over 1 ms, $\theta$ and $z\in[-4.5,4.5]$, for a range of wall bias voltages. Arrows indicate the ion (IDD) and electron (EDD) diamagnetic directions.}
\end{figure}

A snapshot of the turbulent density at $t=6$ ms is provided in figure~\ref{fig:lapdM0comp}(a). Note that this simulation assumed a grounded wall, but as in previous versions of Gkeyll~\cite{Shi2017thesis}, we can bias the wall to reproduce the suppression of turbulence by increasing the azimuthal flow and its shear~\cite{Schaeffer1998}. We do so by restarting the unbiased simulation at $t=3$ ms with varying biasing potentials and running the simulation for an additional 3 ms. In figure~\ref{fig:lapdM0comp}(b)-(c) we show a snapshot of the simulations with the biasing voltages $\phi_{\mathrm{wall}}=-7.28$ V and $\phi_{\mathrm{wall}}=9.83$ V at $t=6$ ms (noting that the biasing was turned on at $t=3$ ms). At these relatively large biases, strong azimuthal flows form and hinder the formation of radially aligned fluctuations. As a result, the biased cases are less turbulent. The changes in azimuthal flow, specifically the azimuthal component of the $E\times B$ drift, for a range of biases are given in figure~\ref{fig:lapdvEcomp}. We see that in the absence of any biasing $\phi_{\mathrm{wall}}=0$ the plasma has a natural weak rotation in the ion diamagnetic direction, which is enhanced as a negative bias strengthens. But with sufficiently large positive bias the flow reverses to the electron diamagnetic direction. These trends qualitatively agree with previous biasing experiments on LAPD~\cite{Schaeffer1998}.

In order to assess the accuracy and performance of the grounded simulation with the nonuniform velocity map simulation, we attempted to run two uniform velocity grid simulations with the same velocity-space cell lengths as the minimum $\Dvpar$ and $\Dmu$ in the the nonuniform simulation. The first of these simulations used $6\times16$ cells in velocity space and $\vparmax=3v_{t0}$, which is where the linearly mapped region of the nonuniform simulation ended. This simulation was unsuccessful because $\vparmax$ is too low, leading to accumulation of $f_s$ near the velocity boundary that produces large gradients and diminishing $\Dt$. A second uniform simulation used the same $\vparmax$ as the nonuniform case, but with $12\times16$ cells in velocity space. This case took 104 hours to reach $t=6$ ms on 4 NVIDIA A100 GPUs, while the nonuniform velocity case can accomplish this on the same hardware in 4.8 hours, entailing a 22X speed-up. Plasma profiles for these two simulations are very similar, see blue solid and dashed orange lines in figure~\ref{fig:lapdProfiles}.

\begin{figure}[h]
  \centering
  \begin{subfigure}[h]{0.45\textwidth}
    \includegraphics[width=\textwidth]{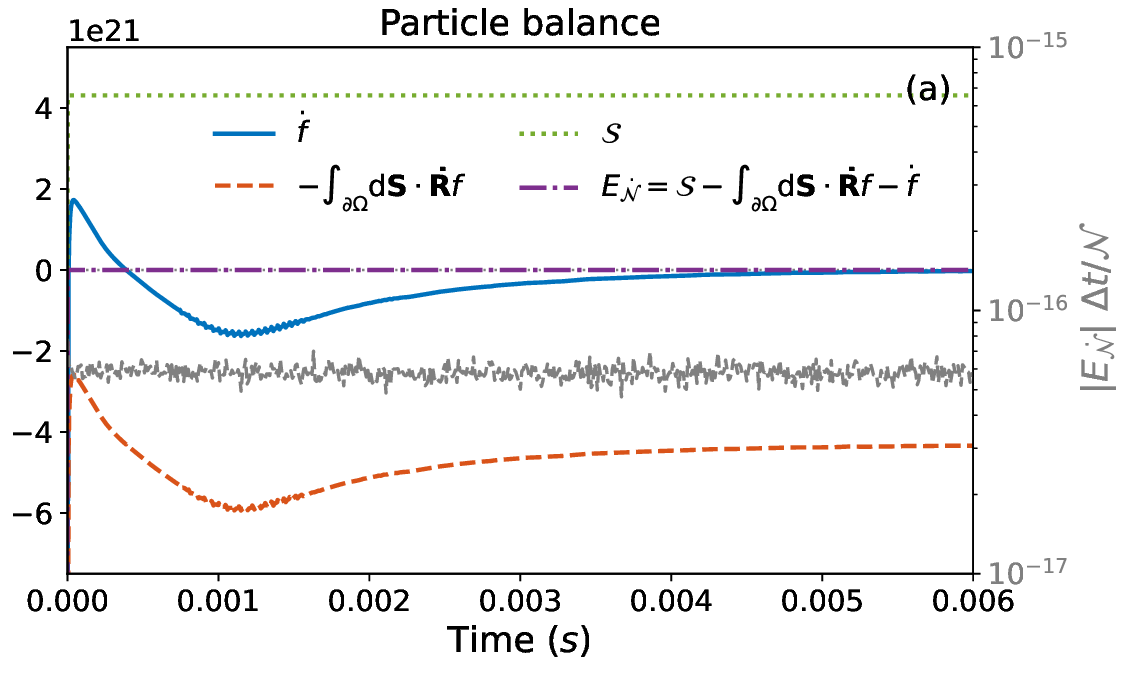}
  \end{subfigure}
  \begin{subfigure}[h]{0.45\textwidth}
    \includegraphics[width=\textwidth]{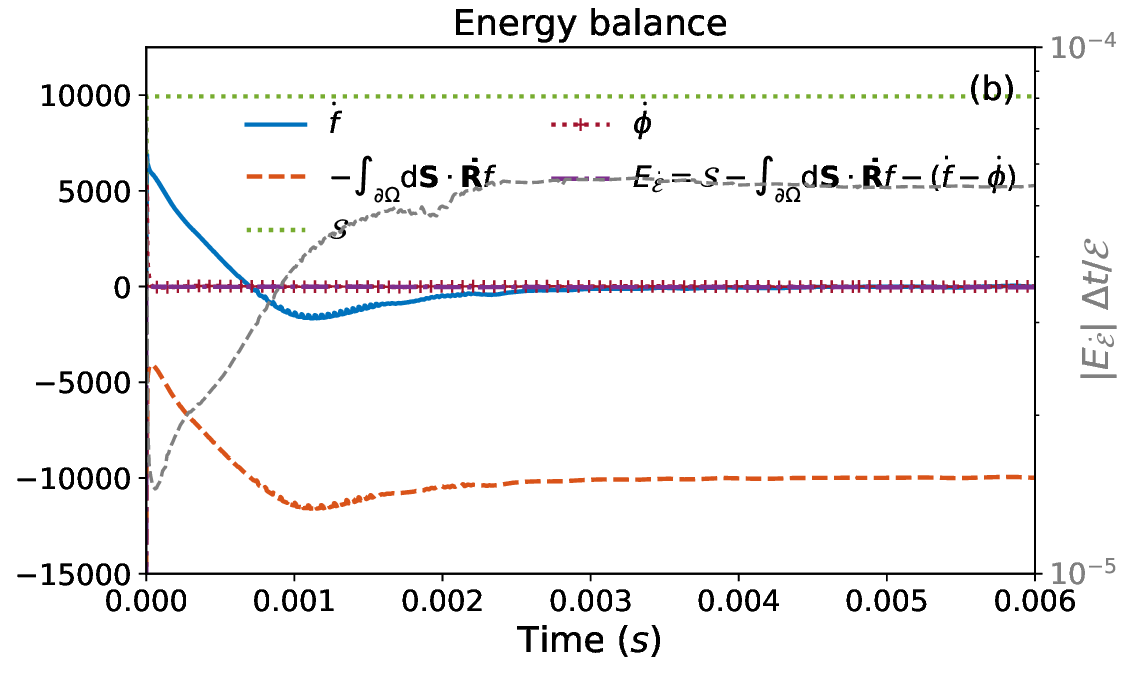}
  \end{subfigure}
\caption{\label{fig:lapdBalance}Particle (a) and energy (b) balance of unbiased LAPD simulation nonuniform velocity discretization. The balance includes the change in a single time step (solid blue) due to fluxes through the boundaries (dashed orange) and sources (dotted green), as well as the error in adding these up (purple dash-dot). Dashed grey lines show the error relative to the time step ($\Dt$) and the total particle ($\mathcal{N}$) and energy ($\mathcal{E}$) content, respectively.}
\end{figure}

Lastly, in section~\ref{sec:algo} and~\ref{sec:conserv} we claimed that the algorithm with nonuniform velocity discretization conserves particles and energy exactly. We can diagnose the extent to which this property holds in these 3D turbulent simulations. To do so we revisit the particle balance equation~\ref{eq:Ndot_final}, including the contribution due to the source:
\begin{eqnal} \label{eq:Ndot_final_wsrc}
    &\underbrace{\d{\mathcal{N}}{t}}_{\dot{f}} = \underbrace{- \sum_{i=1}^{\cdim}\frac{2}{\Dxi{i}}\left(\sum_{j=1}^{\Nperp{x^i_{\max}}}\oint_{K_j}\mathrm{d}\v{S_i}\,\dvlog\,\dmulog\,\Ghat^{\pm}_{x^i,j}\Big|_{\xlog{i}=1} - \sum_{j=1}^{\Nperp{x^i_{\min}}}\oint_{K_j}\mathrm{d}\v{S_i}\,\dvlog\,\dmulog\,\Ghat^{\pm}_{x^i,j}\Big|_{\xlog{i}=-1}\right)}_{-\int_{\partial\Omega}\mathrm{d}\v{S}\cdot\v{\dot{R}}\dot{f}} + \underbrace{\mathbb{V}\sum_{j=1}^{N}\int_{K_j}\dvxlog\,\dvlog\,\dmulog\,\jacobTots{j} \mathcal{S}_j}_{\mathcal{S}}.
\end{eqnal}
There should also be a collision term on the right-hand side of this equation but, as we showed in section~\ref{sec:res_collision}, it conserves particles exactly. We can thus plot the time trace of each of these terms (see figure~\ref{fig:lapdBalance}(a)), as well as the error in particle conservation
\begin{eqnal}
    E_{\dot{\mathcal{N}}} = \mathcal{S} - \int_{\partial\Omega}\mathrm{d}\v{S}\cdot\v{\dot{R}}\dot{f} - \dot{f}
\end{eqnal}
to show that the error in energy conservation is essentially zero; to be precise, if normalized to the ratio of the time step and the particle content ($\Dt/\mathcal{N}$) we would find that the error is of order of machine precision. We can likewise perform a similar accounting of total energy, revisiting equation~\ref{eq:EdotFinal}
\begin{eqnal} \label{eq:EdotFinal_wsrc}
    \underbrace{\d{\mathcal{E}_H}{t}}_{\dot{f}} - \underbrace{\d{\mathcal{E_\phi}}{t}}_{\dot{\phi}} &= \underbrace{- \sum_{s}\mathbb{V}_s\sum_{i=1}^{\cdim}\left(\sum_{j=1}^{\Nperp{x^i_{\max}}}\frac{2}{\Dxi{i}}\oint_{\partial K_j}\mathrm{d}\v{S_i}\,\dvlog\,\dmulog\,H_{s,j\pm}\Ghat^{\pm}_{x^i,j}\Big|_{\xlog{i}=1} - \sum_{j=1}^{\Nperp{x^i_{\min}}}\frac{2}{\Dxi{i}}\oint_{\partial K_j}\mathrm{d}\v{S_i}\,\dvlog\,\dmulog\,H_{s,j\pm}\Ghat^{\pm}_{x^i,j}\Big|_{\xlog{i}=-1}\right)}_{-\int_{\partial\Omega}\mathrm{d}\v{S}\cdot\v{\dot{R}}\dot{f}} \\
    &\quad+ \underbrace{\mathbb{V}\sum_{j=1}^{N}\int_{K_j}\dvxlog\,\dvlog\,\dmulog\,H_j\jacobTots{j} \mathcal{S}_j}_{\mathcal{S}},
\end{eqnal}
and plotting each of these contributions as well as the error in energy conservation
\begin{eqnal}
    E_{\dot{\mathcal{E}}} = \mathcal{S} - \int_{\partial\Omega}\mathrm{d}\v{S}\cdot\v{\dot{R}}\dot{f} - \left(\dot{f}-\dot{\phi}\right).
\end{eqnal}
As shown in figure~\ref{fig:lapdBalance}(b) this error is orders of magnitude smaller than all the other terms governing the dynamics and determining the steady state. The energy conservation error is not machine precision due to the non-reversible, explicit Runge-Kutta time integrator used, which introduces a small amount of diffusion~\cite{Hakim2019}; this error is independent of phase-space resolution and is reduced by using a smaller time step (reducing $f_{\mathrm{CFL}}$ in equation~\ref{eq:cflConstraintDef}).

\section{Summary} \label{sec:conclusion}

This article presents an approach to nonuniform velocity-space discretization in discontinuous Galerkin (DG) continuum kinetic models, using conservative discrete mapped coordinates and demonstrating it in the context of a full-$f$ gyrokinetic model for studying magnetized plasmas. The method consists of rewriting the equations in terms of new computational coordinates and specifying a univariate mapping between physical and computational velocity coordinates. The velocity-space mapping is discretized using DG in a way that results in a continuous inter-cell representation. By carefully designing the numerical fluxes we can ensure that the algorithm for collisionless terms conserves the number of particles and total energy to machine precision regardless of resolution in the absence of temporal discretization error. In addition, collision terms conserve particles, momentum and energy exactly without the additional complications incurred by a pure linear basis~\cite{Francisquez2022} thanks to our novel hybrid basis.

Several tests were performed, independently for collisionless and collisional terms, and also including both types of terms in 1D, 2D, and 3D simulations. The tests show the algorithm's capacity to produce reasonable results with the new nonuniform velocity-space discretization, in a variety of magnetically aligned domains, such as high-temperature superconducting (HTS) mirrors, diverted tokamak scrape-off layers, and turbulent cylindrical plasmas. In each of these cases, simulations with nonuniform velocity discretization reproduce previous results and in some cases, such as in HTS mirror simulations, the number of cells was reduced by as much as a factor of 30. As a result, calculations took 22-60 times less wall-clock time than a uniform velocity grid simulation with equivalent resolution.

This new capability will allow future DG gyrokinetic simulations to use fewer degrees of freedom (and thus fewer computational resources) or better represent parts of phase-space. Future work on the topic of better or more flexible velocity-space representations could include using position-space-dependent mappings~\cite{Taitano2016} or using block-structured grids in velocity space with different reference temperatures for each block~\cite{Jarema2017}. A related desirable capability is that, given that the background temperature evolves in time, it could be helpful to have the option to restart with a different reference temperature and thus a different velocity-space grid, requiring interpolation between the two; such option could bring stability and acceleration to long-running full-$f$ simulations. Lastly, we presented a demonstration of 1D, 2D and 3D simulations with the new nonuniform velocity discretization but an in-depth validation against experiments is beyond the scope of this manuscript; such comparisons are ongoing and will be presented as separate publications.

\section*{Acknowledgments}
We thank the rest of the Gkeyll team for stress-testing the nonuniform velocity capability, Ammar Hakim for providing insights into the use of DG for Hamiltonian systems, and Maxwell Rosen for developing new velocity maps for studying HTS mirrors. This work was supported by the U.S. Department of Energy (DOE) as part of the Computational Evaluation
and Design of Actuators for Core-Edge Integration (CEDA) project in the Scientific Discovery Through Advanced Computing (SciDAC) program under contracts DE-AC02-09CH11466 and DE-FG02-
04ER54742, as well as DOE's Distinguished Scientist program and LDRD grants via DOE contract DE-AC02-09CH11466 for the Princeton Plasma Physics Laboratory. This research used resources of the National Energy Research Scientific Computing Center (NERSC), a Department of Energy User Facility using NERSC awards FES-ERCAP m2116 and Science at Scale m5053, as well as the Stellar cluster supported by Princeton Research Computing, a consortium of groups including the Princeton Institute for Computational Science and Engineering (PICSciE) and the Office of Information Technology's High Performance Computing Center and Visualization Laboratory at Princeton University.



\appendix

\section{Nonuniform velocity space in Vlasov-Poisson} \label{sec:vlasov}

The approach to nonuniform velocity space discretization presented in this work is applicable to other Vlasov systems. We sketch out how using a 1D1V Vlasov-Poisson system as an example; generalization to higher dimensions is straightforward. Assuming a unit charge and mass, the Vlasov-Poisson equation is
\begin{equation} \label{eq:vp}
    \pd{f_s}{t} + \pd{}{x}\dot{x}f_s + \pd{}{v}\dot{v}f_s = 0
\end{equation}
where $\dot{x}=v$ and $\dot{v}=-\partial_x\phi$, and $\phi$ is given by the Poisson equation (assuming a unit permittivity):
\begin{equation} \label{eq:vp_poisson}
    -\pdd{\phi}{x} = \rho = \sum_s\int\dv~f_s.
\end{equation}
We can assume a velocity map
\begin{equation}
    v = v(\cvpar),
\end{equation}
such that equation~\ref{eq:vp} becomes
\begin{equation} \label{eq:vp_vmap}
    \pd{f_s}{t} + \pd{}{x}\dot{x}f_s + \frac{1}{v'}\pd{}{\cvpar}\dot{v}f_s = 0,
\end{equation}
and $v'=\partial_\cvpar v$. Let us multiply this equation by $v'$ and introduce the proxy distribution $F_s=v' f_s$. Projecting this equation onto the DG basis $\pb{\ell}_j$, integrating by parts, and using the transformation to logical coordinates $(\xlog{},\vlog)$:
\begin{equation}
    x = x_j + \frac{\Dx}{2}\xlog{},
    \qquad
    \cvpar = \cvpar_j + \frac{\Dcvpar}{2}\vlog,
\end{equation}
we obtain the weak Vlasov-Poisson equation:
\begin{equation} \label{eq:vp_dg}
    \int_{K_j}\dxlog\,\dvlog\,\pb{\ell}_j\pd{F_s}{t} + \frac{2}{\Dx}\int_{-1}^{1}\dvlog\,\pbpm{\ell}_jG_{x,j}^{\pm} + \frac{2}{\Dcvpar}\int_{-1}^{1}\dxlog\,\pbpm{\ell}_jG_{v,j}^{\pm} - \int_{-1}^{1}\dxlog\,\dvlog\,\left(\pd{\pb{\ell}_j}{x}\dot{x} + \pd{\pb{\ell}_j}{v}\frac{\dot{v}}{v'}\right)F_s = 0.
\end{equation}
Substituting in the DG expansion of $F_s$,
\begin{equation}
    F_{s,j} = \sum_{k=1}^{\numbP} F_{s,j}^{(k)}\pb{k}_j,
\end{equation}
and exploting the orthonormality of $\pb{\ell}_j$, one obtains an evolution equation for the expansion coefficients of $F_s$:
\begin{equation} \label{eq:vp_weak}
    \pd{F_{s,j}^{(k)}}{t} + \frac{2}{\Dx}\int_{-1}^{1}\dvlog\,\pbpm{\ell}_jG_{x,j}^{\pm} + \frac{2}{\Dcvpar}\int_{-1}^{1}\dxlog\,\pbpm{\ell}_jG_{v,j}^{\pm} - \int_{-1}^{1}\dxlog\,\dvlog\,\left(\pd{\pb{\ell}_j}{x}\dot{x} + \pd{\pb{\ell}_j}{v}\frac{\dot{v}}{v'}\right)F_s = 0.
\end{equation}
After computing the electrostatic potential $\phi$ via equation~\ref{eq:vp_poisson}, one can follow the same procedure used for the gyrokinetic equation. The volume term is computed analytically with computer algebra given that the integrand is just a product of polynomials. The surface terms are calculated by constructing a surface modal expansion of the numerical fluxes, $G_{x,j}^{\pm}$ and $G_{v,j}^{\pm}$, obtained by evaluating the fluxes in Flax-Friedrichs form at Gauss-Legendre nodes of the surface, and employing a nodal-to-modal transformation.

\section{Full-$f$ gyrokinetic model in field-aligned coordinates} \label{sec:gk_fac}

This appendix details the transformations leading from the gyrokinetic equation
\begin{equation} \label{eq:app_gkeq}
    \pd{\jacobP f_s}{t} + \div{\jacobP \v{\dot{R}} f_s} + \pd{}{\vpar}\jacobP \dot{\vpar} f_s = \jacobP\mathcal{C}_s + \jacobP\mathcal{S}_s + \jacobP\mathcal{D}_s,
\end{equation}
to the model in field-aligned coordinates used in Gkeyll. The advection velocities in equation~\ref{eq:app_gkeq}, $\v{\dot{R}}$ and $\dot{\vpar}$, are defined in terms of the Poisson bracket
\begin{equation} \label{eq:app_gkpb}
    \{A,H_s\} = \frac{\vBs}{m_s\Bspar}\cdot\left(\grad{A}\pd{H_s}{\vpar} - \pd{A}{\vpar}\grad{H_s}\right) + \frac{\bhat}{q_s\Bspar}\times\grad{H_s}\cdot\grad{A},
\end{equation}
and the particle Hamiltonian
$H_s = m_s\vpar^2/2 + \mu B + q_s\phi$ as:
\begin{eqnal} \label{eq:app_characteristics}
    \v{\dot{R}} &= \{\v{R},H_s\} = \frac{\vBs}{m_s\Bspar}\pd{H_s}{\vpar} + \frac{\bhat}{q_s\Bspar}\times\grad{H_s}, 
    \qquad
    \dot{\vpar} = \{\vpar,H_s\} = -\frac{\vBs}{m_s\Bspar}\cdot\grad{H_s},
\end{eqnal}
using the effective magnetic field
\begin{eqnal} \label{eq:app_vBs}
    \vBs &= \vB+\frac{m_s\vpar}{q_s}\curl{\bhat}.
\end{eqnal}
The electrostatic potential $\phi$ is obtained from the long-wavelength limit of the gyrokinetic quasineutrality equation
\begin{equation} \label{eq:app_gkPoisson}
    -\div{\epsilon_\perp\gradperp{\phi}} = \sum_sq_sn_s = \sum_sq_s\int_{\Omega_{\vv}}\dvv\,\jacobP f_s,
    \qquad
    \epsilon_\perp = \sum_s\frac{m_sn_{s0}}{B_0^2}.
\end{equation}
Here we use $\nabla_\perp=-\bhat\times\left(\bhat\times\nabla\right) = \nabla-\bhat\left(\bhat\cdot\nabla\right)$, the shorthand $\dvv=(2\pi/m_s)\,\dvpar\,\dmu$, and $\Omega_{\vv}$ for the domain in velocity space: $\left(\vpar,\mu\right)\in\left[-\infty,\infty\right]\times\left[0,\infty\right]$. We have also defined the guiding center density $n_s$ as the zeroth velocity moment of $\jacobP f_s$. The reference values $n_{0s}$ and $B_0$ are typically set to the global average at $t=0$ and are independent of time.

The collision term on the right-hand side of equation~\ref{eq:app_gkeq},
\begin{equation} \label{eq:app_colls}
    \mathcal{C}_s = \mathcal{C}_s^{\mathrm{el}} + \mathcal{C}_s^{\mathrm{iz}} + \mathcal{C}_s^{\mathrm{cx}} + \mathcal{C}_s^{\mathrm{rec}} + \mathcal{C}_s^{\mathrm{rad}},
\end{equation}
includes elastic collisions between charged species ($\mathcal{C}_s^{\mathrm{el}}$) and inelastic collisions and reactions (with neutrals), which at the moment encompass ionization ($\mathcal{C}_s^{\mathrm{iz}}$), charge exchange ($\mathcal{C}_s^{\mathrm{cx}}$), recombination ($\mathcal{C}_s^{\mathrm{rec}}$) and radiation line ($\mathcal{C}_s^{\mathrm{rad}}$). Our model of elastic charged-species collisions is typically a Dougherty operator~\cite{Dougherty1967,Francisquez2020}:
\begin{eqnal} \label{eq:app_lbo}
    \mathcal{C}^{\mathrm{el}}_s = \sum_r\nu_{sr}\,\left\{\pd{}{\vpar}\left[\left(\vpar-\uparsr\right) +\vtsr^2\pd{}{\vpar}\right] + \pd{}{\mu}2\mu\left(1+\frac{m\vtsr^2}{B}\pd{}{\mu}\right)\right\}f_s,
\end{eqnal}
where the sum over $r$ is over every charged species other than $s$, $\nu_{sr}$ is the frequency of collisions between species $s$ and species $r$, and $\uparsr$ and $\vtsr$ are the cross-species mean flow and thermal speeds that ensure momentum and energy conservation~\cite{Francisquez2022}. Models of interactions with neutral species and line radiation are described in separate publications~\cite{Bernard2022,Roeltgen2025}. Our source model usually has a Maxwellian or bi-Maxwellian form
\begin{eqnal}
    \mathcal{S}_s = \frac{\dot{n}_{s,\mathrm{src}}}{\left(2\pi\right)^{3/2}v_{t\parallel,\mathrm{src}}v_{t\perp,\mathrm{src}}^2}\left[-\frac{\left(\vpar-u_{\parallel,\mathrm{src}}\right)^2}{2v_{t\parallel,\mathrm{src}}^2}-\frac{2\mu B/m_s}{2v_{t\perp,\mathrm{src}}^2}\right],
\end{eqnal}
given a particle source rate ($\dot{n}_{s,\mathrm{src}}$), the source mean flow speed ($u_{\parallel\mathrm{src},s}$) and temperatures parallel ($T_{\parallel\mathrm{src},s}=m_sv_{t\parallel,\mathrm{src}}^2$) and perpendicular ($T_{\perp\mathrm{src},s}=m_sv_{t\perp,\mathrm{src}}^2$) to the background magnetic field. Lastly, the diffusion model used in the absence of self-consistent anomalous cross-field transport (only present in 3D) is simply
\begin{eqnal} \label{eq:app_diffusion}
    \mathcal{D}_s = \div{\left(D_s\cdot\grad{f_s}\right)},
\end{eqnal}
where $D_s$ is a rank-2 tensor. Incorporating initial (ICs) and boundary conditions (BCs), equations~\ref{eq:app_gkeq}-\ref{eq:app_diffusion} represents a closed system that can be integrated with a suitable numerical method.

\subsection*{Model in field-aligned coordinates} \label{sec:famodel}

A field-aligned curvilinear coordinate system~\cite{Beer1995} $\v{x}=\{x^i\}$ (here $i\in[1,\cdim]$ and $\cdim$ is the number of position-space dimensions) with a mapping to the Cartesian gyrocenter position $\v{R}$ given by $\v{R}(\v{x})$ transforms equation~\ref{eq:app_gkeq} to~\cite{Mandell2021thesis}
\begin{equation} \label{eq:app_gkeq_curvilinear}
    \pd{\jacobP f_s}{t} + \frac{1}{\jacobGeo}\pd{}{x^i}\jacobGeo\jacobP\dualv{i}\cdot\v{\dot{R}} f_s + \pd{}{\vpar}\jacobP \dot{\vpar} f_s = \jacobP\mathcal{C}_s + \jacobP\mathcal{S}_s + \jacobP\mathcal{D}_s,
\end{equation}
where $\tangv{i}=\partial\v{R}/\partial x^i$ and 
$\dualv{i}$
are the tangent (covariant) and dual (contravariant) basis vectors of the $\v{x}$ coordinates, defining the metric tensors $g_{ij}=\tangv{i}\cdot\tangv{j}$ and $g^{ij}=\dualv{i}\cdot\dualv{j}$, and the Jacobian $\jacobGeo=\mathrm{det}(g_{ij})^{1/2} = \mathrm{det}(g^{ij})^{-1/2}$~\cite{Dhaeseleer1991}. Repeated indices imply summation unless noted otherwise or where explicit summation appears. The $\v{R}(\vx)$ mapping is field aligned because the coordinates $x^i$ are chosen so that the magnetic field can be written as
\begin{equation}
    \vB = \cmag(\vx)~\ \dualv{1}\times\dualv{2},
\end{equation}
meaning $\vB$ is aligned with the third tangent vector $\tangv{3}$. The scalar function $\cmag(\vx)$ is constrained by the requirement that $\vB$ must be divergence-free:
\begin{eqnal}
    \div{\vB} &= \grad{\cmag(\vx)}\cdot \dualv{1}\times \dualv{2} + \cmag(\vx)\div{\dualv{1}\times\dualv{2}} = \frac{\vB}{\cmag(\vx)}\cdot\grad{\cmag(\vx)} + \cmag(\vx)\div{\left(\grad{x^1}\times\grad{x^2}\right)} = \frac{\vB}{\cmag(\vx)}\cdot\grad{\cmag(\vx)} \\
\end{eqnal}
since $\dualv{1}\times\dualv{2} = \grad{x^1}\times\grad{x^2} = \curl{\left(x^1\grad{x^2}\right)} - x^1\curl{\grad{x^2}}$ and we use the fact that $\curl{\grad{A}}$ and $\div{\curl{\v{A}}}$ are both zero (for some scalar $A$ and vector $\v{A}$ fields). Therefore, as long as $\vB\cdot\grad{\cmag}=0$, the magnetic field remains divergence free. Given that $\dualv{1}\times\dualv{2}=\tangv{3}/\jacobGeo$, the magnetic field is
\begin{equation} \label{eq:app_Bvec}
    \vB = 
\begin{dcases}
    B^i\tangv{i} = B^3\tangv{3} = \cmag\jacobGeo^{-1}\tangv{3} \qquad &\text{contravariant form}, \\
    B_j\dualv{j} = \cmag\jacobGeo^{-1}g_{3j}\dualv{j} \qquad &\text{covariant form},
\end{dcases}
\end{equation}
and the magnetic field amplitude is 
\begin{eqnal} \label{eq:app_bmag}
    B = \sqrt{g_{ij}B^iB^j} = \sqrt{g^{ij}B_iB_j} = \sqrt{B_iB^j} = \sqrt{g_{33}}\frac{\cmag}{\jacobGeo} = B^3\sqrt{g_{33}},
\end{eqnal}
which allows us to write the magnetic field unit vector as
\begin{equation} \label{eq:app_bhatcomps}
    \bhat = \frac{\vB}{B} =
\begin{dcases}
b^i\tangv{i} = g_{33}^{-1/2}\tangv{3} \qquad &\text{contravariant form}, \\
b_j\dualv{j} = g_{3j}g_{33}^{-1/2}\dualv{j} = [g_{3j}\cmag/(\jacobGeo B)]\dualv{j} \qquad &\text{covariant form}.
\end{dcases}
\end{equation}

In order to compute the contravariant components of $\v{\dot{R}}$ (i.e. $\dualv{i}\cdot\v{\dot{R}}$) appearing in equation~\ref{eq:app_gkeq_curvilinear}, in light of equation~\ref{eq:app_characteristics}, it is necessary to express the effective magnetic field in terms of its contravariant components:
\begin{eqnal} \label{eq:app_vBs_contra}
    B^{*i} &= \dualv{i}\cdot\vBs = B^3\delta^i_3+\frac{m_s\vpar}{q_s}\dualv{i}\cdot\curl{\bhat},
\end{eqnal}
where $\delta^k_{j}$ being the Kronecker delta, and we used the contravariant form for $\vB$ from equation~\ref{eq:app_Bvec}. We can now express the advection speeds (equation~\ref{eq:app_characteristics}) in terms of curvilinear coordinates as: 
\begin{eqnal} \label{eq:app_characteristics_curvilinear}
    \dualv{i}\cdot\v{\dot{R}} &= \frac{B^{*i}}{m_s\Bspar}\pd{H_s}{\vpar} + \frac{1}{q_s\Bspar}\frac{\epsilon^{ijk}}{\jacobGeo}b_j\pd{H_s}{x^k},
    \qquad
    \dot{\vpar} = -\frac{B^{*i}}{m_s\Bspar}\pd{H_s}{x^i}. 
\end{eqnal}
Note that the parallel component of $\vBs$ can be expressed in terms of its contravariant components as
\begin{eqnal} \label{eq:app_Bspar_curvilinear}
    \Bspar = \bhat\cdot\vBs = b_iB^{*i} = b_3\frac{\cmag}{\jacobGeo} + b_k\frac{m_s\vpar}{q_s}\frac{\epsilon^{ijk}}{\jacobGeo}\pd{b_j}{x^i} = B + b_k\frac{m_s\vpar}{q_s}\frac{\epsilon^{ijk}}{\jacobGeo}\pd{b_j}{x^i}.
\end{eqnal}

The collisionless advection terms also require the electrostatic potential $\phi$, for which we must also express the gyrokinetic quasineutrality equation (\ref{eq:app_gkPoisson}) in field-aligned coordinates. First, writing out the outer divergence on the left-hand side of equation~\ref{eq:app_gkPoisson} yields
\begin{eqnal}
    -\frac{1}{\jacobGeo}\pd{}{x^i}\jacobGeo\epsilon_\perp\,\dualv{i}\cdot\gradperp{\phi} = \sum_sq_sn_s.  
\end{eqnal}
Note that the inner product on the left side, given that $\gradperp{}=\grad{}-\bhat(\bhat\cdot\grad{})$, is
\begin{eqnal}
    \dualv{i}\cdot\gradperp{} 
    %
    %
    %
    %
    &= g^{ij}\pd{}{x^j}-\left(b^i\right)^2\pd{}{x^i}, \\
\end{eqnal}
and given the contravariant components of $\bhat$ in equation~\ref{eq:app_bhatcomps}, the expression for $\dualv{i}\cdot\gradperp{}$ becomes
\begin{eqnal}
    &\dualv{i}\cdot\gradperp{} = g^{ij}\pd{}{x^j}-\delta^i_3\left(\frac{\cmag}{\jacobGeo B}\right)^2\pd{}{x^3}.
\end{eqnal}
Therefore the gyrokinetic quasineutrality equation in field-aligned curvilinear coordinates is
\begin{eqnal} \label{eq:app_gkPoissonCurvilinear}
    -\frac{1}{\jacobGeo}\pd{}{x^i}\jacobGeo\epsilon_\perp\left[g^{ij}\pd{}{x^j}-\delta^i_3\left(\frac{\cmag}{\jacobGeo B}\right)^2\pd{}{x^3}\right]\phi = \sum_sq_sn_s.  
\end{eqnal}

Lastly, the collision and source terms on the right-hand side of equation~\ref{eq:app_gkeq_curvilinear} do not involve differential operators in position-space and are thus relatively unaffected by the introduction of field-aligned curvilinear coordinates. The diffusive model of anomalous transport, equation~\ref{eq:app_diffusion}, does need to be adjusted. However, rather than using straightforward algebra in curvilinear spaces~\cite{Dhaeseleer1991} for a diffusion tensor $D_s$ expressed in terms of Cartesian basis vectors, we first recognize that this operator is typically used to introduce {\it cross-field} diffusion in the absence of self-consistent anomalous diffusion (i.e. turbulence). 
Thus when we write the diffusion operator in terms of curvilinear coordinates:
\begin{eqnal}
    \mathcal{D}_s &= \frac{1}{\jacobGeo}\pd{}{x^i}\jacobGeo\dualv{i}\cdot\left(D_s\cdot\grad{f_s}\right) = \frac{1}{\jacobGeo}\pd{}{x^i}\jacobGeo\dualv{i}\cdot\left(D_{s,jk}\dualv{j}\dualv{k}\cdot\pd{f_s}{x^\ell}\dualv{\ell}\right), \\
    &= \frac{1}{\jacobGeo}\pd{}{x^i}\jacobGeo\left(g^{ij}D_{s,jk}g^{k\ell}\pd{f_s}{x^\ell}\right) = \frac{1}{\jacobGeo}\pd{}{x^i}\jacobGeo\left(g^{ij}{D_{s,j}}^{k}\pd{f_s}{x^k}\right), \\
    %
\end{eqnal}
we employ the mixed form of the elements of the diffusion tensor (${D_{s,j}}^k$) because these have units of $m^2/s$ and are what scientists typically input (or measure experimentally). Furthermore, since we are only interested in using this as a model of cross-field diffusion, we assume a diagonal tensor ${D_{s,j}}^k = D_{s,j}{\delta_{j}}^k$:
\begin{eqnal} \label{eq:app_diffModel}
    \mathcal{D}_s &= \frac{1}{\jacobGeo}\pd{}{x^i}\jacobGeo\left(g^{ij}D_{s,j}\pd{f_s}{x^j}\right).
\end{eqnal}

\subsection*{Axisymmetric, small $\rho^*$, $\kpar\ll\kperp$ limit}

The system in equations~\ref{eq:app_gkeq_curvilinear}-\ref{eq:app_diffModel} can be further simplified in several ways for certain systems of interest. First, many magnetic field topologies studied with this model are axisymmetric in the sense that the geometry and the magnetic field amplitude ($B$) does not depend on the binormal coordinate $x^2$. In that case the metric coefficients $g_{ij}$ and $g^{ij}$ are independent of $x^2$, and therefore so are the Jacobian $\jacobGeo$, the function $\cmag$(\v{x}), and the components of $\bhat$. Under axisymmetry, the $\partial_{x^2}b_i$ derivatives in $\curl{\bhat}$ of $\vBs$ (equation~\ref{eq:app_vBs_contra}) vanish, and the parallel component of the effective magnetic field is now
\begin{eqnal} \label{eq:app_BsparAxisymmetric}
    \Bspar = B + \frac{m_s\vpar}{q_s\jacobGeo}\left[-b_1\pd{b_2}{x^3} + b_2\left(\pd{b_1}{x^3}-\pd{b_3}{x^1}\right) + b_3\pd{b_2}{x^1}\right].
\end{eqnal}
Note that if $\vpar\sim\mathcal{O}\left(\vt\right)$ and $\curl{\bhat}\sim\mathcal{O}\left(\abs{\grad{B}}/B=1/L_B\right)$ then the magnitude of the second term in equation~\ref{eq:app_BsparAxisymmetric} relative to the first is $(m_s\vpar/q_s)\curl{\bhat}/B \sim m_s\vt/(q_s B L_B) \sim \rho_s/L_B$. In many experiments of interest (e.g. modern day tokamaks with small $\rho^*=\rho_s/a$, where $a$ is the minor radius), the magnetic field scale length $L_B$ is of the order of the major radius $R > a$, hence $(m_s\vpar/q_s)\curl{\bhat}/B \sim \rho_s/R \ll 1$ and we can approximate $\Bspar\approx B$. 

An additional simplification we will make pertains the gyrokinetic quasineutrality equation (\ref{eq:app_gkPoissonCurvilinear}), where the differential operator includes derivatives along $x^3$, i.e. along the magnetic field. However, in the gyrokinetic ordering $\kpar\ll\kperp$ (here $\kpar$ and $\kperp$ are the wavenumbers parallel and perpendicular to $\vB$) and we can simplify this field equation by dropping the derivatives along $x^3$. In this case the field-aligned quasineutrality equation becomes
\begin{eqnal} \label{eq:app_gkPoissonCurvilinearSimp}
    -\frac{1}{\jacobGeo}\pd{}{x^i}\jacobGeo\epsilon_\perp g^{ij}\pd{}{x^j}\phi = \sum_sq_sn_s,
    \qquad
    i,j\in\{1,2\}.
\end{eqnal}
Lastly, for perfectly axisymmetric simulations we can leverage  the same $\kpar\ll\kperp$ ordering to remove the field-aligned gradients from the diffusion operator in equation~\ref{eq:app_diffModel}, allowing for the simpler diffusion model:
\begin{eqnal} \label{eq:app_diffModel_final}
    \mathcal{D}_s &= \frac{1}{\jacobGeo}\pd{}{x^1}\jacobGeo\left(g^{11}D_{s,1}\pd{f_s}{x^1}\right).
\end{eqnal}

\section{Conservation properties} \label{sec:conserv}

\subsection{Collisionless terms} \label{sec:conserv_collisionless}

In this section we prove that the weak form of the gyrokinetic equation~\ref{eq:gkeq_weak_log_coeff}, with surface and volume terms computed as described in section~\ref{sec:algo_collisionless}, and the calculation of the electrostatic potential according to section~\ref{sec:quasineut_discrete}, conserves particles and energy exactly in the limit of infinite temporal resolution. 

\paragraph{Particle conservation}

In order to check that the total number of particles
\begin{equation}
    \mathcal{N} = \int_{\Omega}\dvx\,\dvv\,\jacobGeo\jacobP f = \frac{2\pi}{m}\frac{\Delta\vx\Dcvpar\Dcmu}{2^{\pdim}}\sum_{j=1}^{N}\int_{-1}^{1}\dvxlog\,\dvlog\,\dmulog\,\JTf_j = \mathbb{V}\sum_{j=1}^N\JTf_j^{(0)}
\end{equation}
is conserved ($\mathbb{V}=2\pi\Delta\vx\Dcvpar\Dcmu/(2^{\pdim}m)$ is a dimensional and normalization factor, and $\pdim=\cdim+2$), we can set $\pb{\ell}=1$ in equation~\ref{eq:gkeq_weak_log_coeff} and sum it over all $N=\Nx\Ny\Nz\Nvpar\Nmu$ cells. Moving the resulting surface terms to the right-hand side we obtain:
\begin{eqnal} \label{eq:Ndot}
    &\d{\mathcal{N}}{t} = \mathbb{V}\sum_{j=1}^{N}\pd{F_j^{(0)}}{t} = - \sum_{j=1}^{N}\frac{2}{\Dxi{i}}\int_{-1}^{1}\mathrm{d}\v{S_i}\,\dvlog\,\dmulog\,\Ghat^{\pm}_{x^i,j} - \sum_{j=1}^{N}\frac{2}{\Dcvpar}\int_{-1}^{1}\dvxlog\,\dmulog\,\Ghat^{\pm}_{\vpar,j}.
\end{eqnal}
Since the fluxes $\Ghat^{\pm}_{x^i,j}$ and $\Ghat^{\pm}_{\vpar,j}$ were constructed to be single-valued at cell surfaces, contributions to each of these terms from adjacent cells cancel each other. Only the contributions from cell surfaces on domain boundaries remain. We employ zero-flux BCs in velocity space so the domain-boundary contribution to the second term in equation~\ref{eq:Ndot} is zero. We are hence left with 
\begin{eqnal} \label{eq:Ndot_final}
    &\d{\mathcal{N}}{t} = \mathbb{V}\sum_{j=1}^{N}\pd{F_j^{(0)}}{t} = - \sum_{i=1}^{\cdim}\frac{2}{\Dxi{i}}\left(\sum_{j=1}^{\Nperp{x^i_{\max}}}\int_{-1}^{1}\mathrm{d}\v{S_i}\,\dvlog\,\dmulog\,\Ghat^{\pm}_{x^i,j}\Big|_{\xlog{i}=1} - \sum_{j=1}^{\Nperp{x^i_{\min}}}\int_{-1}^{1}\mathrm{d}\v{S_i}\,\dvlog\,\dmulog\,\Ghat^{\pm}_{x^i,j}\Big|_{\xlog{i}=-1}\right),
\end{eqnal}
where $\Nperp{x^i_{\min}}$ and $\Nperp{x^i_{\max}}$ is the number of cells at the first and last cell along $x^i$, respectively. If one uses periodic boundary conditions the contributions from the two terms in equation~\ref{eq:Ndot_final} cancel each other, or if one uses zero-flux BCs both of these terms are set  to zero. In these cases $\dot{\mathcal{N}}=0$ and particles are conserved exactly.

\paragraph{Energy conservation} \label{sec:algo_collisionless_Econservation}

We show that the total, discrete energy of the system $\mathcal{E}=\mathcal{E}_H-\mathcal{E}_{\phi}$, composed of the particle energy
\begin{equation} \label{eq:Eparticle}
    \mathcal{E}_H = \sum_s\int_{\Omega}\dvR\,\dvv\,H_s\jacobP f_s = \sum_s\frac{2\pi}{m_s}\int_{\Omega}\dvx\,\dcvpar\,\dcmu\,H_s\JTf_s
\end{equation}
and the field energy
\begin{equation} \label{eq:Efield}
    \mathcal{E}_\phi = \int_{\Omega_{\vx}}\dvR\,\frac{\epsilon_\perp}{2}\abs{\gradperp{\phi}}^2 = \int_{\Omega_{\vx}}\dvx\,\jacobGeo\frac{\epsilon_\perp}{2}\abs{\gradperp{\phi}}^2
\end{equation}
is conserved. The time rate of change of the total energy is
\begin{eqnal} \label{eq:Edot}
    \d{\mathcal{E}}{t} &= \underbrace{\sum_s\frac{2\pi}{m_s}\int_{\Omega}\dvx\,\dcvpar\,\dcmu\,H_s\pd{\JTf_s}{t}}_{\dot{\mathcal{E}}_{1}} + \underbrace{\sum_s\frac{2\pi}{m_s}\int_{\Omega}\dvx\,\dcvpar\,\dcmu\,\pd{H_s}{t}\JTf_s - \int_{\Omega_{\vx}}\dvx\,\jacobGeo\epsilon_\perp\left(\gradperp{\phi}\right)\cdot\pd{\gradperp{\phi}}{t}}_{\dot{\mathcal{E}}_{2}}.
\end{eqnal}
The first of these terms ($\dot{\mathcal{E}}_{1}$) can be obtained by substituting $\pb{\ell}_j=(2\pi/m_s)H_{s,j}$ into our weak scheme, equation~\ref{eq:gkeq_weak_log}, and summing over all species and cells:
\begin{eqnal} \label{eq:Edot1}
    &\dot{\mathcal{E}}_{1} = \sum_{s}\mathbb{V}_s\sum_{j=1}^{N}\int_{K_j}\dvxlog\,\dvlog\,\dmulog\,H_{s,j}\pd{\JTf_{s,j}}{t} = - \underbrace{\sum_{s}\mathbb{V}_s\sum_{j=1}^{N}\frac{2}{\Dxi{i}}\int_{-1}^{1}\mathrm{d}\v{S_i}\,\dvlog\,\dmulog\,H_{s,j\pm}\Ghat^{\pm}_{x^i,j}}_{\dot{\mathcal{E}}_{1,\xdotiSurf{i}}} \\
    &\quad- \underbrace{\sum_{s}\mathbb{V}_s\sum_{j=1}^{N}\frac{2}{\Dcvpar}\int_{-1}^{1}\dvxlog\,\dmulog\,H_{s,j\pm}\Ghat^{\pm}_{\vpar,j}}_{\dot{\mathcal{E}}_{1,\vpardotSurf}} + \underbrace{\sum_{s}\mathbb{V}_s\sum_{j=1}^{N}\int_{-1}^{1}\dvxlog\,\dvlog\,\dmulog\,\left(\frac{2}{\Dxi{i}}\pd{H_{s,j}}{\xlog{i}}\xdoti{i} + \frac{2}{\Dcvpar}\pd{H_{s,j}}{\vlog}\frac{\vpardot}{\vpar'}\right)\JTf_{s,j}}_{\dot{\mathcal{E}}_{1,\mathrm{vol}}}.
\end{eqnal}
The volume term (last term in equation~\ref{eq:Edot1}) cancels out due to the antisymmetry of the Poisson bracket. To see this substitute the definitions of $\xdoti{i}$ and $\vpardot$ in equation~\ref{eq:xdot_vpardot}:
\begin{eqnal}
    \dot{\mathcal{E}}_{1,\mathrm{vol}} &= \sum_{s}\mathbb{V}_s\sum_{j=1}^{N}\int_{-1}^{1}\dvxlog\,\dvlog\,\dmulog\,\left\{\frac{2}{\Dxi{i}}\pd{H_{s,j}}{\xlog{i}}\left[\left(g_{33}^{-1/2}\delta^i_3+\frac{m_s\vpar}{q_s}\frac{\dualv{i}\cdot\curl{\bhat}}{B}\right)\vpar + \frac{\epsilon^{i\ell p}}{q_s}\frac{b_\ell}{\jacobGeo B}\frac{2}{\Dxi{p}}\pd{H_{s,j}}{\xlog{p}}\right] \right. \\
    &\left.\quad+ \frac{2}{\Dcvpar}\pd{H_{s,j}}{\vlog}\frac{1}{\vpar'}\left[-\frac{1}{m_s}\left(g_{33}^{-1/2}\delta^i_3+\frac{m_s\vpar}{q_s}\frac{\dualv{i}\cdot\curl{\bhat}}{B}\right)\frac{2}{\Dxi{i}}\pd{H_{s,j}}{\xlog{i}}\right]\right\}\JTf_{s,j}, \\
    %
    %
    &= -\sum_{s}\mathbb{V}_s\sum_{j=1}^{N}\int_{-1}^{1}\dvxlog\,\dvlog\,\dmulog\,\left(\frac{2}{\Dxi{i}}\pd{H_{s,j}}{\xlog{i}}\frac{\epsilon^{i\ell p}}{q_s}\frac{b_\ell}{\jacobGeo B}\frac{2}{\Dxi{p}}\pd{H_{s,j}}{\xlog{p}}\right)\JTf_{s,j}, \\
    &= -\sum_{s}\mathbb{V}_s\sum_{j=1}^{N}\int_{-1}^{1}\dvxlog\,\dvlog\,\dmulog\,\frac{1}{q\Bspar\jacobGeo}\left[\frac{2}{\Dxi{i}}\pd{H_{s,j}}{\xlog{i}}\left(\epsilon^{\ell i1}b_\ell\frac{2}{\Dxi{1}}\pd{H_{s,j}}{\xlog{1}} + \epsilon^{\ell i2}b_\ell\frac{2}{\Dxi{2}}\pd{H_{s,j}}{\xlog{2}} + \epsilon^{\ell i3}b_\ell\frac{2}{\Dxi{3}}\pd{H_{s,j}}{\xlog{3}}\right)\right]\JTf_{s,j}, \\
    %
    &= 0,
\end{eqnal}
where we used $\vpar=(\partial_{\cvpar}H)/(m\vpar')=2(\partial_{\vlog}H)/(m\vpar'\Dcvpar)$. The two surface terms in equation~\ref{eq:Edot1} ($\dot{\mathcal{E}}_{1,\xdotiSurf{i}}$ and $\dot{\mathcal{E}}_{1,\vpardotSurf}$) make use of the $C^0$-continuity of the Hamiltonian ($H_{s,j\pm}$) and single-valued numerical surface fluxes $\Ghat^{\pm}_{x^i,j}$ and $\Ghat^{\pm}_{\vpar,j}$, to cancel contributions from adjacent cells at common boundaries. We also make use of zero-flux BCs at $\vpar$ boundaries, such that the $\dot{\mathcal{E}}_{1,\vpardotSurf}$ term vanishes and equation~\ref{eq:Edot1} becomes
\begin{eqnal} \label{eq:Edot1_final}
    &\dot{\mathcal{E}}_{1} = - \sum_{s}\mathbb{V}_s\sum_{i=1}^{\cdim}\left(\sum_{j=1}^{\Nperp{x^i_{\max}}}\frac{2}{\Dxi{i}}\int_{-1}^{1}\mathrm{d}\v{S_i}\,\dvlog\,\dmulog\,H_{s,j\pm}\Ghat^{\pm}_{x^i,j}\Big|_{\xlog{i}=1} - \sum_{j=1}^{\Nperp{x^i_{\min}}}\frac{2}{\Dxi{i}}\int_{-1}^{1}\mathrm{d}\v{S_i}\,\dvlog\,\dmulog\,H_{s,j\pm}\Ghat^{\pm}_{x^i,j}\Big|_{\xlog{i}=-1}\right).
\end{eqnal}
Physically speaking, these terms are the surface integral of the particle energy flux through position-space boundaries.

The second and third terms in equation~\ref{eq:Edot} ($\dot{\mathcal{E}}_{2}$) involve exchange of energy between particles and fields. Since $\partial_t H_s=q_s\partial_t\mathcal{P}\left(\phi\right)$ is independent of velocity space, these terms can be written as
\begin{eqnal}
    \dot{\mathcal{E}}_{2} &= \sum_sq_s\int_{\Omega_{\vx}}\dvx\,\pd{\left(\mathcal{P}\left(\phi\right)\right)}{t}\left(\frac{2\pi}{m_s}\int_{\Omega_{\vv}}\dcvpar\,\dcmu\,\JTf_s\right) - \int_{\Omega_{\vx}}\dvx\,\jacobGeo\epsilon_\perp\left(\gradperp{\phi}\right)\cdot\pd{\gradperp{\phi}}{t}, \\
    &= \int_{\Omega_{\vx}}\dvx\,\left(\sum_sq_s\jacobGeo n_s\right)\mathcal{P}\left(\pd{\phi}{t}\right) - \int_{\Omega_{\vx}}\dvx\,\jacobGeo\epsilon_\perp\left(\gradperp{\phi}\right)\cdot\pd{\gradperp{\phi}}{t}.
\end{eqnal}
At this point we use the self-adjoint property of $\mathcal{P}$ (equation~\ref{eq:fem_parproj_adjoint}) to transfer it from $\dot{\phi}$ to the charge density:
\begin{eqnal} \label{eq:Edot2_tmp}
    \dot{\mathcal{E}}_{2} 
    &= \int_{\Omega_{\vx}}\dvx\,\pd{\phi}{t}\mathcal{P}\left(\sum_sq_s\jacobGeo n_s\right) - \int_{\Omega_{\vx}}\dvx\,\jacobGeo\epsilon_\perp\left(\gradperp{\phi}\right)\cdot\pd{\gradperp{\phi}}{t}, \\
    &= \int_{\Omega_{\vx}}\dvx\,\pd{\phi}{t}\overline{\varrho} - \int_{\Omega_{\vx}}\dvx\,\varepsilon^{ij}\pd{\phi}{x^i}\left(\pd{}{x^j}\pd{\phi}{t}\right), \qquad i,j\in\{1,2\}.
\end{eqnal}
We can employ the quasineutrality equation to cancel these two terms. First, we break up the $x^3$ integrals into cell-wise contributions:
\begin{eqnal} \label{eq:Edot2_tmp_v2}
    \dot{\mathcal{E}}_{2} &= \sum_{k=1}^{\Nz}\int_{x^3_{k-1/2}}^{x^3_{k+1/2}}\dx^3\left[\int_{\Omega_{x^1,x^2}}\dx^1\,\dx^2\,\pd{\phi_{k}}{t}\overline{\varrho}_{k} - \int_{\Omega_{x^1,x^2}}\dx^1\,\dx^2\,\left(\pd{}{x^j}\pd{\phi_k}{t}\right)\varepsilon^{ij}_k\pd{\phi_k}{x^i}\right].
\end{eqnal}
Then we employ the weak quasineutrality equation~\ref{eq:gkPoissonCurvilinearSimpJmulWeak} in the first term by replacing $\febxy{m}_k=\partial_t\phi_{k}$ (since $\phi_k$ is in the space spanned by the $\febxy{m}_k$ basis set) to obtain
\begin{eqnal} \label{eq:Edot2_tmp_v4}
    \dot{\mathcal{E}}_{2} &= \sum_{k=1}^{\Nz}\int_{x^3_{k-1/2}}^{x^3_{k+1/2}}\dx^3\left[-\int_{\Omega_{x^2}}\dx^2\,\pd{\phi_k}{t}\varepsilon_k^{1q}\pd{\phi_k}{x^q}\Bigg|_{x^1=x^1_{\min}}^{x^1=x^1_{\max}} - \int_{\Omega_{x^1}}\dx^1\,\pd{\phi_k}{t}\varepsilon_k^{2q}\pd{\phi_k}{x^q}\Bigg|_{x^2=x^2_{\min}}^{x^2=x^2_{\max}} \right. \\
    &\left.\quad+\int_{\Omega_{x^1,x^2}}\dx^1\,\dx^2\,\left(\pd{}{x^p}\pd{\phi_k}{t}\right)\varepsilon_k^{pq}\pd{\phi_k}{x^q} - \int_{\Omega_{x^1,x^2}}\dx^1\,\dx^2\,\left(\pd{}{x^j}\pd{\phi_k}{t}\right)\varepsilon^{ij}_k\pd{\phi_k}{x^i}\right].
\end{eqnal}
For time-independent homogeneous BCs, the first two terms vanish; the last two terms cancel each other.

We can therefore conclude that the discrete energy change of our DG gyrokinetic system is
\begin{eqnal} \label{eq:EdotFinal}
    \d{\mathcal{E}}{t} &= - \sum_{s}\mathbb{V}_s\sum_{i=1}^{\cdim}\left(\sum_{j=1}^{\Nperp{x^i_{\max}}}\frac{2}{\Dxi{i}}\int_{-1}^{1}\mathrm{d}\v{S_i}\,\dvlog\,\dmulog\,H_{s,j\pm}\Ghat^{\pm}_{x^i,j}\Big|_{\xlog{i}=1} - \sum_{j=1}^{\Nperp{x^i_{\min}}}\frac{2}{\Dxi{i}}\int_{-1}^{1}\mathrm{d}\v{S_i}\,\dvlog\,\dmulog\,H_{s,j\pm}\Ghat^{\pm}_{x^i,j}\Big|_{\xlog{i}=-1}\right),
\end{eqnal}
that is, the only energy change in our system is due to energy fluxes through our domain boundaries.

\subsection{Collisional terms} \label{eq:conserv_collisions}

The discrete Dougherty operator described in section~\ref{sec:algo_collisions} can conserve particle, momentum and energy density exactly (independently of phase-space resolution) provided that the primitive moments $\upar$ and $\vt^2$ satisfy discrete constraints described in this section. This approach to computing $\upar$ and $\vt^2$ was described for like-species~\cite{Francisquez2020} and multi-species~\cite{Francisquez2022} collisions in earlier publications, and only involves minor modifications to accommodate the mapped velocity coordinates introduced in this work.

\paragraph{Particle conservation}

In order to show that the change in particle number density (times the Jacobian of $\v{R}(\vx)$)
\begin{eqnal} \label{eq:gklbo_nonuni_m0}
    \jacobGeo M_0 &= \frac{2\pi}{m}\int_{v_{\parallel\min}}^{v_{\parallel\max}}\int_0^{\mu_{\max}}\dvpar\,\dmu\,\jacobGeo\jacobP f = \frac{2\pi}{m}\int_{\cvpar_{\min}}^{\cvpar_{\max}}\int_{\cmu_{\min}}^{\cmu_{\max}}\dcvpar\,\dcmu\,\JTf \\
    &= \frac{2\pi}{m}\sum_{i,j=1}^{\Nvpar,\Nmu}\int_{\cvpar_{i-1/2}}^{\cvpar_{i+1/2}}\int_{\cmu_{j-1/2}}^{\cmu_{j+1/2}}\dcvpar\,\dcmu\,\JTf = \frac{2\pi}{m}\sum_{i,j=1}^{\Nvpar,\Nmu}\frac{\Dcvpar\Dcmu}{4} \int_{-1}^{1}\dvlog\,\dmulog\,\JTf_{i,j}
\end{eqnal}
over time due to the discrete Dougherty operator
\begin{eqnal}
    \pd{\left(\jacobGeo M_0\right)}{t}\Bigg|_{\mathcal{C}^{\mathrm{el}}} &
    = \frac{2\pi}{m}\sum_{i,j=1}^{\Nvpar,\Nmu}\frac{\Dcvpar\Dcmu}{4}\int_{K_{i,j}}\dvlog\,\dmulog\,\jacobTot\mathcal{C}^{\mathrm{el}}
\end{eqnal}
is zero in each position-space cell, replace $\pb{\ell}_j=(2\pi/m)\Dcvpar\Dcmu/4$ in equation~\ref{eq:gklbo_nonuni_weak_2ibp_log} and sum over all velocity space cells (we omit the position-space index since the collision operator only involves advection and diffusion in velocity-space):
\begin{eqnal}
    \pd{\left(\jacobGeo M_0\right)}{t}\Bigg|_{\mathcal{C}^{\mathrm{el}}} &= \frac{2\pi}{m}\frac{\Dcmu}{2}\sum_{i,j=1}^{\Nvpar,\Nmu}\oint_{\partial K_{i,j}}\dvxlog\,\dmulog\,\nu\left\{\left[\left(\vpar-\upar\right)_{\pm}\widehat{\frac{\JTf}{\vpar'}} + v_t^2\mu'\pd{}{\vpar}\widetilde{\frac{\JTf}{\jacobVel}}\Bigg|_{\pm}\right]\right\} \\
    &\quad+ \frac{2\pi}{m}\frac{\Dcvpar}{2}\sum_{i,j=1}^{\Nvpar,\Nmu}\oint_{\partial K_{i,j}}\dvxlog\,\dvlog\,\nu\left[ 2\mu_{\pm}\left(\widehat{\frac{\JTf}{\mu'}} + \frac{m\vt^2}{B}\vpar'\pd{}{\mu}\widetilde{\frac{\JTf}{\jacobVel}}\Bigg|_{\pm}\right)\right] = 0.
\end{eqnal}
The final equality is justified by our $C^0$-continuous advection or drag fluxes, and the use of a recovered $\JTf/\jacobVel$ with continuous derivatives at cell boundaries, such that the surface terms pairwise cancel when summed over velocity-space cells. The surface terms at the boundaries of velocity space vanish due to our zero-flux BCs (equation~\ref{eq:lbo_zeroflux_BCs}).

\paragraph{Momentum conservation}

The particle momentum density (multiplied by $\jacobGeo/m$)
\begin{eqnal} \label{eq:gklbo_nonuni_m1}
    \jacobGeo M_1 &= \frac{2\pi}{m}\int_{v_{\parallel\min}}^{v_{\parallel\max}}\int_0^{\mu_{\max}}\dvpar\,\dmu\,\vpar\jacobGeo\jacobP f = \frac{2\pi}{m}\int_{\cvpar_{\min}}^{\cvpar_{\max}}\int_{\cmu_{\min}}^{\cmu_{\max}}\dcvpar\,\dcmu\,\vpar\JTf \\
    &= \frac{2\pi}{m}\sum_{i,j=1}^{\Nvpar,\Nmu}\int_{\cvpar_{i-1/2}}^{\cvpar_{i+1/2}}\int_{\cmu_{j-1/2}}^{\cmu_{j+1/2}}\dcvpar\,\dcmu\,v_{\parallel i}\JTf = \frac{2\pi}{m}\sum_{i,j=1}^{\Nvpar,\Nmu}\frac{\Dcvpar\Dcmu}{4} \int_{-1}^{1}\dvlog\,\dmulog\,v_{\parallel i}\JTf_{i,j}
\end{eqnal}
can also be shown to have no temporal change due to the Dougherty operator
\begin{eqnal}
    \pd{\left(\jacobGeo M_1\right)}{t}\Bigg|_{\mathcal{C}^{\mathrm{el}}} &
    = \frac{2\pi}{m}\sum_{i,j=1}^{\Nvpar,\Nmu}\frac{\Dcvpar\Dcmu}{4}\int_{K_{i,j}}\dvlog\,\dmulog\,v_{\parallel i}\jacobTot\mathcal{C}^{\mathrm{el}}
\end{eqnal}
by setting $\pb{\ell}_j=(2\pi/m)\left(\Dcvpar\Dcmu/4\right)\vpar$ in equation~\ref{eq:gklbo_nonuni_weak_2ibp_log} and summing over all velocity-space cells:
\begin{eqnal} \label{eq:gklbo_nonuni_weak_2ibp_m1conserve}
    &\pd{\left(\jacobGeo M_1\right)}{t}\Bigg|_{\mathcal{C}^{\mathrm{el}}} = \frac{2\pi}{m}\sum_{i,j=1}^{\Nvpar,\Nmu}\frac{\Dcmu}{2}\oint_{\partial K_j}\dvxlog\,\dmulog\,\nu\left\{v_{\parallel j\pm}\left[\left(\vpar-\upar\right)_{\pm}\widehat{\frac{\JTf}{\vpar'}} + \vt^2\mu'\pd{}{\vpar}\widetilde{\frac{\JTf}{\jacobVel}}\Bigg|_{\pm}\right] - \frac{\vt^2\jacobVel}{\vpar'}\widetilde{\frac{\JTf}{\jacobVel}}\Bigg|_{\pm}\right\} \\
    &\quad-\frac{2\pi}{m}\sum_{i,j=1}^{\Nvpar,\Nmu}\frac{\Dcvpar\Dcmu}{4} \int_{K_j}\dvxlog\,\dvlog\,\dmulog\,\nu\left(\vpar-\upar\right)\JTf + \frac{2\pi}{m}\sum_{i,j=1}^{\Nvpar,\Nmu}\frac{\Dcvpar}{2}\oint_{\partial K_j}\dvxlog\,\dvlog\,\nu v_{\parallel j\pm} 2\mu_{\pm}\left[\widehat{\frac{\JTf}{\mu'}} + \frac{m\vt^2}{B}\vpar'\pd{}{\mu}\widetilde{\frac{\JTf}{\jacobVel}}\Bigg|_{\pm}\right].
\end{eqnal}
Inside the velocity-space domain, the surface flux terms cancel due to continuity and pairwise cancellation of fluxes on either side of an inter-cell surface. At the velocity-space domain boundary zero-flux BCs set the surface terms in square brackets to zero. Therefore, we are left with
\begin{eqnal}  \label{eq:gklbo_nonuni_weak_2ibp_m1conserve_new}
    \pd{\left(\jacobGeo M_1\right)}{t}\Bigg|_{\mathcal{C}^{\mathrm{el}}} &= \frac{2\pi}{m}\sum_{j=1}^{\Nmu}\frac{\Dcmu}{2}\int_{-1}^{1}\dvxlog\,\dmulog\,\nu\left(- \frac{\vt^2\jacobVel}{\vpar'}\frac{\JTf}{\jacobVel}\right)_{i=\Nvpar,\vlog=1} - \frac{2\pi}{m}\sum_{j=1}^{\Nmu}\frac{\Dcmu}{2}\int_{-1}^{1}\dvxlog\,\dmulog\,\nu\left(- \frac{\vt^2\jacobVel}{\vpar'}\frac{\JTf}{\jacobVel}\right)_{i=1,\vlog=-1} \\
    &\quad- \frac{2\pi}{m}\sum_{i,j=1}^{\Nvpar,\Nmu}\frac{\Dcvpar\Dcmu}{4} \int_{K_j}\dvxlog\,\dvlog\,\dmulog\,\nu\left(\vpar-\upar\right)\JTf.
\end{eqnal}
We can ensure that equation~\ref{eq:gklbo_nonuni_weak_2ibp_m1conserve_new} vanishes by enforcing the constraint
\begin{eqnal}  \label{eq:gklbo_nonuni_weak_m1constraint}
    \frac{2\pi}{m}\sum_{j=1}^{\Nmu}\frac{\Dcmu}{2}\int_{-1}^{1}\dmulog\,\nu\left(\frac{\vt^2}{\vpar'}\JTf\right)_{i=\Nvpar,\vlog+} - \frac{2\pi}{m}\sum_{j=1}^{\Nmu}\frac{\Dcmu}{2}\int_{-1}^{1}\dmulog\,\nu\left(\frac{\vt^2}{\vpar'}\JTf\right)_{i=1,\vlog-} + \nu\left(\jacobGeo M_1-\upar \jacobGeo M_0\right) = 0.
\end{eqnal}

\paragraph{Energy conservation}

We can show that the particle kinetic energy density multiplied by $\jacobGeo$ and divided by $m/2$
\begin{eqnal} \label{eq:gklbo_nonuni_m2}
    \jacobGeo M_2 &= \frac{2\pi}{m}\int_{v_{\parallel\min}}^{v_{\parallel\max}}\int_0^{\mu_{\max}}\dvpar\,\dmu\,\left(\vpar^2+\frac{2\mu B}{m}\right)\jacobGeo\jacobP f = \frac{2\pi}{m}\int_{\cvpar_{\min}}^{\cvpar_{\max}}\int_{\cmu_{\min}}^{\cmu_{\max}}\dcvpar\,\dcmu\,\left(\vpar^2+\frac{2\mu B}{m}\right)\JTf \\
    &= \frac{2\pi}{m}\sum_{i,j=1}^{\Nvpar,\Nmu}\int_{\cvpar_{i-1/2}}^{\cvpar_{i+1/2}}\int_{\cmu_{j-1/2}}^{\cmu_{j+1/2}}\dcvpar\,\dcmu\,\left(v_{\parallel i}^2+\frac{2\mu_j B}{m}\right)\JTf = \frac{2\pi}{m}\sum_{i,j=1}^{\Nvpar,\Nmu}\frac{\Dcvpar\Dcmu}{4} \int_{-1}^{1}\dvlog\,\dmulog\,\left(v_{\parallel i}^2+\frac{2\mu_j B}{m}\right)\JTf_{i,j}
\end{eqnal}
undergoes no change due to the Dougherty operator
\begin{eqnal}
    \pd{\left(\jacobGeo M_2\right)}{t}\Bigg|_{\mathcal{C}^{\mathrm{el}}} & 
    = \frac{2\pi}{m}\sum_{i,j=1}^{\Nvpar,\Nmu}\frac{\Dcvpar\Dcmu}{4}\int_{K_{i,j}}\dvlog\,\dmulog\,\left(v_{\parallel i}^2+\frac{2\mu_j B}{m}\right)\jacobTot\mathcal{C}^{\mathrm{el}} = 0,
\end{eqnal}
if we replace $\pb{\ell} = (2\pi/m)\left(\Dcvpar\Dcmu/4\right)\left(\vpar^2+2\mu B/m\right) =(2\pi/m)\left(\Dcvpar\Dcmu/4\right)v^2_{i,j}$ in equation~\ref{eq:gklbo_nonuni_weak_2ibp_log} and sum over all velocity space cells. Such substitution is enabled by our use of a hybrid basis that span a space containing $\vpar^2$~\cite{Hakim2020,Francisquez2020}. Doing so results in
\begin{eqnal} \label{eq:gklbo_nonuni_weak_2ibp_m2conserve}
    &\pd{\left(\jacobGeo M_2\right)}{t}\Bigg|_{\mathcal{C}^{\mathrm{el}}} = \frac{2\pi}{m}\sum_{i,j=1}^{\Nvpar,\Nmu}\frac{\Dcmu}{2}\oint_{\partial K_j}\dvxlog\,\dmulog\,\nu\left\{v^2_{i,j\pm}\left[\left(\vpar-\upar\right)_{\pm}\widehat{\frac{\JTf}{\vpar'}} + v_t^2\mu'\pd{}{\vpar}\widetilde{\frac{\JTf}{\jacobVel}}\Bigg|_{\pm}\right] - \frac{2}{\Dcvpar}\pd{v_{\parallel i}^2}{\vlog}\frac{v_t^2\jacobVel}{\vpar'^2}\widetilde{\frac{\JTf}{\jacobVel}}\Bigg|_{\pm}\right\} \\
    &\quad- \frac{2\pi}{m}\sum_{i,j=1}^{\Nvpar,\Nmu}\frac{\Dcvpar\Dcmu}{4}\int_{K_j}\dvxlog\,\dvlog\,\dmulog\,\nu\left[\frac{2}{\Dcvpar}\pd{v_{\parallel i}^2}{\vlog}\left(\vpar-\upar\right)\frac{\JTf}{\vpar'} - \left(\frac{2}{\vpar'\Dcvpar}\right)^2\pdd{v_{\parallel i}^2}{\vlog}v_t^2\JTf\right] \\
    &\quad+ \frac{2\pi}{m}\sum_{i,j=1}^{\Nvpar,\Nmu}\frac{\Dcvpar}{2}\oint_{\partial K_j}\dvxlog\,\dvlog\,\nu\left[v^2_{i,j\pm} 2\mu_{\pm}\left(\widehat{\frac{\JTf}{\mu'}} + \frac{m\vt^2}{B}\vpar'\pd{}{\mu}\widetilde{\frac{\JTf}{\jacobVel}}\Bigg|_{\pm}\right) - 4\mu\vt^2\frac{\jacobVel}{\mu'}\widetilde{\frac{\JTf}{\jacobVel}}\Bigg|_{\pm}\right] \\
    &\quad- \frac{2\pi}{m}\sum_{i,j=1}^{\Nvpar,\Nmu}\frac{\Dcvpar\Dcmu}{4}\int_{K_j}\dvxlog\,\dvlog\,\dmulog\,\nu \left[\frac{4B}{m}\mu\JTf - 4\left(\frac{2}{\mu'\Dcmu}\right)^2\pd{}{\mulog}\left(\pd{\mu_j}{\mulog}\mu\right)\vt^2\JTf\right].
\end{eqnal}
Since $v_{i,j}^2$ is continuous at cell boundaries in velocity-space, the surface flux terms containing $v_{i,j}^2$ vanish due to pairwise cancellations upon summation and due to zero-flux BCs at domain boundaries. We also make use of the fact that
\begin{eqnal}
    \frac{2}{\Dcvpar}\pd{v_{\parallel i}^2}{\vlog} = \pd{v_{\parallel i}^2}{\cvpar} = 2v_{\parallel i}'v_{\parallel i}, \qquad \left(\frac{2}{\Dcvpar}\right)^2\pdd{v_{\parallel i}^2}{\vlog} = \pdd{v_{\parallel i}^2}{\cvpar} = 2v_{\parallel i}'^2,
\end{eqnal}
to reduce equation~\ref{eq:gklbo_nonuni_weak_2ibp_m2conserve} to
\begin{eqnal} \label{eq:gklbo_nonuni_weak_2ibp_m2conserve_v2}
    &\pd{\left(\jacobGeo M_2\right)}{t}\Bigg|_{\mathcal{C}^{\mathrm{el}}} = \frac{2\pi}{m}\sum_{i,j=1}^{\Nvpar,\Nmu}\frac{\Dcmu}{2}\oint_{\partial K_{i,j}}\dvxlog\,\dmulog\,\nu\left(- 2v_{\parallel i}\frac{v_t^2\jacobVel}{\vpar'}\widetilde{\frac{\JTf}{\jacobVel}}\Bigg|_{\pm}\right) - \frac{2\pi}{m}\sum_{i,j=1}^{\Nvpar,\Nmu}\frac{\Dcvpar\Dcmu}{4}\int_{K_{i,j}}\dvxlog\,\dvlog\,\dmulog\,\nu\left[2v_{\parallel i}\left(\vpar-\upar\right)\JTf - 2v_t^2\JTf\right] \\
    &\quad+ \frac{2\pi}{m}\sum_{i,j=1}^{\Nvpar,\Nmu}\frac{\Dcvpar}{2}\oint_{\partial K_{i,j}}\dvxlog\,\dvlog\,\nu\left(- 4\mu\vt^2\frac{\jacobVel}{\mu'}\widetilde{\frac{\JTf}{\jacobVel}}\Bigg|_{\pm}\right) - \frac{2\pi}{m}\sum_{i,j=1}^{\Nvpar,\Nmu}\frac{\Dcvpar\Dcmu}{4}\int_{K_{i,j}}\dvxlog\,\dvlog\,\dmulog\,\nu \left[\frac{4B}{m}\mu\JTf - 4\vt^2\JTf\right].
\end{eqnal}
The recovered distribution, $\widetilde{\JTf/\jacobVel}$, is continuous across boundaries, and so are its multiplicative factors. Thus, the additional surface terms from the second integration by parts also pairwise cancel upon summation and only their contribution from the velocity boundaries remain. We can ensure that equation~\ref{eq:gklbo_nonuni_weak_2ibp_m2conserve_v2} equals zero by enforcing the constraint:
\begin{eqnal}
    &\frac{2\pi}{m}\sum_{j=1}^{\Nmu}\frac{\Dcmu}{2}\int_{-1}^{-1}\dmulog\,\nu\left( v_{\parallel i}\frac{v_t^2}{\vpar'}\JTf\right)\Bigg|_{i=1,\vlog=-1}^{i=\Nvpar,\vlog=1} + \frac{2\pi}{m}\sum_{i,j=1}^{\Nvpar,\Nmu}\frac{\Dcvpar\Dcmu}{4}\int_{K_{i,j}}\dvlog\,\dmulog\,\nu\left[v_{\parallel i}\left(\vpar-\upar\right)\JTf - \vt^2\JTf\right] \\
    &\quad+ \frac{2\pi}{m}\sum_{i=1}^{\Nvpar}\frac{\Dcvpar}{2}\int_{-1}^{1}\dvlog\,\nu\left( 2\mu\vt^2\frac{\jacobVel}{\mu'}\widetilde{\frac{\JTf}{\jacobVel}}\Bigg|_{j=1,\mulog=-1}^{j=\Nmu,\mulog=1}\right) + \frac{2\pi}{m}\sum_{i,j=1}^{\Nvpar,\Nmu}\frac{\Dcvpar\Dcmu}{4}\int_{K_{i,j}}\dvlog\,\dmulog\,\nu \left[\frac{2\mu B}{m}\JTf - 2\vt^2\JTf\right] = 0.
\end{eqnal}
Reorganizing terms and using the definitions of $M_0$, $M_1$ and $M_2$ we can write this constraint as
\begin{eqnal} \label{eq:gklbo_nonuni_weak_m2constraint}
    %
    %
    &\frac{2\pi}{m}\sum_{j=1}^{\Nmu}\frac{\Dcmu}{2}\int_{-1}^{-1}\dmulog\,\nu \frac{v_{\parallel i}v_t^2}{\vpar'}\JTf\Bigg|_{i=1,\vlog=-1}^{i=\Nvpar,\vlog=1} + \frac{2\pi}{m}\sum_{i=1}^{\Nvpar}\frac{\Dcvpar}{2}\int_{-1}^{1}\dvlog\,\nu\frac{2\mu\vt^2}{\mu'}\JTf\Bigg|_{j=1,\mulog=-1}^{j=\Nmu,\mulog=1} \\
    &\quad+ \nu\left(\jacobGeo M_2 - \upar\jacobGeo M_1 - 3\vt^2\jacobGeo M_0\right) = 0.
\end{eqnal}
If we compute the primitive moments $\upar$ and $\vt^2$ by solving the coupled linear equations~\ref{eq:gklbo_nonuni_weak_m1constraint} and~\ref{eq:gklbo_nonuni_weak_m2constraint} in each position-space cell we will ensure that the discrete Dougherty operator conserves momentum and energy exactly.


\bibliographystyle{elsarticle-num} 
\bibliography{main.bib}






\end{document}